\documentclass[%
 reprint, superscriptaddress, %groupedaddress, %unsortedaddress, %runinaddress, %frontmatterverbose, 
 nofootinbib, amsmath,amssymb, aps, prx %prb, %pra, %prb, %rmp, %prstab, %prstper, %floatfix, longbibliography
 ]{revtex4-2}
\usepackage{graphicx}
\usepackage{multirow}
\usepackage{comment}
\usepackage[usenames]{color}
\usepackage{graphicx}
\usepackage{diagbox}
\usepackage{tabularray}
\usepackage{bm}
\usepackage[colorlinks=true, allcolors=blue]{hyperref}
\usepackage{amsmath}
\usepackage{xcolor}
\usepackage{graphicx}% Include figure files
\usepackage{dcolumn}% Align table columns on decimal point
\usepackage{bm}% bold math
\usepackage{hyperref}% add hypertext capabilities
\usepackage{xr}
\graphicspath{ {./ } }

\newif\iffigures
\figurestrue
\begin{document}

\title{Honeybee-like collective decision making in a Kilobot swarm}

\author{David March-Pons}
\email{david.march@upc.edu}
\affiliation{Departament de Física, Universitat Politècnica de Catalunya, Campus Nord B4, 08034 Barcelona, Spain}

\author{Julia M\'ugica}
%\email{juliajmg@gmail.com}
\affiliation{Departament de Física, Universitat Politècnica de Catalunya, Campus Nord B4, 08034 Barcelona, Spain}

\author{Ezequiel E. Ferrero}
%\email{ezequiel.ferrero@ub.edu}
\affiliation{Departament de Física de la Matèria Condensada, Universitat de Barcelona, Martí i Franquès 1, 08028 Barcelona, Spain.}
\affiliation{Institute of Complex Systems (UBICS), Universitat de Barcelona, Barcelona, Spain
}
\affiliation{Instituto de Nanociencia y Nanotecnolog\'{\i}a, CNEA--CONICET, Centro At\'omico Bariloche, R8402AGP S. C. de Bariloche, R\'{\i}o Negro, Argentina.}

\author{M. Carmen Miguel}
%\email{carmen.miguel@ub.edu}
\affiliation{Departament de Física de la Matèria Condensada, Universitat de Barcelona, Martí i Franquès 1, 08028 Barcelona, Spain.}
\affiliation{Institute of Complex Systems (UBICS), Universitat de Barcelona, Barcelona, Spain
}

\begin{abstract}

Drawing inspiration from honeybee swarms' nest-site selection process, we assess the ability 
of a Kilobot robot swarm to replicate this captivating example of collective decision-making.
Honeybees locate the optimal site for their new nest by aggregating information about potential locations and exchanging it through their waggle-dance.
The complexity and elegance of solving this problem relies on two key abilities of scout honeybees: self-discovery and imitation, symbolizing {\em independence} and {\em interdependence}, respectively. 
We employ a mathematical model to represent this nest-site selection problem and program 
our Kilobots to follow its rules. 
Our experiments demonstrate that the Kilobot swarm can collectively reach consensus 
decisions in a decentralized manner, akin to honeybees. 
However, the strength of this consensus depends not only on the interplay between independence
and interdependence but also on critical factors such as swarm density and the motion of Kilobots. 
These factors enable the formation of a percolated communication network, through which 
each robot can receive information beyond its immediate vicinity.
By shedding light on this crucial layer of complexity --the crowding and mobility conditions
during the decision-making--, we emphasize the significance of factors typically overlooked 
but essential to living systems and life itself.

\end{abstract}

\maketitle

\section{Introduction}

%%%%%%%
%%% Opening. 
%%%%%%%

Collective decision-making is the process by which a group of agents makes a 
choice that cannot be directly attributed to any individual agent but rather 
to the collective as a whole~\cite{bose2017}.
This phenomenon is observed in both natural and artificial systems, and it is 
studied across various disciplines, including sociology, biology, and physics~\cite{dyer2009,sasaki2018}.
In particular, social insects have long been recognized for their fascinating 
behaviors, and collective decision-making is no exception. 
An intriguing example of this can be found in the way honeybees choose their nest
sites~\cite{frisch1954,honeybee_democracy,CouzinNature2005,DongScience2023}. 
This specific problem has been the focus of numerous models of collective decision-making
in honeybees~\cite{britton2002, passino2006, list2009, pais2013, reina2017, gray_multiagent_2018} and serves as 
an inspiration for our study.

%%%%%%%
%%% Generalities of collective decision making. Definition of consensus, quiality and cost.
%%%%%%%

The process of collective decision-making can encompass a virtually 
infinite number of choices. 
For instance, in flocking dynamics, individuals often have to converge 
on a common direction of motion~\cite{cavagna2010scale,Rosenthal2015, Chen2016, Calovi2018,mugica2022}. 
In such cases, achieving consensus in favor of an option is the result of a 
continuous process.
Another category of collective decision-making involves a 
finite and countable set of choices. 
Typical models in this category require a group of individuals to collectively 
determine the best option out of a set of $n$ available choices~\cite{seeley2001, valentini2017, reina2017, reina_voter_2023, bizyaeva_nonlinear_2023}.
Here, the consensus-reaching process becomes a discrete problem.
In real-life scenarios, examples of such decision-making processes include 
selecting foraging patches, travel routes or candidates in a democratic election. 
Within this countable set of choices, consensus is achieved when a {\it large majority}
of individuals in the group favor the same option. 
The threshold for what constitutes a `large majority' is typically defined by the 
experimenter but generally signifies a cohesive collective decision with more than 
50\% agreement among the individuals~\cite{valentini2017}.
In these collective decision-making models, each option is characterized by attributes 
that determine its relative desirability. 
For instance, in the context of selecting a potential nesting site for honeybees, 
these attributes could include size, distance, and vegetation type. 
Measures of {\it quality} and {\it cost} for each potential option encompass these attributes~\cite{seeley2001, honeybee_democracy, garnier2009}.
These two properties can be configured in multiple ways. 
The simplest scenario occurs when options' qualities and costs are the same for 
all available options, referred to as {\it symmetric}. 
In such a scenario, the group faces the challenge of breaking symmetry when 
selecting an option, often resulting from the amplification of random fluctuations~\cite{garnier2009,hamann_analysis_2012,reina2017}. 
In all other scenarios, the decision-making process is influenced by the 
specific combination of qualities and costs among different options. 
For example, in cases where the cost varies among options, making it {\it asymmetric},
but the qualities are symmetric, the option with the minimum cost is 
typically considered the best choice~\cite{schmickl2007}. 
Conversely, in situations with asymmetric qualities but symmetric costs, 
the option with the maximum quality tends to be chosen~\cite{valentini2014, reina2017, zakir_robot_2022, reina_voter_2023}.

%%%%%
%% Generalities about opinion dynamics modelling
%%%%%

Opinion dynamics models are proposed and developed to examine how individuals 
communicate and make decisions within groups. 
These models consist of a group of agents, each with their own instantaneous 
opinion or `state'. 
Individuals interact with one another and revise their opinions based on the 
opinions of others. 
Among the simplest models used to study collective decision-making are the 
well-known {\it voter model}~\cite{holley1975,castellano2009} or the 
{\it majority rule model}~\cite{galam2002}. 
These models are limited in that they assume that all agents are equally 
likely to adopt either opinion. 
In reality, however, individuals may have different preferences, beliefs, or 
biases that can influence their decision-making. 
To address this limitation, more complex models have been developed that 
incorporate features such as stubbornness, partisanship, and heterogeneity~\cite{galam2008,deffuant2000, redner_reality-inspired_2019, de_marzo_emergence_2020, bizyaeva_nonlinear_2023}. 
These models generally help to understand how opinion diversity and 
polarization can arise in groups, and how these dynamics might be 
influenced by different factors~\cite{franci_analysis_2021, Leonard2024}.
%%% Particularidad de los honeybee inspired models
In recent years, the collective decision-making process observed among nest-hunting honeybees has triggered the investigation of numerous collective decision-making models. In these models, agents can adopt a wide variety of individual and social behaviors~\cite{list2009, valentini2014, reina_desing_pattern, gray_multiagent_2018}. They engage in exploration to discover options, and once committed, they recruit uncommitted peers. Conversely, they may also retract their commitment and return to a neutral state or engage in cross-inhibition against opinions different from their own. Importantly, they assess the qualities of the different options, which in turn may modulate each of the aforementioned behaviors~\cite{passino_swarm_2008, honeybee_democracy}.

%%%%%
%% Use of robots and problem specification
%%%%%

Taking inspiration from a simple honeybee-like collective decision-making model~\cite{list2009}, our primary focus is to determine whether autonomous mini-robots, specifically Kilobots, can achieve high levels of consensus for the best quality option. These robots have been extensively used in the field to design and evaluate decision-making models, such as the naming game \cite{trianni2016_naming_game}, as well as other comparable 
honeybee-inspired models \cite{valentini2016,Reina2018_spatiality, zakir_robot_2022}.
These models adopt different approaches to modeling the collective decision-making process; these variations and their implications will be elaborated in the following sections.

As it will emerge from our analysis, the interaction topology of agents plays a crucial role in realistic representations of collective decision-making, adding complexity beyond inherent opinion dynamics. Recent studies have shown that scale-free networks offer better accuracy compared to networks based on agent proximity \cite{khaluf_interaction_models_2018}. However, agents with constrained communication capabilities can only generate limited interaction patterns, restricting the emergence of complex network features such as clustering or small-world properties. By introducing mobility, agents can enhance their communication capabilities \cite{dimidov2016}, motivating a quantitative study of the communication patterns established with this approach.
Our novel aspect resides in exhaustively evaluating the interplay between the evolution of the group consensus, under the simple assumptions of our model, and its relation with the dynamic communication network.

Aside from the general insights we can gain from studying opinion dynamics 
with moving individuals, this scenario possesses the distinctive feature 
of being closer to the behavior observed in the social animal world, 
where a diversity of intriguing signalling mechanisms have been 
previously identified~\cite{johnstone1997}.
In particular, crowding and clustering effects stand out as highly relevant
from our study. 
As they were shown to be crucial for honeybees in trophallaxis~\cite{Fard2020},
they might also be decisive in communication and collective decision making.

In the following, we motivate our work inspired by the honeybees' house 
hunting problem in Sec.~\ref{sec:beesmodel}, where we also present the 
particular discrete opinion-dynamics model under scrutiny.
In Sec.~\ref{sec:Kilobots}, we introduce our experimental study system, 
a {\it Kilobots} swarm, and their emulator, {\it Kilombo}.
Sections \ref{sec:consensus_reaching} and \ref{sec:swarm_analysis} present our 
main results, which combine experiments on Kilobots, numerical results on the 
Kilombo emulator, numerical simulations of the model on specific geometries, 
and their comparison to analytical results of the opinion-dynamics problem.
Finally, in Sec.~\ref{sec:discussion}, we provide a discussion of our results 
and perspectives.
Technical details about the model and the experimental setup are provided in
Appendix~\ref{sec:methods}, and additional complementary analysis and data 
are included in Appendices~\ref{app:botsseen} and \ref{app:networks}.

\begin{figure*}[t!]
\centering
\iffigures
\includegraphics[width=0.95\textwidth]{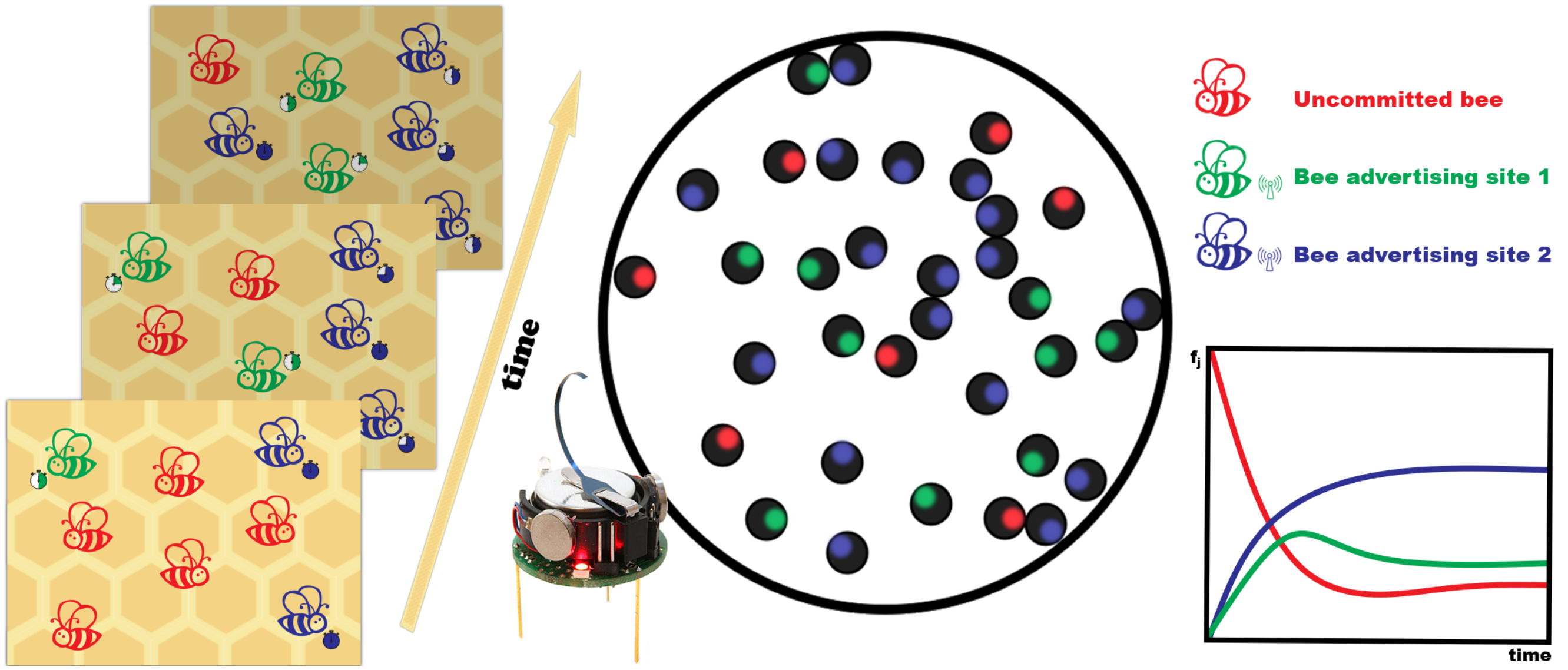}
\fi
  \caption{
  {\it Schematics of the collective decision dynamics under scrutiny.}
  The left column represents the dynamics of the decision model: 
  at a given instant of time committed bees are advertising their state 
  to convince undecided bees to commit to their option. 
  The time they spend advertising their option will be directly related to 
  the quality they perceive for that option. 
  As time advances a build up for the better option is expected since it is 
  benefited from longer advertisement periods. 
  The central column shows a schematic representation of our Kilobot experiments. 
  Each Kilobot has a LED that indicates its state, either uncommitted or 
  committed to (and advertising for) site 1 or 2.
  The right column displays schematically how the different proportions of bees
  on each state evolve in time, starting from a totally uncommitted population. 
}
\label{fig:schematics}
\end{figure*}

\section{Modeling the nest site selection problem in honeybees}
\label{sec:beesmodel}

%%% Scout bees, the waggle dance and collective decision-making

Honeybees are social insects that reside in large colonies. 
The way scout honeybees select new nest sites represents an interesting example 
of a collective decision-making process that involves a combination of individual 
and group behaviors. 
In recent years, researchers have made significant progress in understanding the 
mechanisms behind this behavior, which is crucial for the survival and reproduction 
of honeybee colonies~\cite{seeley2012, beekman2018, DongScience2023}.

The process of collective decision-making in honeybees has primarily been 
studied in the species \textit{Apis mellifera}. 
Towards the end of spring, honeybee colonies split, with approximately 
two-thirds of the colony leaving the nest along with the queen in search of a new 
nesting site. 
During this process, a fraction of the swarm scouts the surroundings to gather 
information about potential new sites, assessing their quality based on traits 
such as size, food availability, or the degree of concealment
%~\cite{seeley2010, seeley2012}.
~\cite{honeybee_democracy, seeley2012}.
When a scout bee discovers a promising new nesting site, it returns to the swarm and communicates information about the site fitness and location to other bees through the intricate waggle dance~\cite{seeley2001, dyer2002}. Doing so, she may recruit other scout bees that remained in the swarm to also explore - and subsequently advertise - the same location.
The duration of the waggle dance is correlated with the honeybee's perception 
of the site's quality. 
A longer and more animated dance indicates a more suitable nest site, while a shorter
and less dynamic dance corresponds to a less desirable site~\cite{seeley1997}. 
Consequently, high-quality sites receive longer and more frequent advertising, 
while low-quality sites see reduced attention, resulting in an overall increase 
in the number of bees visiting and dancing for high-quality sites and a decrease 
in those doing so for low-quality ones.

In addition to their quality, each site also has its associated cost, 
which represents the likelihood that a scout bee will discover the site, 
considering factors such as distance or concealment. 
This leads to the possibility that some high-quality options may go 
unnoticed due to their associated costs.
Over time, the dances performed by the honeybees tend to converge on a 
single site, and once a potential nest site has attracted a sufficient 
number of bees, a quorum is formed, ideally in favor of the best 
available option. 
This entire process ensures that the migrating part of the colony moves 
together to their new home~\cite{beekman2018}.

\subsection*{Model of a fully-connected scout bee network}
Several mathematical models of the honeybee nest-site selection problem have 
been proposed in the
literature~\cite{britton2002,myerscough2003,passino2006,perdriau2007,list2009,pais2013,valentini2014, reina_desing_pattern, reina2017, gray_multiagent_2018}).
In particular, List, Elsholtz, and Seeley~\cite{list2009} introduced an agent-based
model inspired by the decision-making process of honeybees. 
This model integrates both an individual's self-discovery of potential nest sites 
and the existing interdependencies, 
which encompass interactions among bees, leading to imitation and the adoption of 
sites presented by other bees. 
The model explicitly incorporates various parameters, including the number of sites, 
site quality, site self-discovery probabilities, and group interdependence.
Here, we adopt a very similar approach to describe the rules that govern the behavior 
of our Kilobot robots.

%%%%%
%%LES model presentation
%%%%%

The nest-site choice model proposed in~\cite{list2009} consists of a 
swarm of $N$ scout bees that reach consensus and collectively decide on 
one of the potential nest sites, labeled $1,2,3,...,k$. 
Each site $j$ has an intrinsic quality $q_j \geq 0$ that determines the 
time a bee spends advertising site $j$ through the waggle dance.
At time $t$, a bee can either be dancing for one of the $k$ sites (i.e., promoting it) 
or not dancing for any site, indicating that it is still searching for a site, 
observing other bees' dances, or simply resting. 
Formally, a vector $x_{i,t} = (s_{i,t}, d_{i,t})$ represents the state of bee $i$ 
at time $t$, where $s_{i,t} = 0$ if bee $i$ is not dancing for any site. 
When a bee is dancing, $s_{i,t}=1,\ldots,k$, indicating the site the bee is promoting, 
and $d_{i,t}$ represents the remaining duration of bee $i$'s waggle dance.
In each discrete time-step (which sets the unit of time in the model), the states of bees 
are updated in parallel.
While a bee is dancing for a site, its dance duration $d_{i,t}$ decreases by one time 
unit in each time-step until it reaches zero, at which point the bee stops advertising, 
and its state returns to the non-dancing value $s_{i,t}=0$.
Non-dancing bees have a probability $p_{i,t+1}$ of starting to dance for site $j$ 
at time $t+1$. 
When this occurs, $s_{i,t+1}=j$, and the duration of the new dance is set equal 
to or proportional to the site quality (we have simplified the 
original model; for more nuanced details on this step, refer to ~\cite{list2009}). Here we use $d_{j,t+1}=q_j$. 
The probability $p_{j,t+1}$ estimates the likelihood of a bee finding site $j$ 
and committing to advertising it. 
It is calculated as follows:

\begin{equation}
	p_{j,t+1} = (1 - \lambda)\pi_j + \lambda f_{j,t} .
	\label{eq:list_probs}
\end{equation}
Here, $\pi_j$ represents the a priori self-discovery probability of site $j$, 
i.e., the success rate of a scout bee targeting an option in the environment is simply incorporated into the system through this model parameter.
$\lambda$ denotes the bees' interdependence, and $f_{j,t}$ represents the proportion 
of bees already dancing for site $j$ at time $t$. As probabilities, the condition $\sum_{j=0}^k p_{j,t+1} = 1$ must be satisfied. Similarly, the fractions $f_{j,t}$ must satisfy a normalization condition $\sum_{j=0}^k f_{j,t} = 1$. Moreover, it's important to note that when $\lambda = 0$, the probability transitions are solely determined by $\pi_j$. Hence, the condition $\sum_{j=0}^k \pi_j \leq 1$ must also be met. This inequality accounts for the possibility that bees may fail to commit to any site and remain neutral for another time step (i.e., $p_{0,t+1} \neq 0$).
The interdependence parameter $\lambda$ ranges between $0$ and $1$, 
determining the extent to which bees rely on each other to decide to dance for a site. 
When $\lambda = 0$, the probability of committing to site $j$ depends solely on the 
self-discovery probability $\pi_j$, regardless of the proportion of bees dancing for it.
Conversely, as $\lambda$ approaches $1$, the probability of committing to site $j$ 
at time $t+1$ becomes almost entirely dependent on the proportion of bees already 
dancing for it at time $t$, denoted as $f_{j,t}$. 
In other words, a higher value of $\lambda$ means that committing probability 
relies more on imitation of other 
bees. For further details, please refer to Appendix~\ref{app:analytical}.
The self-discovery probabilities of available sites $\pi_j$ are chosen in a way that 
ensures $\sum_{j=1}^k \pi_j < 1$, and in general, the sum does not exceed the maximum 
value of approximately 0.6, which corresponds to a 60\% probability of independent 
commitment to any available nest site. 

It's worth emphasizing that in this model, every bee can observe the dancing 
state of all other bees in the swarm, regardless of their relative separation.
In this regard, the model developed by List et al. represents a mean-field 
stochastic agent-based model.
Galla~\cite{galla2010} formulated a master equation for the commitment probabilities 
within the same model as presented in~\cite{list2009}. 
In this formulation, he replaces the fixed duration $d_{j}=q_j$ of the waggle dances 
with stop-dancing rates $r_j=1/q_j$. 
Following this approach, it is possible to derive a set of non-linear differential equations 
that describe the evolution of the average values $\langle f_{j,t}\rangle$. 
These equations closely align with the results obtained from the original 
stochastic model. 
Furthermore, in the long time limit, one can analytically determine the 
stationary values of $\langle f_{j,t}\rangle$ using this mean-field approximation. Appendix~\ref{app:analytical} provides a brief description of this approach. 
Our model simulations implement the same stochastic method.

The non-linear differential equations of this model resemble those describing other honeybee-inspired models~\cite{pais2013, valentini2014, reina_desing_pattern, reina2017}, which have also been tested using Kilobots~\cite{valentini2016, Reina2018_spatiality, zakir_robot_2022}. However, two principal differences result in distinct behaviors that must be highlighted. First, the commitment probabilities, whether from independent discovery or recruitment do not explicitly depend on the options' qualities. Second, our model does not include cross-inhibition interactions, which are stop signals exchanged between agents holding different opinions to prompt them to revert to a neutral state and reassess their opinions.

\subsection*{Model of quenched bee configurations}
Our analysis will also consider the limiting case of random static, and generally 
non fully-connected, configurations of agents, the {\it quenched configuration limit}, 
on which we run the same collective bee-like decision model.
In this limit, the global proportions $f_{j,t}$ of agents in $j$ state appearing 
in Eq.~\ref{eq:list_probs}, as considered in the original model of List 
{\em et al.}~\cite{list2009}, are replaced by local proportions of agents computed 
from a fixed list of `neighbors' for each individual $i$ in the group. 
Lists of neighbors are computed after introducing a finite communication radius 
$r_{\tt int}$ around each agent in a random quenched configuration to identify 
other agents in this circular area of influence. 
These lists are calculated only once for each random configuration, and remain 
unchanged during the decision dynamics. 

\section{Experimental Kilobot swarm}
\label{sec:Kilobots}

%%% Kilobots as independent interacting individuals %. Decentralized decisions

In this study, we use a Kilobot swarm as our experimental system 
for investigating consensus reaching and exploring the interplay between the 
two most important factors in the honeybee-like nest-site selection model proposed 
by List {et al.}, i.e. independent discovery and imitation. 

Kilobots are compact open-source swarm robots, measuring 3.3 cm in diameter 
and 3.4 cm in height, purpose-built for the study of collective 
behavior~\cite{rubenstein2012}. 
Our primary goal is to experimentally investigate how the introduction of 
restricted robot communication capabilities (or local interactions), robot 
locomotion and spatial constraints, impact consensus reaching in comparison 
to the bee-like models introduced in Sec.~\ref{sec:beesmodel}. 
These mini-robots have previously been used to study collective 
decision-making~\cite{trianni2016_naming_game, valentini2016, Reina2018_spatiality, talamali2021_less_more, zakir_robot_2022, raoufi2023}, pattern formation~\cite{gauci2018},
morphogenesis~\cite{slavkov2018}, space exploration~\cite{dimidov2016, talamali2020_foraging}, collective transport of objects~\cite{rubenstein2013} in different experimental setups, and morphological computation and decentralized learning~\cite{BenZion2023}.

\subsection*{Kilobots' locomotion, decentralized control and information exchange}

Kilobots feature three slender, metallic legs—one in the front and two at the back.
With calibrated lateral vibrating motors, they can effectively overcome 
static friction, enabling self-propulsion.
Moreover, they have the capability to rotate either clockwise or counter-clockwise 
by selectively activating one of the two vibrating motors.
Kilobots are equipped with an Arduino controller, memory storage, 
and an infrared transmitter and receiver for bidirectional communication. 
Within an interaction radius of up to $10$ cm, as tested in complete darkness conditions (see App.~\ref{app:botscomm}), Kilobots can exchange
messages with nearby robots, with each message carrying up to $9$ bytes 
of information. 
During communication, the receiving robot assesses the intensity of 
incoming infrared light, enabling it to calculate relative distances 
to neighboring robots. Due to the observed variability in the measured communication range, we limited Kilobots to consider messages within a $7$ cm range. This was done to ensure homogeneity among the swarm's communication capabilities.

Each Kilobot in the swarm can execute various user-programmed instructions
and functions, with each processing cycle (or {\it loop iteration}) 
representing a unit of time in their dynamics.
During our experiments, Kilobots exist in discrete states, and their 
current state is visually conveyed through RGB LED lights. 
This makes Kilobots an ideal experimental system for studying collective 
decision-making, combining decentralized activity, 
limited communication capabilities and locomotion —effectively making them 
`programmable insects'.

We place up to $35$ Kilobots in a circular {\it arena} with a radius $R$ 
delimited by rigid walls. 
Using an azimuthal camera, we capture the  Kilobot activity in accordance 
with the guidelines of the nest-site selection model outlined in the 
previous section (see also Fig.~\ref{fig:schematics}).
Further details about the robots technical features and about the experimental 
setup are presented in Appendix~\ref{app:experimentalsetup}.

\subsection*{Kilombo: the Kilobot swarm emulator}

A useful tool to work alongside physical experiments is the Kilobot-specific 
simulator software \textit{Kilombo}\cite{jansson2015}. 
This is a C-based simulator that allows the code developed for simulations to 
be run also on the physical robots, removing the slow and error prone step of 
converting code to a different platform. 
In this way we can also perform simulations using Kilombo, to test our
experimental setup and to support our results - and to complement them 
whenever it has not been possible to perform further experiments. 

\begin{figure*}[!t]
	\centering
 \iffigures
    \includegraphics[width=0.92\textwidth]{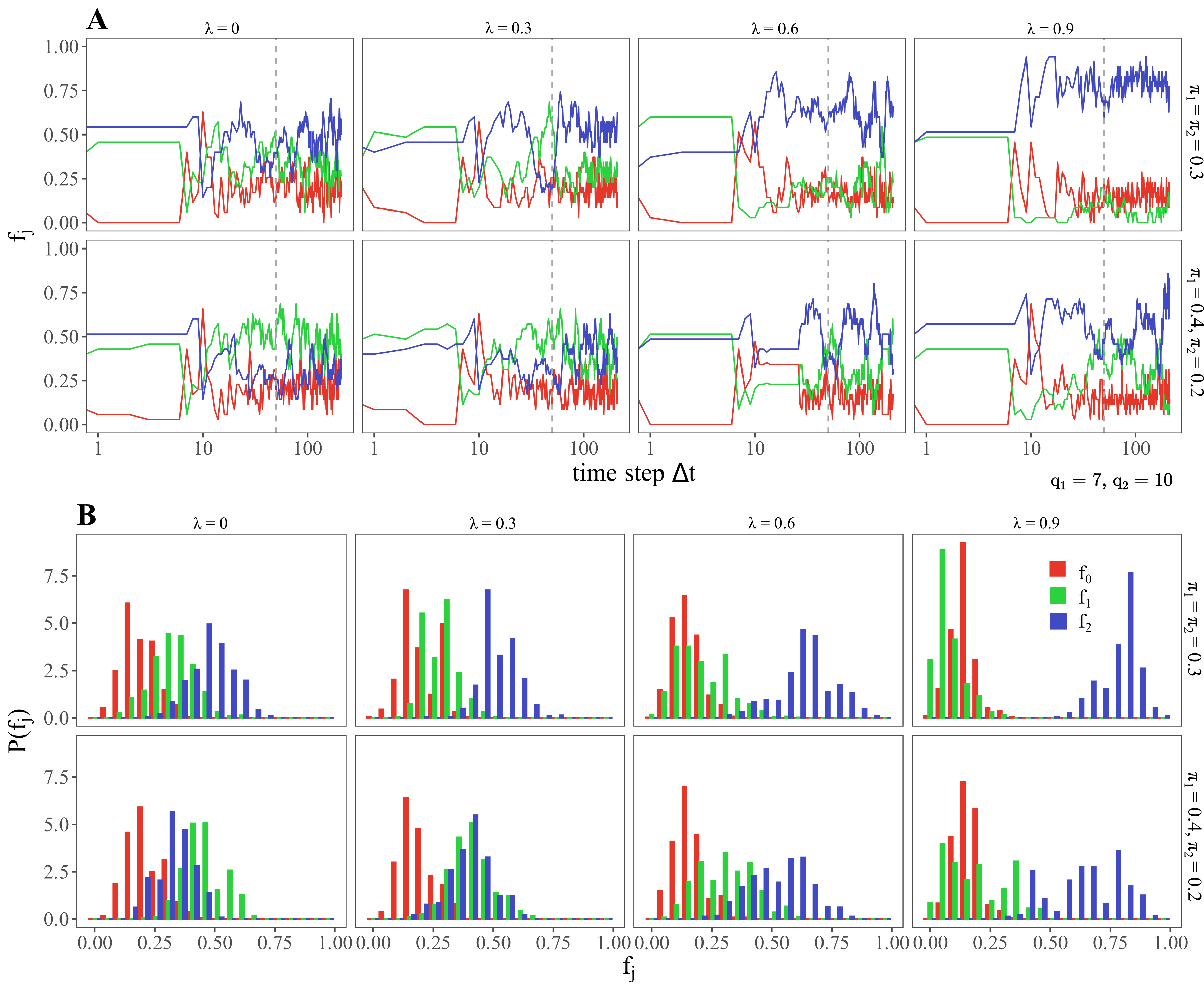}
 \fi
  	\caption{
   \textbf{A}: {\it Proportions of Kilobots $f_j$ dancing for the different states}
   as a function of time. 
   Red, green and blue represent the proportion of uncommitted ($f_0$) 
   and dancers for low-quality ($f_1$) and high-quality ($f_2$) sites, 
   respectively. 
   \textbf{B}: {\it Probability density function of $f_j$ values} in the stationary state, {$P(f_j)$. 
   Plots correspond to symmetric (up, $\pi_1 = \pi_2 = 0.3$}) an asymmetric (bottom, $\pi_1 = 0.4, \; \pi_2 = 0.2$) a priori 
   discovery probabilities, with qualities $q_1 = 7$, $q_2$ = 10 and interdependence parameter values 
   $\lambda ={0, 0.3, 0.6, 0.9}$ (left to right).
   Five repetitions of the experiment for each $\lambda$ were performed. The data were collected after a transient time interval of $50$ time steps, indicated by the vertical gray dashed line in the temporal evolution plots.
   }\label{fig:fjtime_Kilobots}
\end{figure*}

\section{Consensus reaching in a bee-like Kilobot swarm}
\label{sec:consensus_reaching}

In this section we describe, both experimentally and theoretically, how the 
complex decision-making problem of reaching strong consensus for the best-available 
option is solved by our bee-like Kilobot ensemble under different conditions.
Essentially, we analyze the temporal dynamics of the proportion of bees (bots) 
that either advocate for one of the possible sites or remain uncommitted. 
We examine how these proportions evolve and eventually stabilize, while also 
exploring the criteria that signify the attainment of a consensus in this 
steady state. 
We compare our experimental results and Kilombo simulations with mean-field 
theoretical results finding intriguing resemblances in sufficiently crowded 
conditions, or after long enough exploration times, but also hints towards 
important divergences in other plausible conditions.
By implementing the nest-site selection model within a physical Kilobot system, 
we gain the capacity to explore the consequences of more realistic robot 
interactions and the role of space and locality on consensus formation.

\subsection*{Collective decision-making in Kilobots}
\label{sec:consensus_Kilobots}

We start by running our experiments in a group of $N=35$ Kilobots. 
We deploy the Kilobots within a circular arena of radius $R=20$ cm and task 
them with assuming the role of scout bees. 
These Kilobots engage in a dynamic process defined by the List {\em et al.} model, 
as elaborated in Sec.~\ref{sec:beesmodel}. 
Each Kilobot holds an internal state and displays it with a color LED.
Throughout the course of the experiment, 
Kilobots adjust their internal states based on the probabilities outlined 
in Eq.~\ref{eq:list_probs}, but having only partial and individual information 
of the population of bees advertising each site, $f_{i,t}$. 
The computation of these values takes place after a given time-step, $\Delta t$, 
of the decision process relying on the information that each uncommitted Kilobot 
can gather from its immediate surroundings.

While the typical outcomes of the List {\em et al.} model dynamics have been 
documented in prior research~\cite{list2009,galla2010}, previous investigations 
have primarily scrutinized these developments either at the mean-field 
(fully-connected) level~\cite{list2009,galla2010} or within the context 
of nearest neighbor interactions within a square lattice~\cite{galla2010}. 
In contrast, the present approach involves committed Kilobots that move 
in the circular arena as persistent random walkers during their engagement 
in the advertising
% , or dancing, 
phase of the consensus-searching dynamics. 
This advertising movement results in a varying number of neighbors that they 
can communicate with. Such behavior mirrors that of real bees, which are known to interact with a limited number of neighbors during activities such as dancing or observing a waggle dance~\cite{Sumpter2006,judd1994}.
In order to correctly quantify the relevance of these factors, we first 
benchmark the communication capabilities of individual bots, as 
reported in App.~\ref{app:botsseen}.

We narrow the focus of our experiments to the case of two sites: 
a high-quality site (designated as site $2$) and a lower-quality site 
(referred to as site $1$).
The binary decision problem was chosen as the focus of our study due to its simplicity, which allows for a clearer understanding of the interplay between model parameters and the time-varying communication network. Additionally, our analysis in~\cite{marchpons2024consensus} revealed that introducing more than two options does not significantly impact the final results.
At each time-step Kilobots are either dancing for site $1$, site $2$, 
or not advertising any site. 
As an extra feature towards emulating the natural behavior of honeybees, 
we introduce an explicit difference in the dynamics of committed and 
uncommitted Kilobots. 
While Kilobots advertising an option perform a persistent random walk (PRW) (see details on App.~\ref{app:experimentalsetup}), uncommitted Kilobots, instead, come to a halt to wacth other Kilobot advertisements. This feature is partially inspired by the differentiation between advertising bees, which perform the waggle dance, and uncommitted bees, which adopt a more passive role.

We study both the time evolution and the steady state of the decision-making 
experiment on our bee-like Kilobot system.
The proportion of Kilobots not advertising a site and the proportions of Kilobots 
dancing for sites $1$ and $2$, also referred to as the {\em dance frequencies} 
$f_0$, $f_1$ and $f_2$, respectively, are monitored as a function of time until 
they stabilize and fluctuate around mean values.
The duration required for the system to attain this steady state varies 
depending on the model's parameters.
Generally, large interdependence $\lambda$ leads to stronger majorities, 
while dragging out the evolution up to that state, as extensively analyzed across various scenarios in our simulations~\cite{marchpons2024consensus}.
Moreover, when the competing sites have similar qualities the swarm 
takes longer to reach a consensus.
In our experiment, site qualities are fixed to $q_1=7$ and $q_2=10$; 
and thus, advertising times for site $1$ and $2$ are equal to $q_1 \Delta t$ 
and $q_2 \Delta t$, respectively.
This choice of qualities allows us to analyze a dispute between the two available nest-site options without entering into an excessively time-consuming transient dynamics phase. Note that, as in real honeybees colonies, the Kilobot swarm must make the decision for the best option within a reasonable time span, hopefully before their battery power is exhausted.

In addition, our analysis encompasses the evolution under different levels of 
interdependence $\lambda$ and considers two distinct scenarios for 
the discovery probabilities: 
a symmetric case where $\pi_1=\pi_2$ and an asymmetric case where 
$\pi_1>\pi_2$, favoring the lower-quality site. The specific values of $\pi_{i,j}$ will determine the conditions under which the system can achieve consensus at different levels of interdependence. For example, in the symmetric case, this relationship is illustrated in Fig.~\ref{fig:fj_time_meanfield}. A comprehensive analysis of this interplay is provided in~\cite{marchpons2024consensus}.

Figure \ref{fig:fjtime_Kilobots}A illustrates the dynamics of $f_0$, $f_1$, and $f_2$ 
over time, for different $\lambda$ and for both the symmetric and asymmetric cases.
The initial conditions are set as follows: $f_1(t=0)=f_2(t=0)\simeq 0.43$, and 
$f_0(t=0)\simeq 0.14$ (i.e. approximately $15$ Kilobots dancing for site 1, $15$ 
dancing for site 2 and $5$ uncommitted). 
Other initial conditions have also been tested but, as expected, stationary results 
do not depend on the particular choice used in the experiments (see Supp. Fig.~\ref{SM-suppfig:different_initial_conditions} for an example in quenched simulations).
The time evolution of each population is jerky and fluctuating. 
This behavior is also expected and is an inherent consequence of the model dynamics, 
as Kilobots promoting a site revert to an uncommitted state after their dancing 
period concludes. 
Additionally, due to the limited system size, these fluctuations are 
particularly noticeable.
Nevertheless, systematic behaviors can be grasped when examining mean values 
and full distributions.
Across all values of $\lambda$, there exists a transient phase in which the 
larger population oscillates between the three states $j$, resulting in 
significant variations in the values of $f_j$. 
However, roughly after $\sim50$ time-steps, a steady state is achieved, 
and each $f_j$ fluctuates around its mean value.
In the scenario with symmetric $\pi$ values, irrespective of $\lambda$, 
$f_2$ eventually becomes the dominant population in the steady state. 
Moreover, increasing the interdependence parameter $\lambda$ amplifies 
the difference between the proportion of Kilobots dancing for the 
high-quality site ($f_2$) and the low-quality site ($f_1$).
Interestingly, when we shift towards asymmetric self-discovery 
probabilities ($\pi_1>\pi_2$), it is observed that if $\lambda$ 
is not sufficiently large, the steady state can be dominated by $f_1$ 
or present a stalemate between $f_1$ and $f_2$. 
However, increasing $\lambda$, the system gains the capability to 
favor the less accessible yet higher-quality option $2$. This result aligns with other honeybee-inspired collective decision-making models, where agents break the symmetry between non-equivalent options and commit to the highest quality option by enhancing the strength of social feedback~\cite{reina2017, bizyaeva_nonlinear_2023}.

Across all scenarios and parameter sets, there is notable dispersion 
in the values of $f_j$, resulting in broad probability distributions 
$P(f_j)$ for all three $j$, as exhibited in Fig.~\ref{fig:fjtime_Kilobots}B. 
When $\lambda$ is low, the distributions $P(f_j)$ for all three states 
tend to overlap. 
However, with an increase in $\lambda$, the distributions $P(f_1)$ 
and $P(f_2)$ gradually separate from each other, eventually exhibiting 
significantly distinct mean values when the interdependence is at its 
highest, $\lambda = 0.9$. We quantitatively confirm this trend by measuring the Jesnen-Shanon Diverenge (JSD) of these probability distributions (see Supp. Fig. ~\ref{SM-suppfig:JSD_histos}). Specifically, the JSD tends to increase with $\lambda$ in the symmetric discovery scenario, while in the asymmetric discovery case, it exhibits a minimum at $\lambda = 0.3$, after which $f_2$ takes the lead.

While our primary focus has been on analyzing the stationary average values and their distributions around the mean, it is also valuable to examine the outcome of the decision process in more detail. For this purpose, we quantify how frequently each possible outcome (simple/strong majority for each option or draw) occurs in the experiments' stationary states. These results are displayed in Supp. Fig.~\ref{SM-suppfig:pie_charts_kbs}, where the effect of $\lambda$ becomes more evident. In the symmetric discovery scenario, the most frequent outcome is a simple majority for option 2 when $\lambda = 0.0, \; 0.3$. Further increasing the interdependence allows the system to reach strong consensus (defined in the following section). Similarly, in the asymmetric discovery scenario, we observe a shift from mostly reaching a simple majority for the bad option (when $\lambda = 0$) to a strong consensus for the good option ($\lambda \geq 0.6$). When $\lambda = 0.3$, where we observed overlapping histograms, it is equally likely for option 1 to win by a simple majority as for option 2.

\subsection*{Consensus in numerical and analytical approaches}

%%%%%%%
%%% First, time evolution in stochastic simulations and integrated exact solution
%%%%%%%

\begin{figure}[!t]
\centering
\iffigures
\includegraphics[width=1\columnwidth]{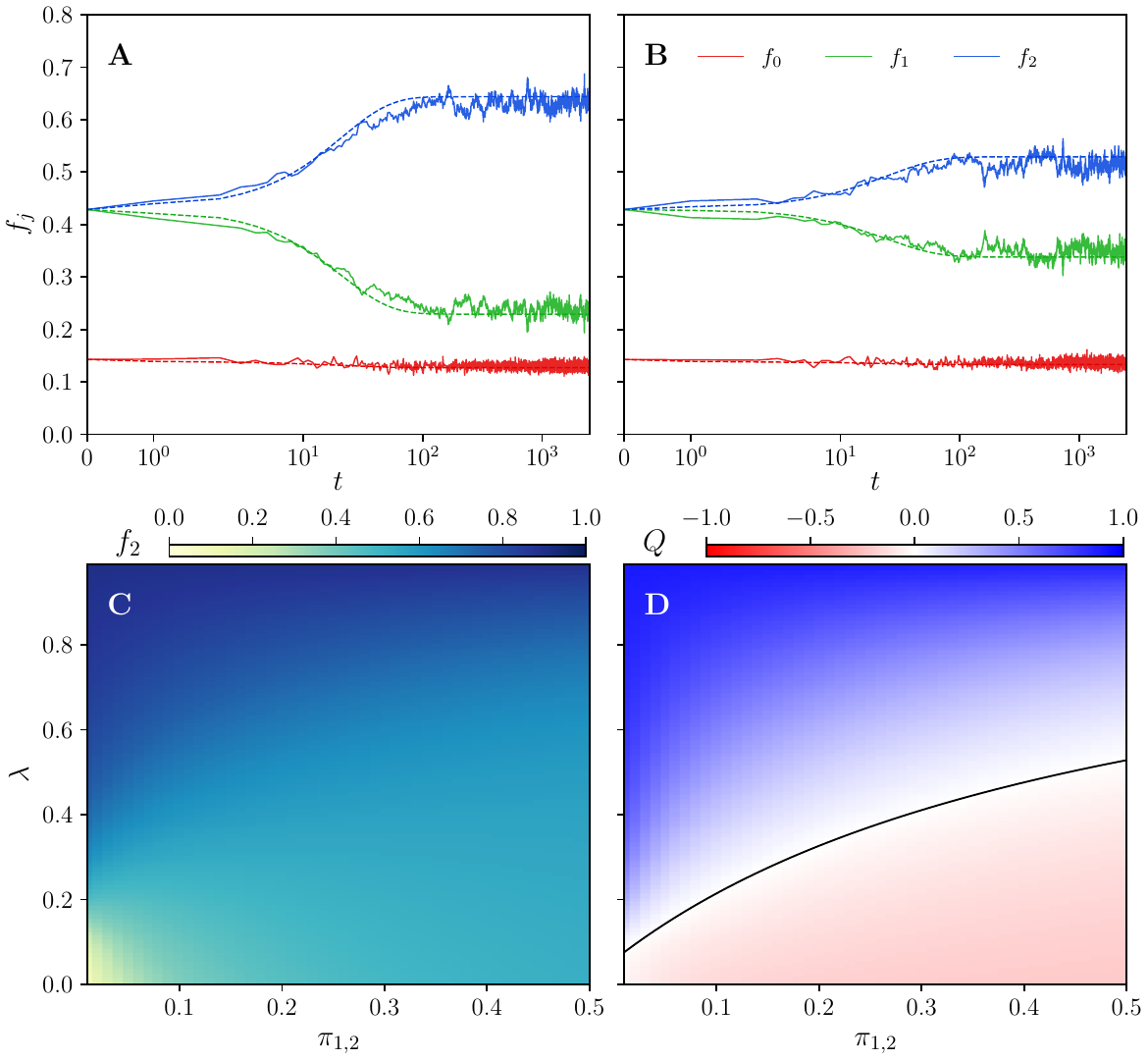}
\includegraphics[scale=0.45]{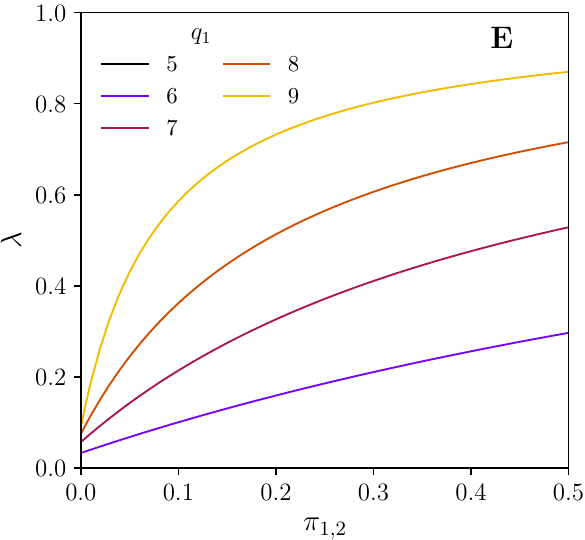}
\fi
  \caption{
  \textbf{A-B}:
  {\it Dance frequencies $f_0$ (red), $f_1$ (green) and $f_2$ (blue)
  as a function of time} obtained from numerical simulations of the 
  stochastic model, and from the numerical integration of the mean-field 
  deterministic equations (smooth superimposed curves in the same color). 
  A: Symmetric discovery scenario with probabilities $\pi_1 = \pi_2 = 0.3$. 
  B: Asymmetric discovery scenario with probabilities $\pi_1 = 0.4$ and $\pi_2 = 0.2$. 
  Other parameters are: $N=35$, $q_1 = 7$, $q_2 = 10$, and $\lambda = 0.6$.
  Simulations were averaged over $100$ realizations.
  \textbf{C-D}: {\it Stationary values of $f_2$ (C) and $Q$ (D)} in the parameter 
  space $(\pi_{1,2}, \lambda)$ obtained from solutions of the deterministic 
  equations of the model. 
  The black line on (D) corresponds to the theoretical crossover line where $Q=0$. 
  Parameters are $q_1 = 7$, and $q_2 = 10$.
  \textbf{E}:
  {\it Consensus crossover lines} or $\lambda$-thresholds for consensus, 
  $\lambda^*$, i.e. $Q(\lambda^*)=0$ as a function of $\pi_{1,2}$ 
  (in the symmetric scenario, $\pi_1 = \pi_2$). 
  Colors represent different values of the low-quality site $q_1$, 
  while $q_2 = 10$ is maintained constant.
  }\label{fig:fj_time_meanfield}
\end{figure}

In the following paragraphs we proceed with the numerical analysis of the nest-site 
selection problem within a fully connected system.
Figure~\ref{fig:fj_time_meanfield}A-B presents the values of the dance  
frequencies $f_0$, $f_1$ and $f_2$ as a function of time. 
These values are obtained from both stochastic simulations and the numerical 
integration of the deterministic mean-field equations, as detailed in 
App.~\ref{app:analytical}.
The observed curves are qualitatively similar to the ones displayed by the 
evolution of the model in Kilobots (as shown in Fig.~\ref{fig:fjtime_Kilobots}). 
It is worth noting that, as for the Kilobots, we use $q_1 = 7$ and $q_2 = 10$.
In both studied cases, whether symmetric ($\pi_1 = \pi_2 = 0.30$) or asymmetric 
($\pi_1 = 0.4$, $\pi_2=0.2$), with the same value of $\lambda = 0.6$, we observe 
the high-quality nest site taking the lead in the steady state.

Simulations of the stochastic model display finite-size fluctuations around the 
mean values, whereas the numerical integration of the analytical solution produces 
smooth evolution curves.
Notice that the analytical curves do not accurately represent the transient state at 
very short times, but they accurately describe the average stationary value for each 
parameter set.
This reassuring result allows us to perform a parametric exploration of the model 
without the need of resource-intensive simulations.

%%%%%%%
%%% Parametric exploration using the integrated solution
%%%%%%%

Employing the analytical solution, we delve into the exploration of the parameter 
space defined by $\pi_j$ and $\lambda$. 
We do not only asses the stationary dance frequencies, $f_j$, but also a strong majority
definition of {\it consensus}:
\begin{equation}
	Q  = f_2 - 2f_1 \text{.}
\end{equation}

This definition implies that there must be twice as many bees dancing for the
high quality site than for the low quality site for the condition $Q>0$ to be met. 
Consequently, this represents a \textit{large majority} consensus, i.e., a majority 
by a factor of $2/3$ in the case there were no uncommitted bees in the system. 

\begin{figure*}[t!]
	\centering
 \iffigures
    \includegraphics[width=0.95\textwidth]{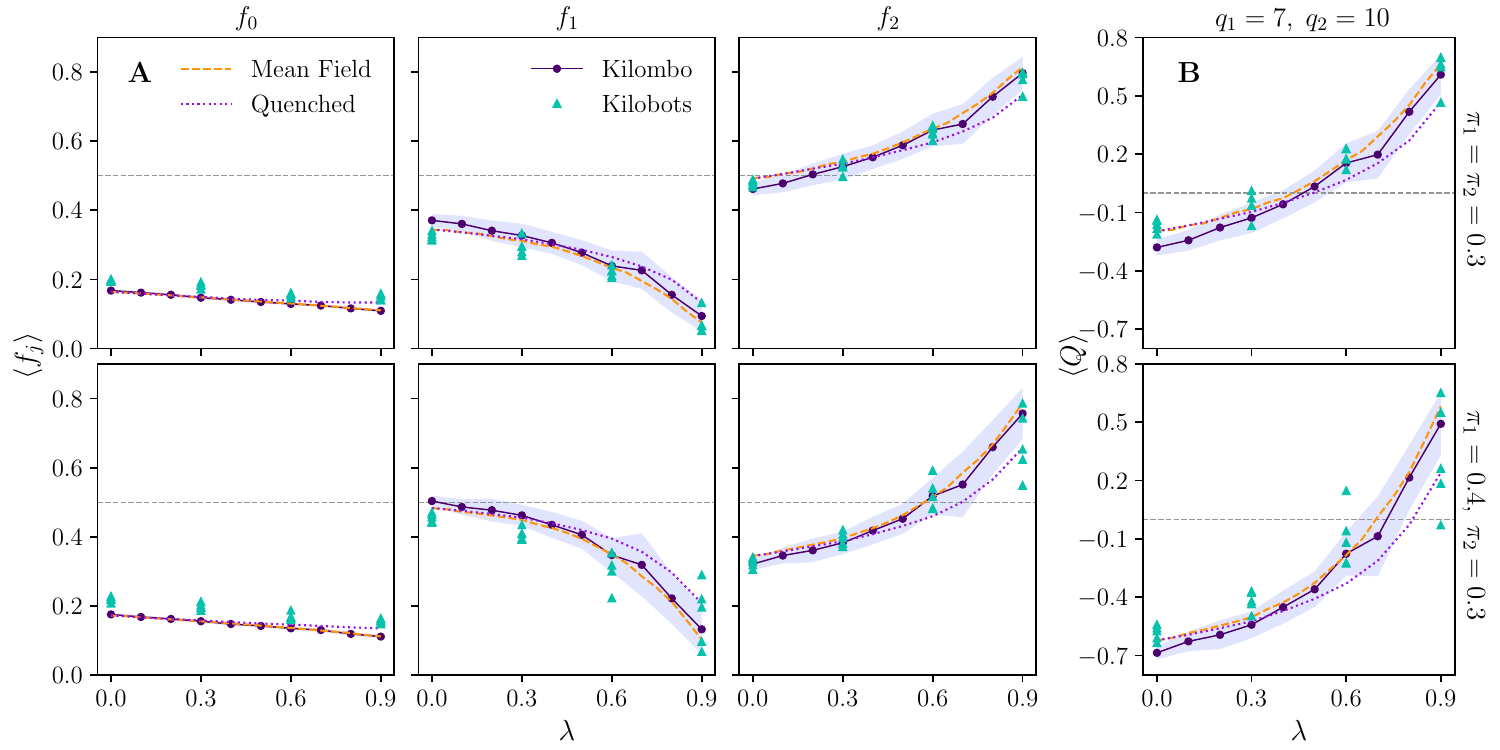}
 \fi
	\caption{
    {\it Stationary dance frequencies and consensus as a function of interdependence} 
    \textbf{A}: Average frequencies $\langle f_0\rangle$, $\langle f_1\rangle$ and 
    $\langle f_2\rangle$ and 
    \textbf{B}: Average consensus $\langle Q\rangle$ as a function of $\lambda$ for 
    symmetric (up) and asymmetric (bottom) $\pi_j$ for physical Kilobots, and for 
    simulations using Kilombo, quenched configurations and fully-connected networks.
    \textbf{Triangles}: Mean values from experiments of $N=35$ Kilobots 
    (5 repetitions per parameter combination). 
    \textbf{Solid purple line with points}: Mean values from Kilombo simulations 
    with $N=35$ bots averaged over $50$ repetitions. Shaded area: Kilombo standard deviation.
    \textbf{Dashed orange line}: Mean values from simulations of the fully connected model 
    with $N=35$ bots averaged over $100$ repetitions. 
    \textbf{Dotted pink line}: Mean values from quenched configurations with $N=35$ bots
    averaged over $100$ repetitions.
 }
	\label{fig:fj_stat_Kilobots}
\end{figure*}

In Fig.~\ref{fig:fj_time_meanfield} (C-D) we present the outcome of the decision process 
in the symmetric scenario, $\pi_1 = \pi_2$. We asses both the stationary value of $f_2$
(Fig.~\ref{fig:fj_time_meanfield}C) and the consensus, $Q$ (Fig.~\ref{fig:fj_time_meanfield}D).
The exploration of the parameter space is done by varying the values of the interdependence
$\lambda$ (y-axis) for each value of the self-discovery probabilities $\pi_{1,2}$ (x-axis) 
(a similar parameter space exploration for the asymmetric scenario is presented in Supp. Fig.~\ref{SM-suppfig:phase_piasymm}). 
The color charts illustrate how these two metrics vary across the parameter space.
We observe a smoothly varying trend indicating that the proportion of individuals 
dancing for the high quality site $f_2$ increases along with the interdependence 
parameter $\lambda$, regardless of the specific value of $\pi_1$ and $\pi_2$. 
This phenomenon arises from bees placing more reliance on the opinions of their peers, 
and as $\lambda$ increases, it enhances the reinforcement for the best possible option.
Contrarily, when both independent discoveries ($\pi_1=\pi_2$) increase simultaneously, 
there is a relative decrease in the number of bees dancing for site $2$. 
This means that as $\pi$ values increase while keeping $\lambda$ constant, more 
advertisements are motivated by independent discoveries rather than by opinion sharing. 
Consequently, when $\pi_1 = \pi_2$, $f_1$, the population of bees dancing for site $1$ 
increases at the expense of site $2$, hindering overall consensus.
In the extreme case of $\pi_1=\pi_2=0.5$, this translates in consensus never being 
achieved (i.e. $Q < 0$) if the interdependence parameter is lower than $\sim 0.5$. 
The black line depicted in Fig.~\ref{fig:fj_time_meanfield}D corresponds to the 
strong consensus crossover $Q=0$, computed using the analytical solution.
A system with a set of parameters given by points below this line will never find 
strong consensus, but it will do for a set of parameters given by a point above.
In other words, after surpassing a $\pi_{1,2}$-dependent $\lambda$ {\it threshold}, 
$\lambda^*$, the system crosses over to a strong consensus state, $Q > 0$, 
and the strength of this consensus intensifies with increasing interdependence.
The region of non-consensus expands as $\pi_{1,2}$ values grow, thereby narrowing 
the range of interdependence values that lead to consensus.

%%%%%
%% The variation of $\lambda^*$ with q1
%%%%%

Certainly, the `critical' line $\lambda^*(\pi_1,\pi_2)$ is contingent upon
the values of qualities $q_1$ and $q_2$ as well.
In Fig.~\ref{fig:fj_time_meanfield}E, we represent $\lambda^*$ as a function 
of $\pi_1 = \pi_2$, maintaining a fixed value of $q_2 = 10$ while exploring 
different choices of $q_1$.
We observe that when $q_1$ is markedly smaller than $q_2$, 
particularly when $q_1 \leq q_2 / 2$,
the region of no strong consensus basically disappears,
i.e., the swarm is able to choose the high quality site 
with strong majority, allowing for consensus even when $\lambda = 0$.
On the other hand, when $q_1$ increases and approaches the value of $q_2$, 
the competition between sites intensifies. 
Consequently, a higher value of $\lambda$ becomes necessary to counteract 
the influence of the discovery probabilities.
% 
% ABOUT QUALITY SENSITIVITY AND WEBER'S LAW
For a given $\lambda$ and $\pi_{1,2}$, the quality difference necessary for the system to achieve consensus is directly proportional to the mean quality of the options~\cite{marchpons2024consensus}, and thus the $\lambda^*$ crossover lines depend solely on the options' relative quality difference.
This observation is in agreement with Weber's Law of perception, as reported in previous studies on similar collective decision-making models~\cite{pais2013, reina_psychophysical_2018}. Unlike these models, ours does not require incorporation of discovery or recruitment rates that depend on the options' qualities to adhere to this fundamental law.
% ABOUT SYMMETRY BREAKING AND CROSS INHIBITION
Finally, when the qualities and discovery probabilities of available options are equal, both options become equivalent, transforming the decision problem into a symmetry-breaking task between identical options. Here, the main distinction from other honeybee-inspired models~\cite{reina_desing_pattern, Reina2018_spatiality} --the presence or absence of cross-inhibition interactions-- becomes crucial. In scenarios where external information is continually introduced through independent commitment transitions, cross-inhibition has been identified as a vital mechanism for breaking decision deadlocks~\cite{zakir_robot_2022, reina_cross_inhibition_2023}. In contrast, our model system can resolve deadlocks differently by adapting its behavior upon deadlock detection, either by halting exploration (effectively setting $\pi_\alpha = 0$) or by maximizing social interactions (i.e. $\lambda = 1$). A detailed analysis of these parameter limits is provided in~\cite{marchpons2024consensus}.

\subsection*{Comparison between experiments and simulations}

%%%%%
%% Presentation of the aims of the section (comparison among different approaches)
%%%%%

We would like to quantitatively compare the steady-state averages in 
Kilobots with those obtained from the modeling approaches.
Our goal is to distinguish the emerging properties of the real system, 
which involves restricted communication capabilities, and moving 
individuals, in comparison to the fully-connected approximations made in mean-field solutions.
To enhance this comparison, we utilize Kilombo~\cite{jansson2015}, 
the Kilobot's emulator, to conduct complementary simulations under 
the same experimental conditions. 
On the other hand, alongside the fully-connected stochastic model simulations, 
we also incorporate the steady-state results obtained from simulations with 
quenched configurations of bots running the same collective decision-making model.

%%%%%%%
%%% Descrition of the figure general trends
%%%%%%%

Figure \ref{fig:fj_stat_Kilobots} displays the average stationary values of 
the dance frequencies $\langle f_0\rangle$, $\langle f_1\rangle$, $\langle f_2\rangle$, 
and the consensus $\langle Q\rangle$ as functions of the interdependence parameter 
$\lambda$ for both symmetric and asymmetric scenarios defined by the same self-discovery
probabilities $\pi_i$, and site-quality values $q_1=7$ and $q_2=10$, considered previously. 
Consistently, the primary trend is similar across all approaches:
at small values of interdependence $\lambda$, the majority of the population gravitates 
towards the high-quality site ($2$) in the symmetric case. 
Conversely, it aligns with the low-quality site ($1$) when asymmetric self-discovery
probabilities favor it. 
In this case, there is no consensus for the higher quality option, and $Q$ assumes 
negative values.
As $\lambda$ increases, $\langle f_1\rangle$ decreases while $\langle f_2\rangle$ 
increases for both the symmetric and asymmetric scenarios. 
Moreover, $\langle f_1\rangle$ and $\langle f_2\rangle$ exhibit similar functional 
trends but in opposite directions, resulting in the stationary value of 
$\langle f_0\rangle$ remaining nearly independent of $\lambda$.
When examining the stationary average of the strong consensus parameter $\langle Q\rangle$, 
we observe a smooth transition from non-consensus to consensus as $\lambda$ varies. 
This is rather a crossover more than a phase transition, but it can be precisely identified. 
Notably, in scenarios with symmetric discovery probabilities, consensus is achieved 
at smaller $\lambda$ values compared to asymmetric scenarios. 
This is because there is less introduction of independent discoveries for the 
low-quality site, leading to less misleading information that needs to be 
discarded through communication.

%%%%%%%
%%% Quantitative comparison among different approaches
%%%%%%%

Now, let's delve into a quantitative comparison of the different approaches, 
going beyond the observed consistency in the data.
First, it's noteworthy that the stationary values obtained from physical Kilobots 
closely match those from Kilombo simulations, even within the standard deviation
(represented by shaded areas). 
This alignment underscores the reliability of the emulator in complementing real 
measurements.
More remarkably, both experimental and emulator results closely align with 
the mean-field results. 
This coincidence might be less expected, and we will discuss the reasons 
behind it for the chosen set of parameters.
These findings suggest that mobile Kilobots, as they interact with their 
local environment, can effectively sample the system's state and transmit 
information throughout, almost as if they were fully connected, achieving 
collective decision-making comparable to mean-field fully-connected individuals. 
We will test this hypothesis in the next section.
However, it's worth noting that this agreement is not perfect. 
While $\langle f_j\rangle$ and $\langle Q\rangle$ values are nearly identical for 
all approaches at low values of $\lambda$, some differences become noticeable after 
an interdependence value of approximately $\lambda\sim 0.5$. 
In this range, quenched configurations exhibit higher $\langle f_1\rangle$ and 
lower $\langle f_2\rangle$ values than the fully-connected system. 
Furthermore, for high $\lambda$ 
we observe a broader scattering of the experimental results, especially in the asymmetric discovery scenario.
As a consequence of these variations, we observe a lower consensus value 
compared to the fully-connected case. 
In the asymmetric scenario, particularly for high interdependence ($\lambda \gtrsim 0.7$), quenched simulations show a crossover to consensus at a substantially higher value of $\lambda$ compared to the fully-connected case. Additionally, experiments exhibit significant fluctuations in stationary consensus, with some realizations failing to achieve consensus or closely resembling the quenched results.

In summary, in a high enough interdependent system, the experimental results 
displayed in Fig.~\ref{fig:fj_stat_Kilobots} demonstrate that mobile individuals, 
such as Kilobots moving in space and integrating information over time, are 
capable of achieving high consensus values comparable to fully-connected systems. 
The exact threshold depends on model parameters such as quality differences and 
self-discovery probabilities.
Quenched configurations, characterized by fixed neighbor lists for interactions, 
help us assess the importance of Kilobots' movement and mixing under experimental
conditions. 
Despite the limited communication range of Kilobots, our hypothesis is that their 
mobility enables information to spread in a manner that the outcome
of the decision-making problem becomes similar to mean-field results.
However, in quenched configurations conditions, local consensus for the low-quality 
option may emerge, diminishing the chances of attaining strong consensus for the 
best-quality option. 
In agreement with Raoufi et al.~\cite{raoufi2023}, we observe that the network structure significantly influences the accuracy of the final decision. A greater communication range results in a better-connected network, characterized by a higher average degree, leading to improved accuracy. Furthermore, by varying the weighting factor of social interactions, i.e., the interdependence parameter, we recognize its significance in collective decision-making processes with limited communication.

\section{Information spreading in the Kilobot swarm}
\label{sec:swarm_analysis}

In this section we provide a more detailed analysis of the information spreading taking place in our Kilobot swarm as they play the bee-like decision-making process. 
Various models of state-contagion dynamics have been explored using self-propelled particles~\cite{peruani_dynamics_2008, levis_synchronization_2017, starnini_emergence_2016, paoluzzi_information_2020}. The relationship between agents' states, their mobility, and packing fraction gives rise to different physical phenomena. For instance, motility-induced phase separation occurs at low densities~\cite{paoluzzi_information_2020}, while segregation between agents with opposing opinions is observed in~\cite{starnini_emergence_2016}. We take a simpler approach by considering a model that does not include correlations between movement patterns and the underlying opinion dynamics. Instead, 
our aim is to investigate under which circumstances consensus reaching in 
motile physical Kilobots can be almost as effective as in an idealized mean-field-like 
communicating system. This is achieved by allowing them to gather local information over a limited temporal window.

When analyzing the data presented in Figure~\ref{fig:fj_stat_Kilobots}, 
and observing the consistency between the fully-connected approach and 
the results obtained from Kilobots (both experimental and emulated) 
across a wide range of values for $\lambda$, our primary hypothesis 
centers around the crucial role played by the Kilobot density and the 
Kilobot motion in allowing them to form a connected communication network \cite{trianni2016_naming_game, dimidov2016, khaluf_interaction_models_2018}. 
Through a percolated communication network, each Kilobot can receive information 
beyond its immediate surroundings, set by the infrared sensor capabilities, 
and spread it throughout the system. 

We investigate the occurrence of this percolation transition in the Kilobot 
intercommunication network, to understand its relation with consensus formation.
This intercommunication network 
can be represented as a complex network~\cite{Newman10,BARTHELEMY20111}, 
where nodes correspond to Kilobots, and where two nodes are connected by 
an edge if the corresponding Kilobots are within an Euclidean distance 
smaller 
% than their communication, or interaction, radius.
than their interaction radius.
Since infrared communication in Kilobots is approximately isotropic, the network 
is undirected, meaning that if bot $i$ interacts with bot $j$, bot $j$ 
would also interact with bot $i$. 
Consequently, these networks are based on proximity. Although they do not have the information transfer advantages of long-range networks with scale-free degree distributions~\cite{khaluf_interaction_models_2018}, 
the Kilobots' mobility enhances their communication and consensus reaching capabilities~\cite{dimidov2016}.
Due to the agents' mobility, this proximity network is time-varying (see Supp. Fig.~\ref{SM-suppfig:scheme_kilosims} and Supp. Video 4).

In \cite{trianni2016_naming_game}, the authors used Kilobots and the \textit{naming game} to explore the importance of percolating communication networks. However, despite involving moving robots, the percolation threshold discussed there was related to instantaneous interactions between robots.
Here, we consider time-integrated networks that result from 
information exchanges occurring within a short temporal window $\Delta t$ while our robots move. 
This is why our percolation threshold $\eta^*$ includes a movement' component over this temporal window. Our percolation analysis demonstrates that at robot densities well below the instantaneous percolation threshold, where information percolation would not occur in corresponding static configurations and thus strong consensus would not be achieved, the movement of agents over the time interval $\Delta t$ facilitates consensus formation.

We therefore characterize the percolation transition in the Kilobot 
communication network examining standard quantities such as the mean 
communication cluster size, the emergence of a giant component, 
the cluster size distributions, or the average connectivity, 
as a function of the Kilobots communication radius and their 
advertising time window (some details and complementary analysis are left 
for Appendix~\ref{app:networks}).
%And, more importantly, we study how this transition impacts the outcome of the decision-making problem.
Finally, we study how this transition impacts the outcome of the decision-making problem.

\subsection*{Communicating clusters of Kilobots}

Leveraging numerous spatial configurations generated in Kilombo simulations, 
we examine the cluster structure within the Kilobot communication network. 
A cluster is defined as a connected component in which nodes can be reached 
from one another via continuous paths of adjacent edges~\cite{Newman10}. 
In computational terms, a Kilobot is considered part of the same communication cluster 
as a focal Kilobot if it resides within a circular region centered on the focal Kilobot 
with a radius of $r_{\tt int}$, denoting the effective interaction radius. 
This recursive process identifies all clusters and their sizes in each spatial configuration.

The mean cluster size, denoted as $\langle \mathcal{S} \rangle$, is a key parameter 
in the analysis of cluster structures. 
It shares similarities with transport coefficients like magnetic susceptibility and 
specific heat and plays a crucial role in the geometrical percolation 
process~\cite{stauffer2018}. 
This quantity measures the fluctuations within the cluster size distribution and 
aids in detecting a continuous percolation transition, where the communication network 
shifts from having only small isolated clusters of Kilobots to forming a large, 
connected communicating component. 
The percolation transition occurs as a function of the interaction radius $r_{\tt int}$ 
at a fixed Kilobot density, or as a function of density for fixed values of $r_{\tt int}$. 
Both quantities can be combined into a single control parameter $ \eta = N r_{\tt int}^2/R^2$,
which measures the effective area {\em covered} by Kilobots. 
Note that we are referring to an effective communication area rather than to the physical 
area occupied by the Kilobot swarm. 
Thus, the percolation transition takes place at a threshold value $\eta^*$ of this 
control parameter. 

The mean cluster size is defined as~\cite{stauffer2018}:
\begin{equation}
\langle S \rangle = \frac{\sum's^2 n(s)}{\sum's n(s)},
\end{equation}
where, $n(s)$ represents the number of clusters of size $s$, 
i.e. composed of $s$ Kilobots, and summations $\Sigma^{'}$ exclude the 
{\it giant component}
of the network, $S_{max}$, which is the largest cluster observed in a given configuration. 
In finite systems, $\langle \mathcal{S} \rangle$ exhibits 
a characteristic peak, instead of diverging, at the percolation threshold~\cite{stauffer2018,Newman10}.

\begin{figure}[t!]
	\centering
 \iffigures
    \includegraphics[width=1\columnwidth]{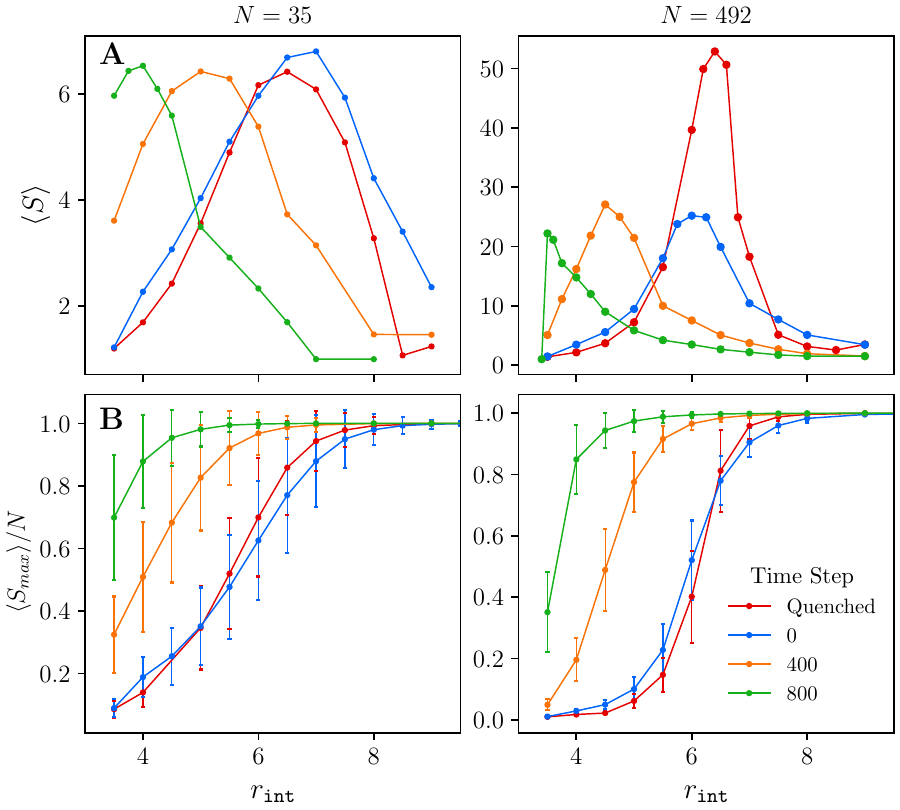}
 \fi
  	\caption{
   \textbf{A}: {\it Mean cluster size} $\langle S \rangle$ and \textbf{B}: {\it Average giant component} normalized by the system size $\langle S_{max} \rangle /N$
   as a function of the interaction radius $r_{\tt int}$ for Kilombo configurations 
   integrated over different time-steps $\Delta t=0, 400,800$ Kilobot loop iterations, 
   and for random quenched configurations.
   The left panel shows the results for $N = 35$ Kilobots, while the right panel shows 
   results for $N = 492$ Kilobots and the same number density $n=0.028$ ${\tt bots}/cm^2$.
   }
	\label{fig:MC_AGC_vs_ri_Kilombo}
\end{figure}

Figure~\ref{fig:MC_AGC_vs_ri_Kilombo} displays the mean cluster size and average 
giant component obtained from Kilombo simulations of Kilobots as they execute 
PRW trajectories.
Specifically, we calculate the mean cluster size characterizing their communication 
network integrated over the exploratory, or advertising, time window $\Delta t$.
This integration considers the total number of communication contacts accumulated 
over the time-step $\Delta t$.
We examine the communicating clusters for various time windows $\Delta t=0,400,800$ 
(measured in Kilobot's loop iterations) or equivalently, for $\Delta t=0, 4.12$, 
and $8.24$ seconds, and for two different system sizes, with $N=35$ and $492$ Kilobots. 
At this point, we want to increase the system size while maintaining a constant 
Kilobot density, so the arena size is adjusted to keep a constant value 
of $n=N/\pi R^2=0.028$ ${\tt bots}/cm^2$.
Later on, we will also vary the Kilobot density for completeness.

\begin{figure}[t!]
	\centering
 \iffigures
 \includegraphics[width=1\columnwidth]{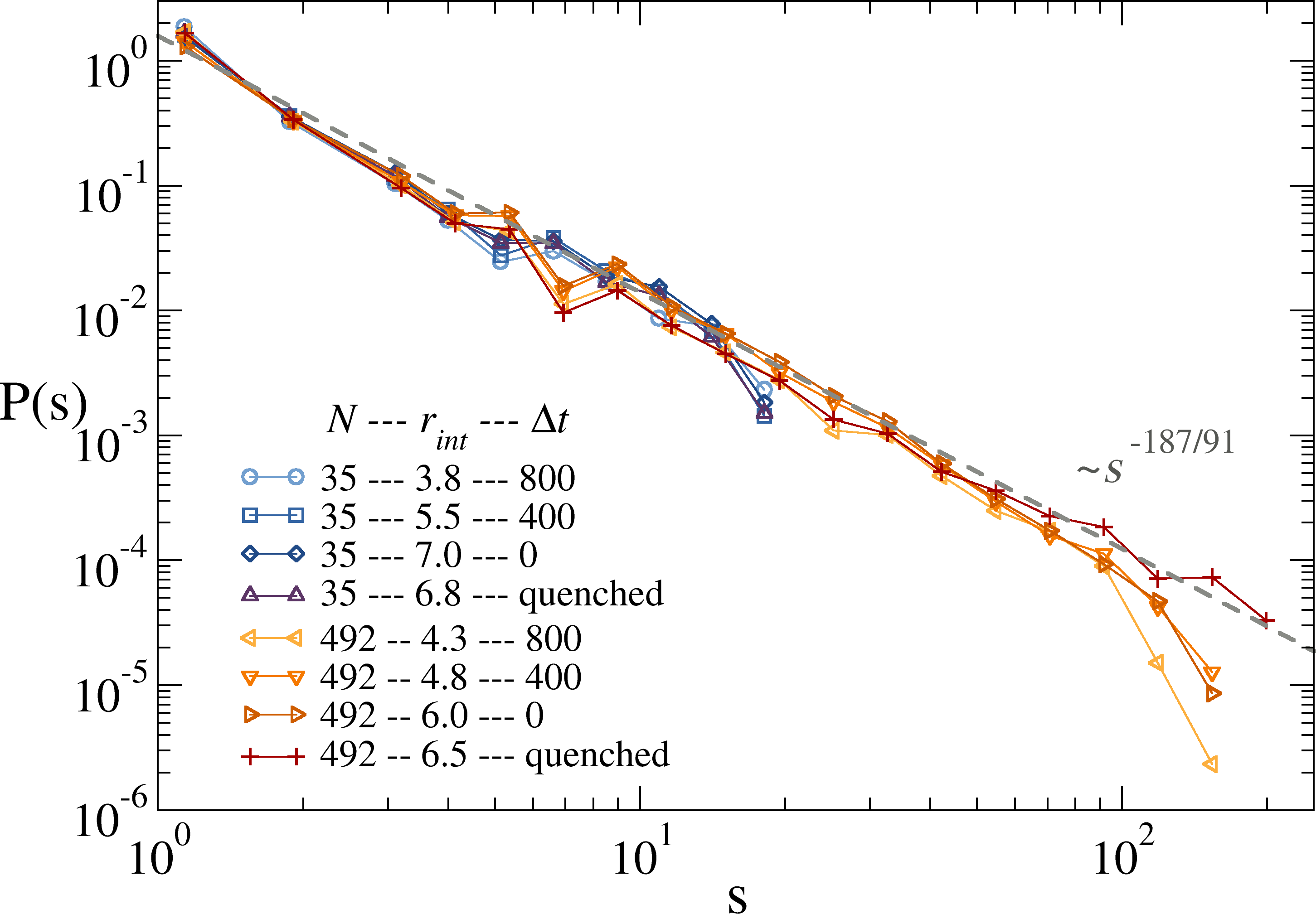}
 \fi
  	\caption{
   {\it Probability distribution of cluster sizes $P(s)$} 
   at the percolation transition for Kilombo and quenched Kilobot simulations.
   The plot shows results for two system sizes with $N = 35$ and $N=492$ and the same Kilobot number density $n=0.028$ bots/cm$^2$. 
   For $N=35$, the corresponding percolation thresholds are: $r_{\tt int}^* \simeq 3.8$ cm (Kilombo configurations integrated over $\Delta t = 800$ loops), $r_{\tt int}^* \simeq 5.5$ cm (Kilombo configurations integrated over $\Delta t = 400$ loops), $r_{\tt int}^* \simeq 7.0$ cm (instantaneous Kilombo configurations with $\Delta t = 0$), and $r_{\tt int}^* \simeq 6.4$ cm (random quenched configurations). 
   For $N = 492$ one finds: $r_{\tt int}^* \simeq 4.3$ cm (Kilombo configurations integrated over $\Delta t = 800$ loops), $r_{\tt int}^* \simeq 4.8$ cm (Kilombo configurations integrated over $\Delta t = 400$ loops), $r_{\tt int}^* \simeq 6.0$ cm (instantaneous Kilombo configurations with $\Delta t = 0$), and $r_{\tt int}^* \simeq 6.5$ cm (random quenched configurations). 
   The gray dashed-line corresponds to a power law decay with an exponent $\tau=187/91$ characterizing the cluster distribution in $2D$ continuous percolation. 
   It emphasizes the compatibility of our results with a percolation transition.
   }
	\label{fig:Ps_vs_s_quenched}
\end{figure}

The mean cluster size $\langle S \rangle$ varies with the interaction 
radius $r_{\tt int}$, as shown in Fig.~\ref{fig:MC_AGC_vs_ri_Kilombo}A,
exhibiting a peak at a threshold radius denoted by $r_{\tt int}^*$.
For comparison and consistency check, we have included a clustering analysis 
for random quenched configurations of Kilobots using the same sizes and 
density conditions. 
Notice that indeed the data corresponding to $\Delta t=0$ (instantaneous snapshots) 
closely resembles the results obtained from the quenched configurations. 
The small discrepancies are due to the existence of short range spatial correlations 
in configurations obtained from the Kilobots' dynamics, which includes collisions. 
Such correlations are absent in the quenched configurations.

The threshold radius $r_{\tt int}^*$ undergoes a notable shift towards 
smaller values as $\Delta t$ increases.
Larger values of $\Delta t$ correspond to increasing intercommunication opportunities
due to the Kilobot's advertising dynamics (see App~\ref{app:botsseen}), and therefore 
to higher number of contacts in the communication network. 
Thus, a larger $\Delta t$ translates into a reduced percolation threshold radius, 
beyond which a giant communication component forms. Fig.~\ref{fig:MC_AGC_vs_ri_Kilombo}B 
shows the average size of the giant component as a function of  $r_{\tt int}$ 
for the same values of $\Delta t$. 
As the giant component approaches saturation the communication network has percolated. 
In particular, after an advertising time window of $\Delta t=800$ loops ($8.24$~s) in 
the experimental system with $N=35$ Kilobots, the percolation threshold shifts from
approximately $6.5 \pm 0.1$~cm to $r_{\tt int}^*=3.75 \pm 0.25$~cm, very close to the 
minimum distance between a Kilobot pair ($r_e\simeq 3.3$~cm) due to excluded volume interactions (see Supp. Fig.~\ref{SM-suppfig:MCS_relation}).

Interestingly, at the finite-size percolation threshold, the normalized distribution
$P(s)=n(s)/n_{\text{tot}}$ of cluster sizes exhibits a power-law decay, 
$P(s)\sim s^{-\tau}$, persisting up to a cutoff value that depends on system size. 
This scaling behavior is shown in Fig.~\ref{fig:Ps_vs_s_quenched}, where we observe a 
fair consistency with an exponent value $\tau \simeq 2.055$
expected for continuous percolation in $D=2$~\cite{stauffer2018}.
Beyond the cutoff, the decay becomes usually much steeper. 
A percolation analysis based on physical contact between particles is reported in \cite{levis_synchronization_2017}, where motile particles interact in order to synchronize their internal oscillators. In that study, physical percolation occurs at higher packing fractions, leading to significant many-body interactions in the system. As a result, they obtain a slightly smaller exponent for the power-law decay of the cluster size distribution at the percolation transition, $\tau \simeq 1.7$. It would be intriguing to explore whether a similar exponent emerges in our experiments under similar crowding conditions. However, such analysis is currently beyond the scope of the present work.

In summary, increasing values of $\Delta t$ yield lower percolation threshold values
for the pairwise interaction radius, as a result of the Kilobot exploratory dynamics. 
By moving, bots increase their average number of communication contacts favouring the 
widespread of information through the system. 
Given that in the experimental set-up discussed in the previous section we have 
$N=35$ Kilobots with an approximate infrared communication radius of $7$~cm, 
exploring their neighborhood for a time window of $\Delta t=800$ loops ($\Delta t=8.24$~s), 
we can conclude that the Kilobot dynamics is effectively generating a percolating 
infrared communication network, which enables information exchange at the system-wide scale, or as in a fully-connected mean-field like scenario.

\subsection*{Crowding effects in consensus reaching}

%%%%%%%
%%% Crowding effects on the relative pupulations in the quenched approximation
%%%%%%%

\begin{figure*}[!t]
	\centering
 \iffigures 
    \includegraphics[width=0.8\textwidth]{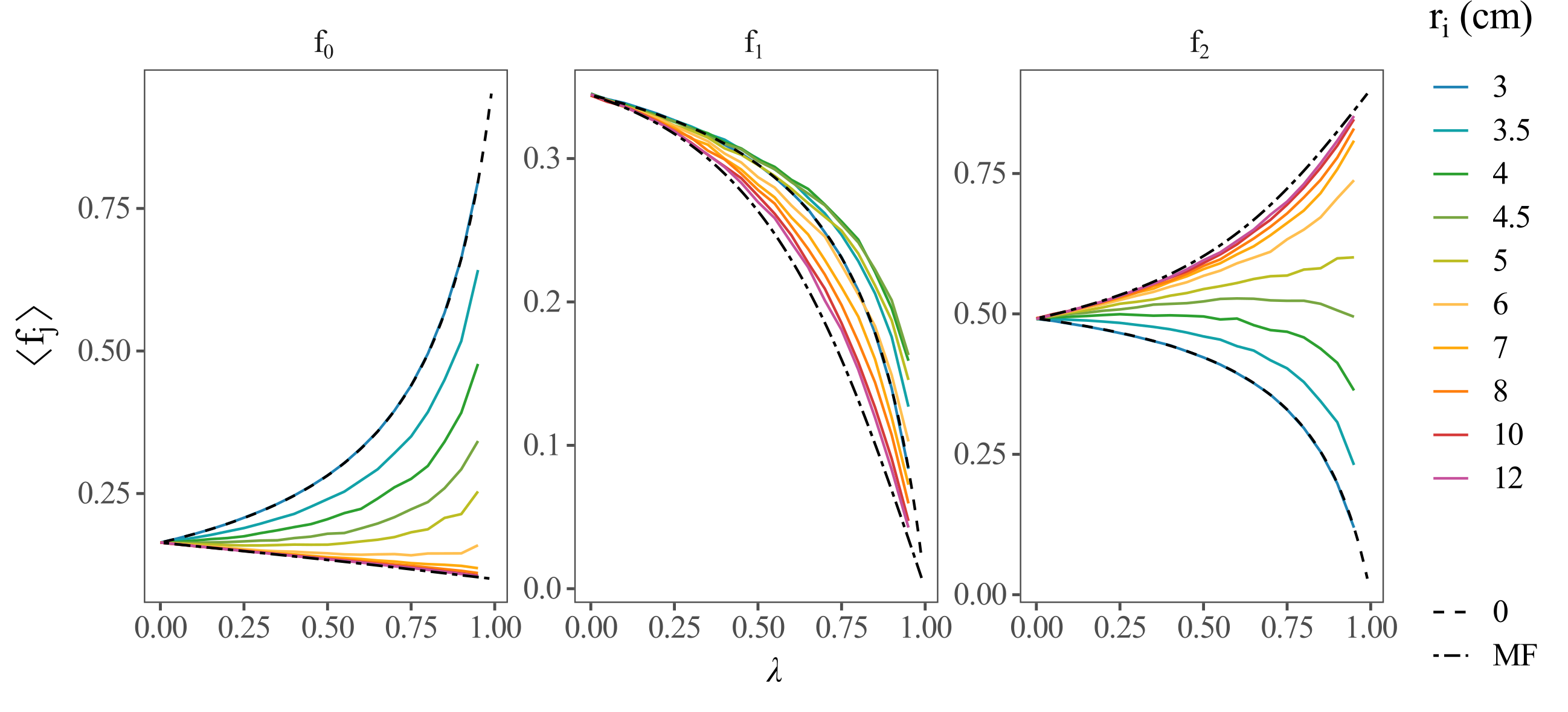}
   \fi 
    \caption{
    {\it Stationary proportions of bots dancing for the different sites} as a 
    function of the interdependence parameter $\lambda$ for different values of the 
    communication radius $r_{\tt int}$. 
    Color curves correspond to simulations of the model on random quenched configurations.
    Dot-dashed lines show the result of the mean-field approximation, and 
    dashed lines display the limit case of isolated bots, i.e. $r_{\tt int} = 0$.}
	\label{fig:ri_comp_quenched}
\end{figure*}

Once we understand the importance of communication among seemingly sparse 
and distant individuals, while they integrate remote sensing over a short 
temporal window as they disperse in space, we can return to the study of 
the main collective decision-making observables in less favorable conditions. 
In particular, it is now evident that the quantitative values of the dance frequencies, 
including uncommitted individuals ($f_0$) and those promoting specific sites 
($f_1$, $f_2$), as well as the consensus parameter $Q$, will depend on the swarm's 
effective crowding.

In this section, we explore how the number density of Kilobots $n=N/\pi R^2$, their  
communication distance $r_{\tt int}$, and their sensing time $\Delta t$ influence 
the consensus outcome, considering different values of the interdependence 
parameter $\lambda$.
Eventually, the dimensionless control parameter $ \eta = N r_{\tt int}^2/R^2$, 
measuring the effective area {\em covered} by Kilobots, will determine the formation 
and strength of consensus as a function of the bee-like model parameters $\lambda$, $q$, 
and $\pi$. 

To illustrate crowding effects, we start by systematically analyzing the influence of 
the communication range $r_{\tt int}$ on the decision-making process in the quenched
configuration approximation.
Figure~\ref{fig:ri_comp_quenched} displays the stationary values of the dance 
frequencies $f_0$, $f_1$ and $f_2$ as a function of the interdependence parameter 
$\lambda$.
The data is presented for various values of the interaction radius 
$r_{\tt int} \in [3,12]$ cm, represented by different colored curves. 
We fix the number density of Kilobots to $n=0.028$ bots/cm$^2$, nest-site qualities 
$q_1=7$ and $q_2=10$, and independent discovery probabilities $\pi_1=\pi_2=0.3$ to match 
the experimental conditions. 
For comparison, we also include the expected stationary values in the mean-field 
approximation, represented by dotted-dashed lines, as well as the limiting 
case where bots remain completely isolated, a trivial limit of the decision making 
process that one can easily work-out analytically, shown as black dashed lines.
The dependency of these zero-interaction curves on $\lambda$ arises 
from the fact that increasing $\lambda$ limits self-discovery 
(as one can deduce from Eq.~\ref{eq:list_probs}) without any information 
exchange taking place.

For the largest interaction radius, the stationary values of $f_j$ tend 
towards the mean-field predictions. 
As discussed previously, beyond an interaction radius of approximately $6.5$ cm, 
we have a percolating communication network that allows information exchange 
among nearly all bots in the system. 
Consequently, for all interaction radii greater than this threshold 
($r_{\tt int}^*\simeq 6.5$), the $f_j$ curves roughly match the mean-field results, 
and completely stabilize around $r_{\tt int}\sim 10$ cm.
Conversely, as the interaction radius decreases from $r_{\tt int}^*$, 
the $f_j$ values significantly deviate from mean-field predictions, 
particularly for higher values of the interdependence parameter $\lambda$, 
when communication capabilities become crucial.
When $r_{\tt int} \leq 3$ cm, due to excluded volume effects, there are no 
bots within the intercommunication distance (since each bot has a diameter of $3.3$ cm, 
touching robots cannot interact), and the stationary values of $f_j$ follow the isolated
Kilobots limit.

\begin{figure*}[t!]
	\centering
    \iffigures
   \includegraphics[width=0.95\textwidth]{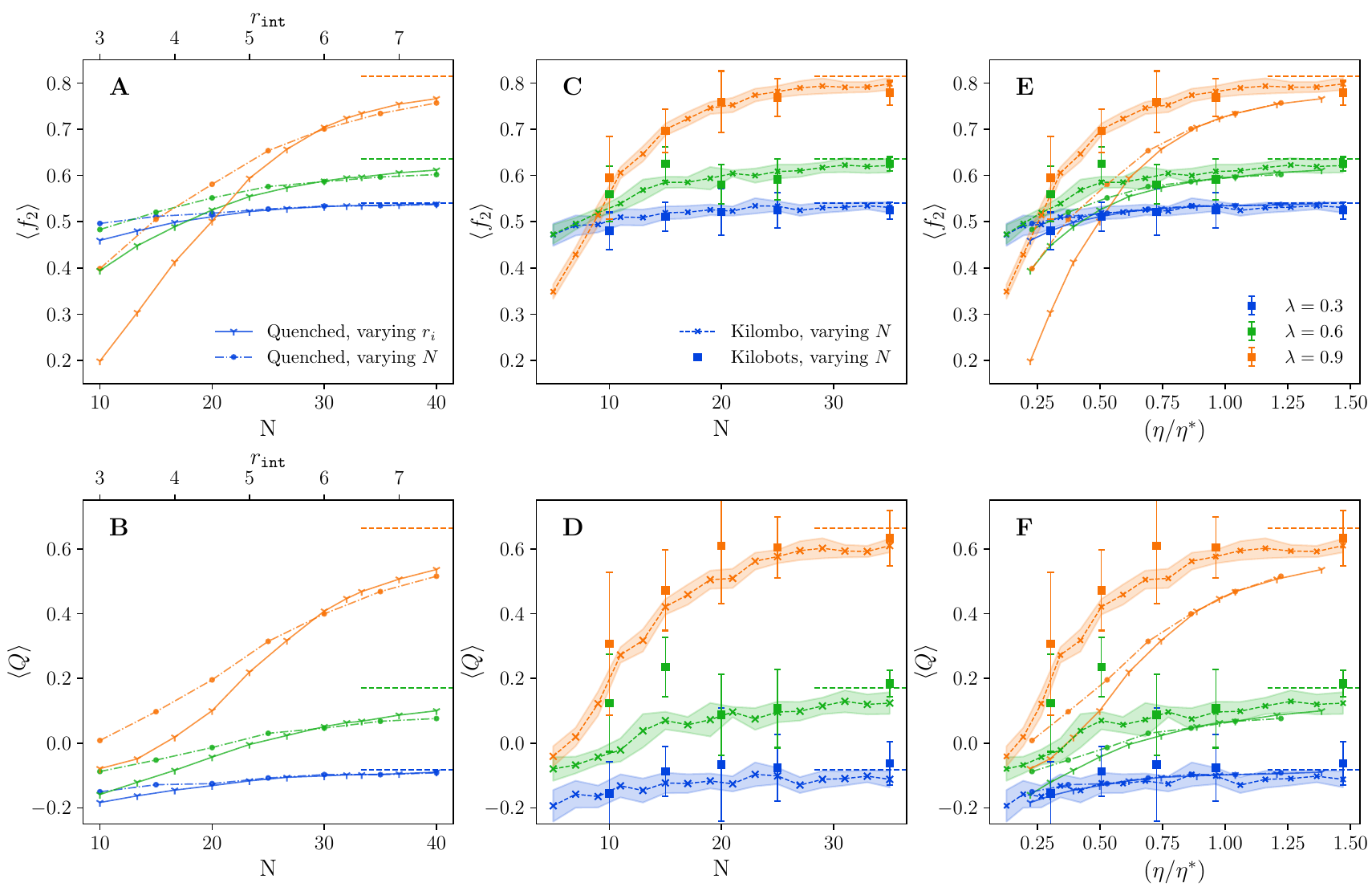}
	\fi
     \caption{
     {\it Stationary dance frequency for the high quality site $\langle f_2\rangle$} (top) and {\it consensus parameter $\langle Q \rangle$} (bottom) as a function of the crowding control variables $N$, $r_{\tt int}$, and $\eta$. 
     Figures \textbf{A}-\textbf{B} show results obtained from 
     (i) different ensemble sizes $N$ and a fixed value of $r_{\tt int}=6.5$ cm 
     (dashed-dot lines with dot symbols), 
     (ii) from different interaction radii $r_{\tt int}$ and a constant number 
     of $N=35$ bots (continuous lines with Y symbols), both in the quenched position 
     approximation;
     Figures \textbf{C}-\textbf{D} show results obtained from
     (iii) Kilombo simulations (x symbols) varying $N$ and (iv) 
     Kilobot experiments for different ensemble sizes,
     $N=10, \; 15, \; 20, \; 25, \; 35$
     (square symbols).
     All simulations/experiments have been conducted for three contrasting values of the parameter 
     $\lambda = 0.3, 0.6, 0.9$ (blue, green and orange lines, respectively). 
     Panels \textbf{E}-\textbf{F} show the stationary values depicted in A-B and C-D, now as a function of the percolation parameter $\eta$ rescaled by the 
     corresponding percolation threshold $\eta^*$. }
	\label{fig:f2andQ_nriscaled_Kilobots}
\end{figure*}

When $\lambda=0$, the value of the interaction radius becomes irrelevant, 
and all curves converge to the same point, known analytically from the mean-field 
approximation solution of the model. 
The same analysis can be carried out by fixing an interaction radius $r_{\tt int}$ 
and changing the number of bots $N$ in a fixed-size arena. 
The final outcome will be the same, as it is shown in Fig.~\ref{fig:f2andQ_nriscaled_Kilobots}.

Fig.~\ref{fig:ri_comp_quenched} reveals an interesting phenomenon occurring at low $r_{\tt int}$: for high values of $\lambda$, the values of $f_1$ exceed the zero-interaction limit bound. This is because in small isolated clusters present in such quenched configurations, at high values of $\lambda$, one opinion may dominate for long periods without the other entering into the discussion, thus being overrepresented. Both populations are influenced by this effect, but it is more noticeable for $f_1$ as it is expected to diminish when $r_{\tt int}$ increases, while $f_2$ is expected to grow. This overshooting effect occurs at the expense of the uncommitted population; hence, the pronounced decrease of $f_0$ for low $r_{\tt int}$ and high $\lambda$ can be attributed to this phenomenon.

%%%%%%%
%%% Crowding effects on the relative pupulations in Kilobots and Kilombo
%%%%%%%

Finally, we investigate the impact of crowding on mobile PRW Kilobots at different 
densities by changing the number of bots $N \in [5,35]$ in the same arena of radius 
$R=20$ cm. 
Based on the analysis presented in the previous section and in 
Fig.~\ref{fig:ri_comp_quenched}, we anticipate that Kilobots, 
with the ability to move and to gather information over a temporal period $\Delta t$, 
will achieve better communication and consequently higher consensus 
levels compared to quenched configurations in the same conditions.

Figure~\ref{fig:f2andQ_nriscaled_Kilobots} provides a comparison of stationary 
averaged values of $f_2$ and $Q$ in experiments, Kilombo simulations, 
and simulations using random quenched configurations. 
We consider three contrasting values of the interdependence parameter $\lambda$ in the 
symmetric probability discovery scenario.
In Fig.~\ref{fig:f2andQ_nriscaled_Kilobots}A-B, we show the impact of working with 
different Kilobot numbers $N$ or restraining their communication radius $r_{\tt int}$.
Plots~\ref{fig:f2andQ_nriscaled_Kilobots}A-B show the comparison between varying $N$ 
or $r_{\tt int}$ in the quenched approximation. 
Furthermore,  plots~\ref{fig:f2andQ_nriscaled_Kilobots}C-D show experimental results 
for varying number $N$ of Kilobots contrasted with Kilombo simulations.
In all cases the average values of $f_2$ and $Q$ increase gradually with $N$ until they 
reach a plateau for $N>N^*\sim 30$ Kilobots. 
The plateau values lie very close to the asymptotic mean-field results for each value of
$\lambda=0.3, 0.6$ and $0.9$. 
This means that even for lower number densities than the ones considered in 
the experiments shown in Sec.~\ref{sec:consensus_reaching} (where $N = 35$), 
the system seems to be able to perceive global information about dancing frequencies 
and to achieve consensus values very similar to mean-field results. 
This is, again, a signature of the formation of a percolating communication network as 
a function of Kilobot density.
On the contrary, the consensus parameter for very small ensemble sizes is below zero for 
the three contrasting values of the interdependence parameter considered, and thus strong
consensus for the best quality option is not achieved in such poorly communicated swarms. 

Within error bars, the stationary values of $f_2$ and $Q$ obtained from experiments with 
$N \in [5,35]$ Kilobots match those from Kilombo simulations. 
Furthermore, moving Kilobots achieve higher levels of consensus than quenched 
configurations with the same number of robots. 
This is particularly evident for small groups of Kilobots and high enough values of
interdependence. 
We have seen that the PRW motion of Kilobots enhances the transmission of information 
and, thus, shifts the percolation transition towards smaller values of the coverage 
parameter $\eta$, by either decreasing the threshold value of $r_{\tt int}^*$ or, 
equivalently, by reducing the number $N^*$ of Kilobots required to observe the 
percolation transition. 
This shift can be clearly seen by comparing Figs.~\ref{fig:f2andQ_nriscaled_Kilobots}A-B 
and Figs.~\ref{fig:f2andQ_nriscaled_Kilobots}C-D, for both experimental and Kilombo 
simulation results of moving Kilobots that establish new contacts over a given time 
window of $\Delta t = 800$ loops.

In Figs.~\ref{fig:f2andQ_nriscaled_Kilobots}E-F, the same data is now represented in 
terms of the dimensionless control parameter $ \eta = N r_{\tt int}^2/R^2$, 
measuring the effective area {\em covered} by Kilobots, rescaled by the corresponding
percolation threshold $\eta^*$ calculated in each case. 
Despite the rescaling by $\eta^*$ yields qualitatively similar dependencies on the 
control parameter for all the cases considered, moving Kilobots slightly outperform 
the consensus reached in quenched conditions at high values of interdependence. 
This is probably due to the existence of enhanced spatial correlations below the 
percolation transition. 
Such correlations appear as a result of their characteristic dynamics, which includes 
the possibility of exhibiting collisions and jams and, therefore, the formation of 
slightly larger clusters.
This fact improves the formation of strong consensus for the best quality site around 
the percolation transition, when $\lambda$ is high enough, as large clusters chiefly 
vote for the same option, enhancing $f_2$ while hindering $f_1$.

% OLD:
%Overall, these results highlight the importance of information communication and interdependence in the Kilobots’ collective decision-making process, and demonstrate the capabilities of swarm robots to successfully achieve such a complex task in far from ideal conditions. Figure~\ref{fig:f2andQ_nriscaled_Kilobots} shows that high enough interdependence, is indeed crucial for building up strong consensus for the best-available option in poorly communicating swarms. 
% 
Overall, these results underscore the significance of information spreading facilitated by agents' mobility, interaction time, and interdependence. They highlight the capability of robot swarms to achieve consensus successfully even under less-than-ideal conditions. Identifying the percolation transition allows us to pinpoint the specific combination of parameters that lead to nearly mean-field performance. In contrast to~\cite{trianni2016_naming_game}, our study relies on  mobility networks integrated over a short temporal communication window, emphasizing that agents' mobility and communication capabilities within this timeframe play a pivotal role in fostering strong consensus in natural systems.

To conclude this section, we would like to mention that the percolation threshold is 
affected by finite size effects. 
In particular, the percolation correlation length for finite systems can only attain 
a maximum value 
\begin{equation}
    \xi_{max} \sim (\eta^{*}(N) - \eta^{*}_{\infty})^{-\nu}\sim N^{1/d_f}
\end{equation}
where $\nu = 4/3$ and $d_f = 91/48$ are the critical exponents for the correlation 
length and the fractal dimension~\cite{stauffer2018}, respectively, for continuous percolation in $D=2$. 
Therefore, for the percolation threshold one expects that,

\begin{equation}
    \eta^{*}(N) \sim \eta^{*}_{\infty} + C N^{-\frac{1}{\nu d_f}},
\label{eq:pstar}
\end{equation}
where $C$ is an arbitrary constant. 
From this expression, we expect that the percolation radius for different Kilobot  densities scales as
\begin{equation}\label{eq:scalin_rint}
(r_{\tt int}^{*}(N))^2 \sim (r_{\tt int}^{*\infty})^2 + C' N^{-\left(\frac{1}{\nu d_f} + 1\right)}
\end{equation}
with $(r_{\tt int}^{*\infty})^2 \sim 0$. Indeed, in the case of quenched configurations 
we obtain $r_{\tt int}^{*2} \sim N^{-1.21}$, while in the case of Kilobots we obtain 
$r_{\tt int}^{*2} \sim N^{-1.26}$ 
%(see Supplementary Material~\cite{suppmat}) 
(see Supp. Fig.~\ref{SM-suppfig:MCS_relation}. 
These exponents, although slightly smaller than the continuous percolation expectation, 
are another indicator of a percolation process taking place in the communication network.

\section{Discussion}
\label{sec:discussion}

We have investigated the problem of the nest-site selection process 
of honeybee swarms using Kilobots, i.e. minimalist robots that can mimic their 
consensus-reaching behavior.
Kilobots engage in a honeybee-like collective decision model 
while they move in the experimental arena, and we analyze how adding 
space and local interactions affects consensus reaching in 
(a simplified variant of) the model proposed by List and 
coworkers in 2009~\cite{list2009}.
In order to rationalize our experimental results, we use an analytical approach, 
obtained from the deterministic differential equations governing the dynamics 
in the mean-field approximation~\cite{galla2010, marchpons2024consensus}, as well as numerical simulations in both 
fully connected and random quenched Kilobot configurations.
Furthermore, we complement the limited statistics of our experiments with 
simulations using the Kilombo emulator of the Kilobot dynamics.

The problem of reaching consensus decisions in a decentralized manner 
--i.e., the need of targeting the best among many available options when 
many agents participate in the decision process and none of them exerts 
particular influence--  displays great complexity and beauty and relies 
on information pooling and on communication. 
In our experiments, self-discovery and imitation, i.e. {\it independence} 
and {\it interdependence}, are both essential ingredients in collective 
decision making. 
Despite the differences between our studied model and other honeybee-inspired models such as \cite{valentini2014, pais2013, reina_desing_pattern, reina2017}, we similarly highlight the importance of social interactions in collective decision-making processes. We find that high levels of interaction enable the swarm to more accurately identify the highest-quality option, especially when the qualities of options are close or when discovery probabilities (uncorrelated with site qualities, unlike the aforementioned works) are large.
The primary distinction between our model and others in the literature is the absence of cross-inhibition ~\cite{zakir_robot_2022, reina_cross_inhibition_2023}. As a result, if symmetry breaking is required to resolve deadlocks, it can be achieved through a simple adaptation in behavior—specifically, by ceasing exploration. For an extensive analysis of the parameter space and dynamics of symmetry breaking, refer to~\cite{marchpons2024consensus}.
Finally, in line with recent studies that also use decentralized robot swarms~\cite{trianni2016_naming_game, valentini2016, Reina2018_spatiality, talamali2021_less_more, zakir_robot_2022}, we have demonstrated that Kilobot swarms are capable of reaching such a complex consensus decisions collectively in a decentralized manner even in far from ideal conditions.

The resulting quantitative strength of the final consensus depends not only on the prescriptions of independence and interdependence characterizing the underlying opinion dynamics but also on the resulting communication network produced by the agents' mobility.
This approach allows us to extend previous studies on swarm decision making, which have already considered the topology of interactions~\cite{trianni2016_naming_game, khaluf_interaction_models_2018, raoufi2023}, by explicitly examining the time-varying and time-integrated communication network along with its percolation transition. 
We have observed that, within the bounded decision-making space, the temporal window during which agents can interact, in addition to the communication range and bot density, controls the percolation of the communication network. At percolation, our data for cluster size distributions, as well as for the finite-size scaling of the percolation threshold itself, are consistent with standard continuous percolation critical exponents in $D=2$. 
Besides underscoring mobility as a crucial factor in order to foster consensus within a poorly swarm~\cite{trianni2016_naming_game} (i.e. below an static percolation threshold), our approach acknowledges the significance of agents' mobility within a short temporal communication window in real, natural or robotic, systems for identifying a dynamic communication percolation threshold. Specifically, the interaction patterns generated by mobility reduce the necessary interaction range or system density compared to static networks. 
%We view this as a qualitative advancement compared to previous studies~\cite{trianni2016_naming_game}.

%
Thus, we have concluded that an effective communication coverage, which integrates communication range, agent density, and motility, controls the transition, 
enhancing the effective information transfer of a network based on proximity interactions rather than on long-range scale-free degree distributed interactions~\cite{khaluf_interaction_models_2018}. This facilitates the achievement of 
consensus in the model dynamics.
Without the communication coverage reaching a critical threshold, the consensus 
is poor or nonexistent, and high enough interdependence, or imitation, 
turns out to be crucial for building up strong consensus for the best-available 
option in poorly communicating swarms.
This is even more the case in asymmetric scenarios where self-discovery favors 
sub-optimal options. 

Our study contributes to the understanding of the complexity of 
decentralized decision-making by interacting and moving agents,
establishing the main variables to pay attention at.
Simultaneously, it raises a warning on the interpretation of 
simple agents models solved at the mean-field level or simulated 
in static regular grids. 
This is especially relevant when considering the limited communication capabilities of real systems, such as the social insects that inspire our research.

It's important to note that our analysis assumes a temporally invariant scenario, where options, discovery probabilities, and qualities remain constant. However, it would be interesting to investigate alternative scenarios where agents encounter a changing environment, such as options with fluctuating qualities or the appearance of new options. In such cases, a more constrained topology of interaction may prove to be a beneficial asset for the swarm, as observed in recent studies (e.g., \cite{talamali2021_less_more, aust_hidden_2022}).

We believe that the swarm robotics approach, whether through conducting real experiments or simulations with realistic emulators, is better suited for studying the capabilities of complex real systems. This is because it allows for the introduction of imperfect, yet to some extent uncontrollable, motion, communication, and synchronization capacities.
By extensively describing their advantages and limitations, their 
capabilities and uncertainties, we give a recipe on how to 
address them as a swarm when playing a democratic game.
Alongside the literature on collective decision making in robotic swarms, we hope that our work will encourage further analysis from a physics perspective on {\it mini-robots as programmable social matter}.

\begin{acknowledgments} 
We acknowledge financial support from the Spanish MCIN/AEI/10.13039/501100011033, 
through projects PID2019-106290GB-C21, PID2019-106290GB-C22, PID2022-137505NB-C21 
and PID2022137505NB-C22.
D.M. acknowledges support from the fellowship FPI-UPC2022, granted by 
Universitat Polit\`ecnica de Catalunya.
E.E.F. acknowledges support from the Maria Zambrano program of the 
Spanish Ministry of Universities through the University of Barcelona
and PIP 2021-2023 CONICET Project Nº 0757.
We would like to thank Ivan Paz for professional help in the design of 
the \textit{kilocounter} tracking software and Quim Badosa for technical 
assistance in Kilobot experiments.
\end{acknowledgments}

\appendix
\renewcommand{\thefigure}{A\arabic{figure}}
\setcounter{figure}{0}

\section{Methods}
\label{sec:methods}

\subsection{Deterministic solutions of the Mean Field Model}
\label{app:analytical}

In Ref. \cite{galla2010}, T. Galla provided an analytical approach to List {\em et al.} 
model using a simple alteration of the original model. As mentioned in the main text, 
Galla replaces the state variables $(s_{i,t}, d_{i,t})$ by $s_{i,t}$, and introduces a 
dance abandonment rate $r_j$ for each site such that $<r_j> \sim q_j^{-1}$. 
Following the mathematical details provided in \cite{galla2010} one can arrive to a set 
of deterministic differential equations that describe the time evolution of the system. 
For each site $j=1,...,k$:
\begin{equation}
	\langle\dot{f}_{j,t}\rangle = (1 - \rho(t))[(1-\lambda)\pi_j + \lambda \langle f_{j,t}\rangle] - r_j \langle f_{j,t}\rangle
	\label{eq:fjt_analytical}
\end{equation}
where $\rho (t) = \sum_{\alpha = 1}^{k} \langle f_{\alpha,t}\rangle$.
Eq.~\ref{eq:fjt_analytical} can be numerically integrated for any fixed choice 
of parameters. 
Nevertheless, an expression for the stationary points of these equations can be found as 
the solution of $k$ coupled quadratic equations, obtained by setting 
$\langle\dot{f}_{j,t}\rangle = 0$,
\begin{equation}
	f^*_j = \biggr[\frac{r_j}{1-\rho^*}  - \lambda \biggr]^{-1} (1 - \lambda) \pi_j \quad j = 1,...,k.
	\label{eq:fj_analytical}
\end{equation}

In order to solve this system of equations, that unavoidably depends on the stationary 
value $1 - \rho^{*} = f_0^{*}$, one can combine the $k$ equations to solve first a closed
equation for $f_0^{*}$:
\begin{equation}
	f^*_0 = 1 - (1 - \lambda)\sum^k_{j=1}\biggr[\frac{r_j}{f^*_0}-\lambda\biggr]^{-1} \pi_j. 
	\label{eq:f0_analytical}
\end{equation}
Eq.~\ref{eq:f0_analytical} has $k+1$ roots that can be found by solving the equation 
numerically or by rearranging it as a $(k+1)$-th degree polynomial in $f_0$. 
Some of these roots lead to unphysical solutions with $f^*_0 > 1$. 
From the remaining valid solutions with $f_0^* \leq 0$, only one leads to valid 
($f^*_j \leq 1$) and linearly stable solutions for the rest of dance frequencies. 
Stochastic simulations and the integration of Eq.~\ref{eq:fjt_analytical} confirm 
the stability of this result. 

The extreme cases $\lambda = 0$ and $\lambda \rightarrow 1$ have simpler solutions. 
First, setting $\lambda = 0$ in Eq.~\ref{eq:f0_analytical} leads to a simpler solution,
\begin{equation}
    f_0^{*} = \frac{1}{1 + \sum_{m=1}^k \pi_m q_m}, 
    \label{eq:f0_lambda0}
\end{equation}
that we can use to compute the result for the rest of the dancing frequencies. 
Using Eq.~\ref{eq:fj_analytical}, we obtain
\begin{equation}
    f_j^{*} = \frac{\pi_j q_j}{1 + \sum_{j=m}^k \pi_m q_m}. 
    \label{eq:fj_lambda0}
\end{equation}
When $\lambda \rightarrow 1$, due to the extreme reliance on interdependence, the site with a greater quality will be finally dominating the whole system, leaving no 
agents committed to the other sites and only a small quantity of uncommitted agents. 
Assuming that $q_k > q_{k-1} > ... > q_1$, we can impose that $f_1^* = ... = f_{k-1}^* = 0$, 
and using~\ref{eq:fj_analytical} we find the following stationary solution:
\begin{equation}
    f_0^* = r_k, \quad f_1^* = ... = f_{k-1}^* = 0, \quad f_k^* = 1 - r_k.
    \label{eq:fj_lambda1_buena}
\end{equation}
This result is validated by simulations, or after solving the deterministic equations at 
high values of $\lambda$. 
A linear stability analysis confirms that this solution is the only stable solution in the 
limit $\lambda \rightarrow 1$~\cite{marchpons2024consensus}.

\subsection{Kilobots experimental set-up}
\label{app:experimentalsetup}

\begin{figure}[!t]
	\centering
 \iffigures
 \includegraphics[width=1\columnwidth]{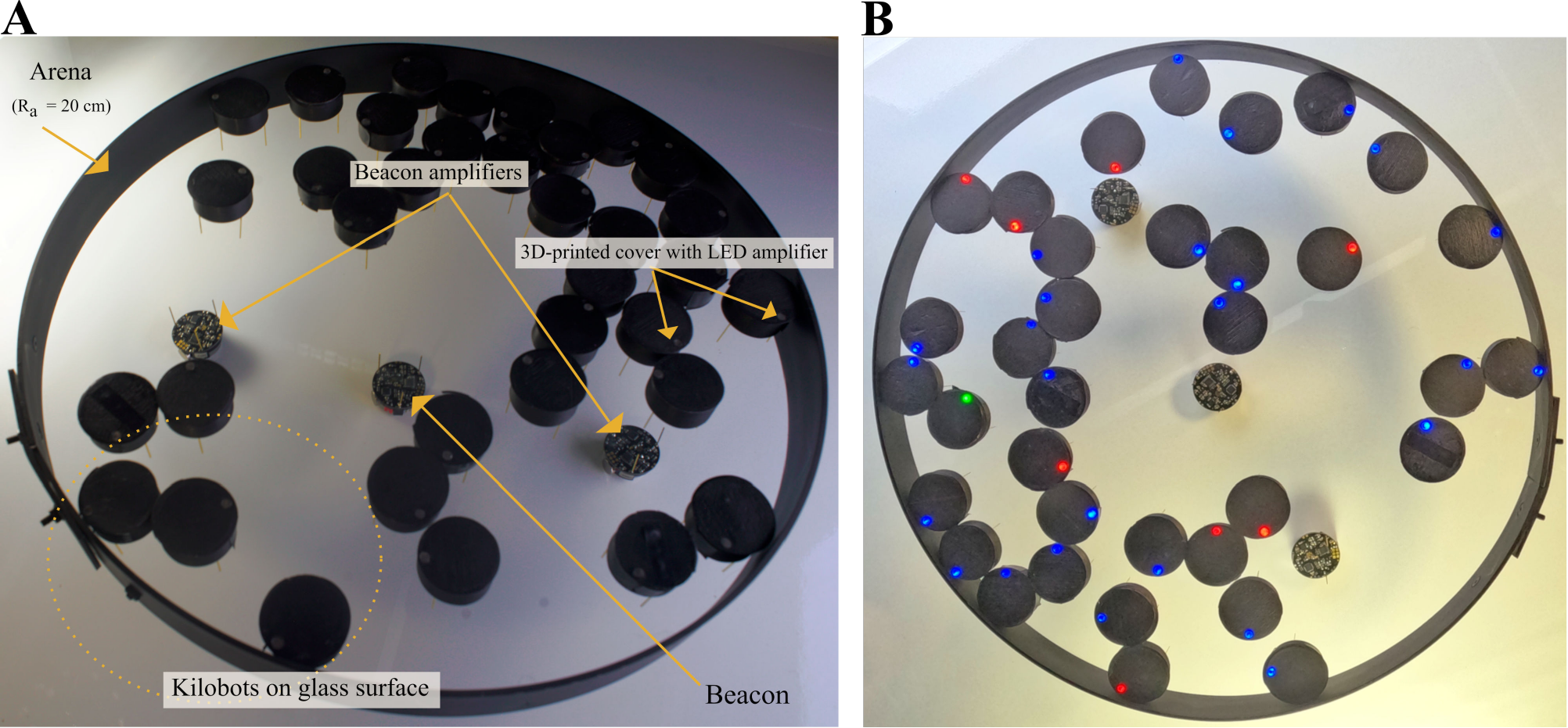}
 \includegraphics[width=1\columnwidth]{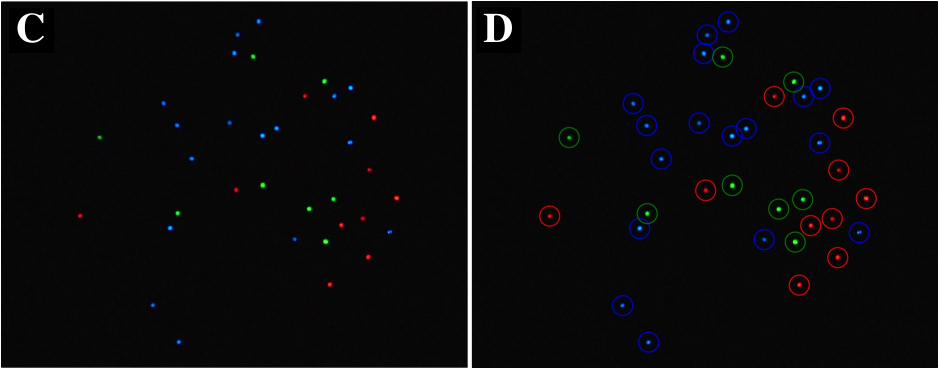}
 \fi
  	\caption{
   \textbf{A}: Kilobots experimental setup. 
   \textbf{B}: Kilobots dynamics as described in Sec.~\ref{sec:beesmodel}
   (see videos 2,3 in~\cite{suppmat}). 
   Each Kilobot is covered by a custom 3D-printed case than enhances visual 
   on the led light and allows for tracking.
   \textbf{C}: A single frame showcasing a group of $35$ Kilobots, each identified 
   by their colored light, dancing for site one (green), site two (blue), 
   or not dancing for any site (red).
   \textbf{D}: The frame in C processed using the kilocounter software.   
   }
	\label{fig:exdesign_Kilobots}
\end{figure}

Kilobots have been instrumental in collective behavior research \cite{rubenstein2012}. 
Kilobots execute user-programmed functions in \textit{loops}, and the loop duration 
varies based on the complexity of the operations. 
In our case, Kilobots perform a persistent random walk while gathering 
information, estimating population frequencies, computing transition probabilities,
and indicating their commitment state during each loop.
For our experiments, we set up a workspace with Kilobots moving on a glass 
surface held $15$ cm above a whiteboard melamine base.
On the melamine base, we place a central Kilobot that acts as a 
beacon to enhance synchronization of the Kilobots' internal clocks, alongside two 
additional Kilobots that amplify the beacon signal.
Our typical setup is illustrated in Fig.~\ref{fig:exdesign_Kilobots}-A -- see also Supp. Video 2.
Synchronization is important to coordinate concurrent processes, such as those involved 
in the collective decision model. 

Our Kilobots operate on a persistent random walk (PRW) dynamics, characterized by straight motion segments with durations that follow an exponential distribution, averaging around $\sim 3.8$ seconds. Upon completing each straight movement, the Kilobots randomly choose to turn left or right. To prevent clustering at the arena's borders, the PRW includes discrete wide turning angles, enabling the Kilobots to turn away from the border more effectively. The Kilobots randomly select between short turns and long turns, with equal probabilities.

Groups of $10$ to $35$ Kilobots move as PRWs in a circular arena with a $20$ cm radius. 
After transmitting and receiving messages, and gathering information from their local
environment,  during a time-step $\Delta t$, typically $800$ loop iterations or 
approximately $8.24$ seconds , Kilobots update their state according to the model dynamics, 
and consequently `dance for' either site 1 (low quality), site 2 (high quality), or for 
no site, displaying such individual state in their LED (red for non-dancing, green if 
dancing for site 1, and blue if dancing for site 2).

To prevent our group of Kilobots from clustering at the wall of the circular observation 
area, their random dynamics was configured with discrete wide turning angles to promote 
quicker turning away from the border. 
Turning times consist of approximately $\sim 2.8$ seconds ($125^{\circ}$) or $\sim 5.8$ 
seconds ($251^{\circ}$), and their moving forward states last approximately $\sim 3.8$ 
seconds.
To better identify the Kilobots' states, we covered each Kilobot with a custom 
3D-printed black casing, leaving only the LED light visible, as seen in Fig.~\ref{fig:exdesign_Kilobots}-B. 

We recorded the Kilobots dynamics and LED states using a digital camera with a spatial
resolution of $1920x1080$ pixels and a temporal resolution of $25$ frames per second.
Each recording session lasts typically 30 minutes. 
To automatically count the number of Kilobots dancing for each site, we extract images 
at each time step $\Delta t$, and we make use of the \textit{kilocounter} software, 
specifically developed for our work~\cite{kilocounter}.
Kilocounter identifies and counts colored blobs in these  
recordings, allowing us to analyze Kilobot behavior, as shown in
Fig.~\ref{fig:exdesign_Kilobots}C-D.
To facilitate the tracking, videos are recorded in a dark setting (see Supp. Video 3 for an snapshot of the experiments).

\begin{table}[t!]
  \centering
  \caption{Number of realizations of the Kilobot experiments for each experimental condition defined by $N$ and $\lambda$.\\
  }  
  % \begin{tblr}{|Q[c,1.25cm]||*{5}{Q[c,1.1cm]|}}\hline
  \begin{tblr}{|Q[c,0.15\columnwidth]|*{5}{Q[c,0.1\columnwidth]|}}
  \hline
  \hline
  \backslashbox{$\mathbf{\lambda}$}{$\mathbf{N}$} & {\bf 10} & {\bf 15} & {\bf 20} & {\bf 25} & {\bf 35} \\ 
  \hline
  $\mathbf{0.3}$ & 10 & 6 & 5 & 5& 5\\ %\hline
  $\mathbf{0.6}$ & 10 & 7 & 5 & 5 & 5\\ %\hline
  $\mathbf{0.9}$ & 10 & 5 & 5 & 5 & 5\\
  \hline
  \hline
  \end{tblr}
 % \newline %\newline
  \label{experiments_stats} 
\end{table}

Table~\ref{experiments_stats} sums up the amount of experiments conducted for each 
condition, defined by the system size $N$ and the interdependence parameter $\lambda$. 
Due to the time consuming process of conducting the experiments, the number of 
realizations is limited.

\section{Kilobots communication capabilities within the experimental setup}
\label{app:botscomm}

In the methods section, we outlined that the Kilobot experiments were conducted under complete darkness to facilitate tracking of the LED lights for subsequent analysis of the system state. We briefly asses the communication capabilities of the Kilobots under these conditions.

Physical Kilobots are typically designed to interact within a range of 7 cm to 10 cm, as noted in previous studies~\cite{rubenstein2012}. However, Valentini et al. observed a communication range of up to 20 cm for two isolated Kilobots when placed on a glass surface under light conditions~\cite{valentini2016}. We confirmed this observation by conducting the same test.

\begin{figure}[b!]
    \centering
    \includegraphics[width=1\columnwidth]{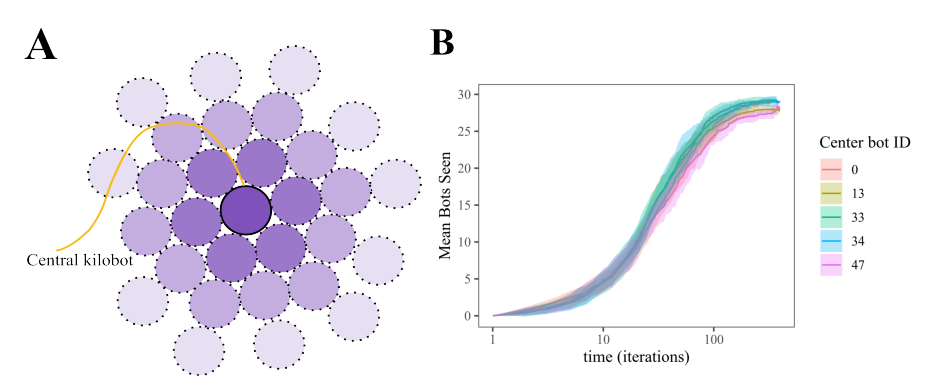}
    \caption{\textbf{A}: Representation of the experimental conditions under which we tested the Kilobots' ability to sense their surroundings up to an approximate distance of 3 Kilobot diameters (approx. 10 cm). \textbf{B}: Mean number of Kilobots seen by a central Kilobot as a function of time. Different Kilobots, identified by their bot ID, where used in the test.}
    \label{fig:botsSeenDark}
\end{figure}

Under more crowded conditions and in complete darkness, we have studied how a central Kilobot can sense the number of Kilobots in its surroundings up to an approximate distance of 3 Kilobot diameters, or approximately 10 cm. Figure \ref{fig:botsSeenDark} depicts the experimental setup (A) and results for different Kilobots (B). We observe that in the long run, Kilobots are able to sense their entire surroundings, consisting of 29 other Kilobots.
In the following section, we further study the Kilobots communication capabilities while performing the same dynamic behavior as in the collective decision making experiments.

\section{Kilobots detected over a time-step}

\label{app:botsseen}

In the study of opinion dynamics, understanding how individuals interact to make 
decisions is crucial. 
Kilobots gather information from their neighbors and then act accordingly.
We want to know how many other bots are detected by a Kilobot during a time-step.

We thus implement an algorithm for a bot to communicate, at each time-step, 
the number of bots seen over the previous time-step.
We will run this algorithm on uncommitted, non-dancing bots, 
while the dancing bots (promoting either site $1$ or site $2$) perform PRWs.
First, we check that the maximum number of Kilobots detected by an uncommitted 
bot (during a time-step $\Delta t = 800$ loops) is around $15$, 
in the 20 cm arena with 35 Kilobots running the nest-site selection model. 
We restrict the count to bots seen within an interaction radius of $\sim 7$ cm,
about $2$ Kilobots' body lengths. 
This is possible by filtering for the infrared signal intensity.
We then divide the number of bots detected during a time-step $\Delta t$ in four 
intervals, and assign each interval a color: 
red ($0-3$ bots), green ($4-7$ bots), blue ($8-11$ bots) and white ($12-15$ bots). 
At each time step, uncommitted bots flush their LEDs according to this color code,  
and we count the numbers with the kilocounter software. 
We perform five repetitions of $210$ time-steps (about $30$ minutes each) to gather 
statistics (over $\sim 20000$ counts) (see Supp. Video 1 for a demonstration of the experiment).

\begin{figure}[!t]
	\centering
 \iffigures
	\includegraphics[width=0.75\columnwidth]{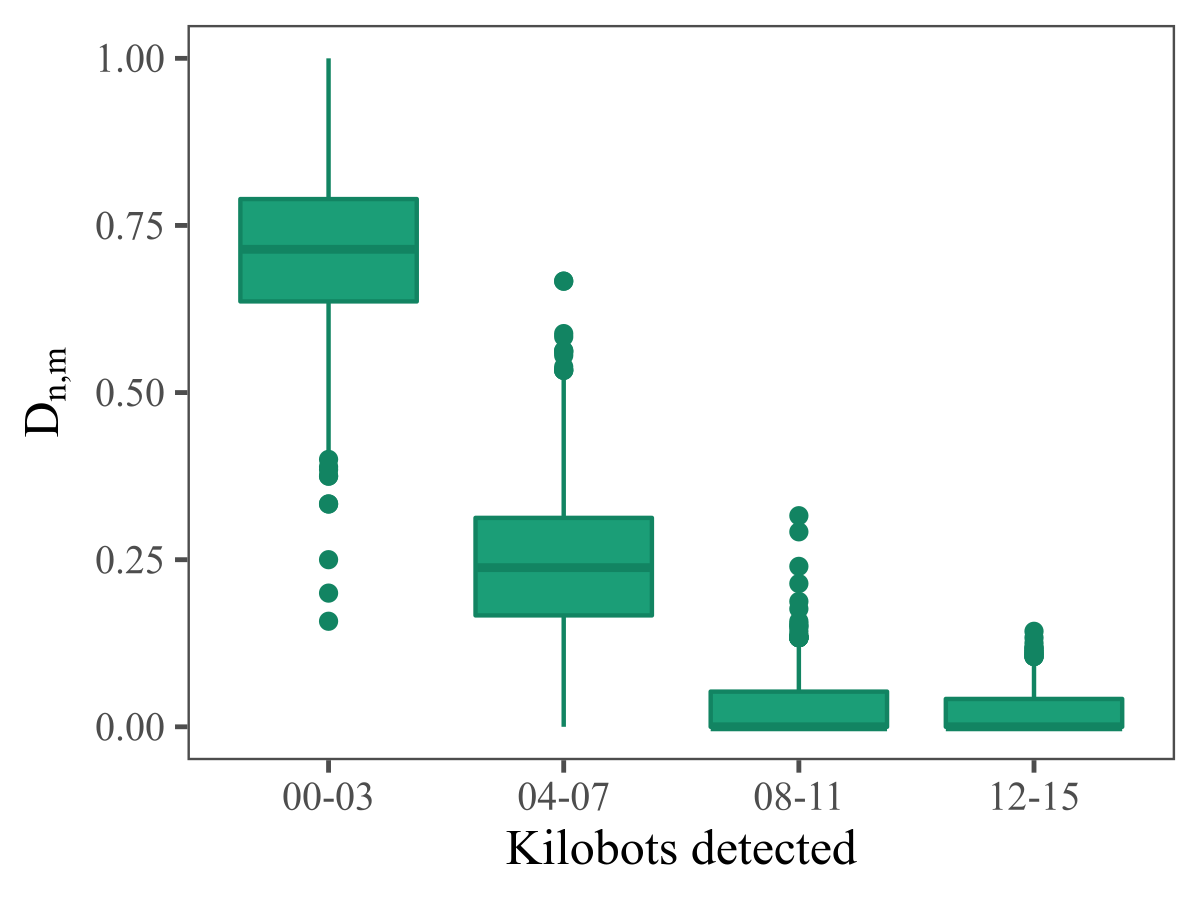}
 \fi
  	\caption{Kilobots detected over a time step $\Delta t$ in experiments. The boxplots show the proportion of Kilobots detecting from $n$ to $m$ neighbors for four different $(n,m)$ ranges.}
	\label{fig:botseenmodel_deltat}
\end{figure}

Figure~\ref{fig:botseenmodel_deltat} shows in a boxplot the ratio $D_{n,m}$,
which represents the proportion of Kilobots detecting from $n$ to $m$ neighbor
Kilobots during $\Delta t$. 
Most Kilobots only detect $0-3$ other bots, or $4-7$, during a time step. 
In other words, undecided bots detect on average $2.92 \pm 2.50$ bots during 
each $\Delta t$.

\begin{figure}[t!]
	\centering
 \iffigures
    \includegraphics[width=1\columnwidth]{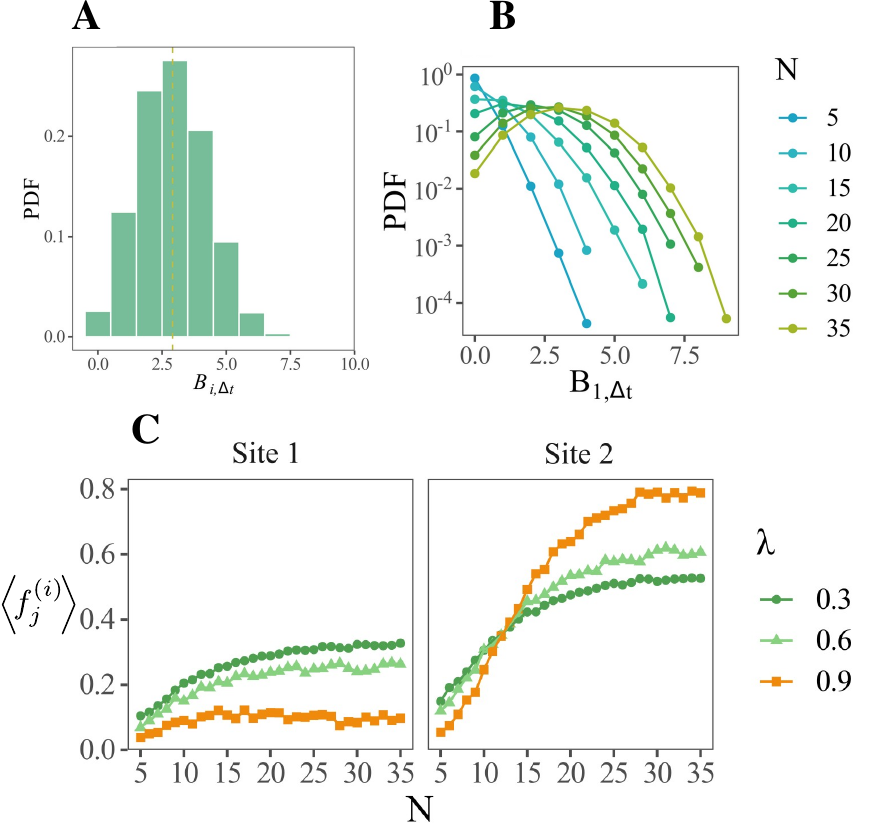}
 \fi
	\caption{
    Kilobots detected during a time-step $\Delta t$ in Kilombo.
    \textbf{A}: Probability distribution, $B_{i,\Delta t}$, of Kilobots detected by a 
    focal Kilobot $i$ in a time-step $\Delta t$.  
    \textbf{B}: Same as A in logscale and for different number $N$ of Kilobots in the arena.
    \textbf{C}: Average ratio of Kilobots dancing for site $j$ detected by a focal 
    Kilobot $i$ in a time-step $\Delta t$, $f^{(i)}_j$, as a function of the number 
    $N$ of Kilobots in the arena, for three different values of $\lambda$. 
    }\label{fig:botseen_Kilombo}
\end{figure}

Additionally, we used the Kilombo simulator, confirming that the number of Kilobots
detected per $\Delta t$ was consistent with our experimental results. 
Figure~\ref{fig:botseen_Kilombo}A shows the distribution of bots 
detected by a focal Kilobot $i$ in $\Delta t$, $B_{i,\Delta_t}$. 
Uncommitted bots in Kilombo detected a mean of $2.91 \pm 1.35$ bots.
With the reassurance that Kilombo fairly mimics quantitatively the real 
system, we also analyze how the number of bots seen vary when 
varying the Kilobot density.
Figure~\ref{fig:botseen_Kilombo}B shows the distribution $B_{i,\Delta_t}$ as the number 
of Kilobots $N$ goes from $N=5$ until $N=35$. 
As $N$ increases, curves shift towards the right but, in all cases, the average 
$B_{i,\Delta_t}$ remains below 5.

The emulator also provides more detailed information about the proportions of 
detected bots that were dancing for each site.
We analyze the ratio of Kilobots seen by the focal Kilobot $i$ separating those 
dancing for each available site $j$, $f^{(i)}_j$, during $\Delta t$ for three 
contrasting values of the interdependence parameter $\lambda$. 
In Fig.~\ref{fig:botseen_Kilombo}C, we represent the average $\langle f^{(i)}_j\rangle$
for $j=1,2$ as $N$ is increased, for $\lambda=0.3, 0.6,0.9$. 
We observe that the $\langle f^{(i)}_j\rangle$ increase gradually with $N$ until they 
reach a plateau for a group of approximately $30$ Kilobots. 
As we discuss in the main text, this feature corroborates the existence of a 
percolating communication network.

\section{Complementary analysis of the communication network}
\label{app:networks}

\begin{figure}[t!]
	\centering
 \iffigures
	\includegraphics[width=1\columnwidth]{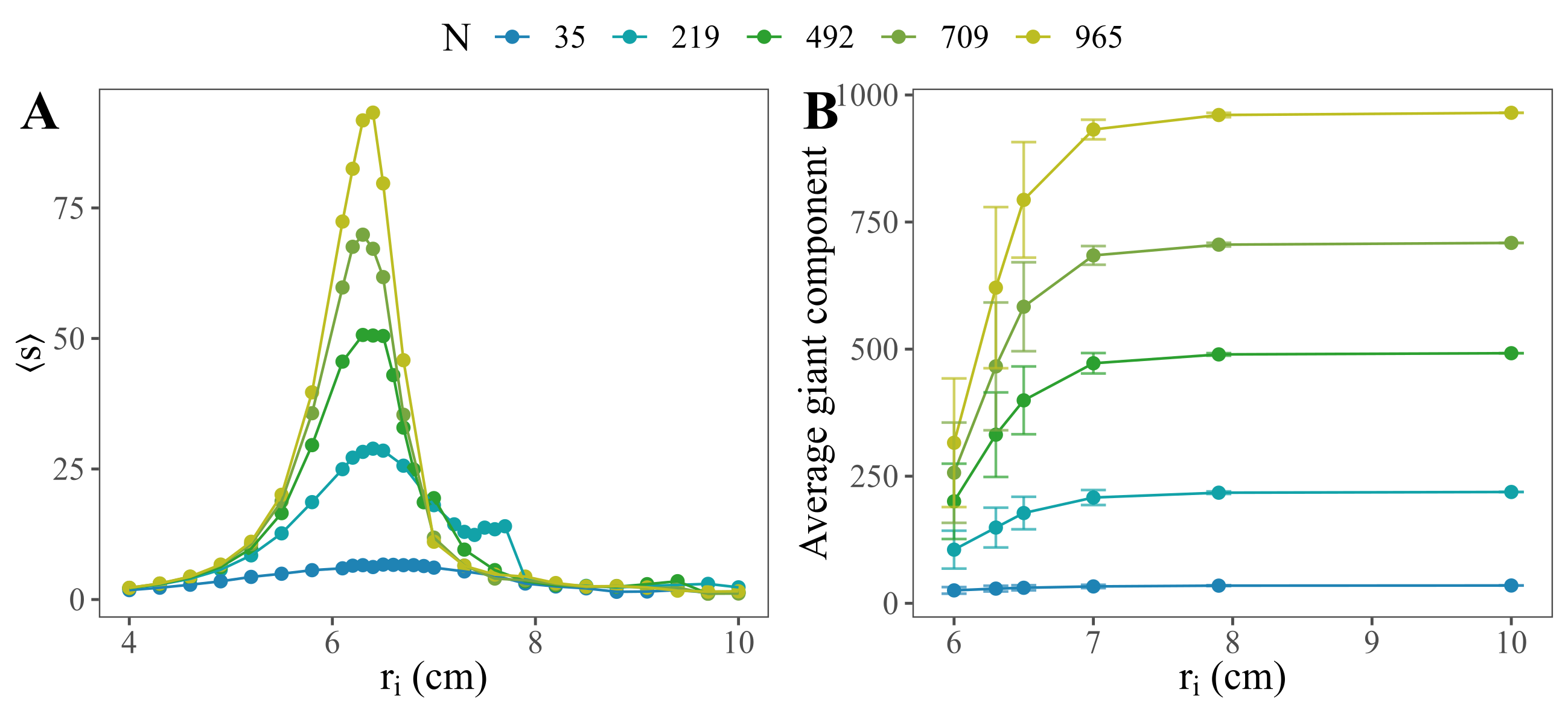}
  \fi	
   \caption{\textbf{A}: Mean cluster size, and \textbf{B}: Average giant component as 
   a function of the communication radius $r_{\tt int}$ for quenched configurations on 
   different system sizes that preserve the same number density. 
   Cluster sizes were averaged over $1000$ configurations per system size.
   }\label{fig:MeanClusterSizeN_quenched}
\end{figure}

\subsection{Finite-size scaling in quenched configurations}
\label{app:mcs_finitesize_quenched}

Here we provide a complementary finite-size scaling analysis of communicating cluster 
formation in quenched configurations of bots randomly located on a circular arena.
In Figure \ref{fig:MeanClusterSizeN_quenched}A, we plot the mean-cluster size 
$\langle S \rangle$ for different system sizes, preserving the same number density
$N/\pi R^2=0.028$ bots$/cm^2$, as a function of the communication radius $r_{\tt int}$, 
which characterizes the maximum extent of message transmission, and thus of information
exchange, through infrared sensors among physical Kilobots.  
We can identify the critical percolation interaction radius at around 
$r_{\tt int}^*=6.5 \pm 0.2$ cm. 

Continuous percolation threshold values for two dimensional discs of 
effective radius $r_{\tt int}$ in a square box of linear dimension 
$L$ with periodic boundary conditions are found in the 
literature~\cite{mertens2012}. 
The critical filling factor in that particular geometry is 
$\eta^*=N\pi r_{\tt int}^2/L^2\simeq 1.128$, or equivalently, 
$r_{\tt int}^*=(L^2\eta^*/(N\pi))^{1/2}$. 
Assuming a similar scaling behavior in our case, with a fixed 
rigid circular wall, would yield a smaller threshold radius 
of approximately $3.59$ cm, indicating that both our circular 
geometry and fixed boundary conditions give rise to packing 
and size effects that cannot be neglected in the quantitative 
determination of this non-universal threshold value. 
On the other hand, such effects should not be relevant regarding 
the behavior of critical exponents.

In Figure~\ref{fig:MeanClusterSizeN_quenched}B, we represent 
the average size of the {\em giant component} (the largest connected 
cluster in the system) as a function of the interaction radius $r_{\tt int}$. 
This quantity attains its maximum value, comparable to the system size, 
after the percolation threshold. 
As the maximum value of the mean cluster size, the size of the 
{\em giant component} at the percolation threshold scales as a 
power-law of the system size. 

\subsection{Communication network degree distribution}
\label{app:networkdegree}

\begin{figure} %[t!]
    \centering
    \iffigures
    \includegraphics[width=1\columnwidth]{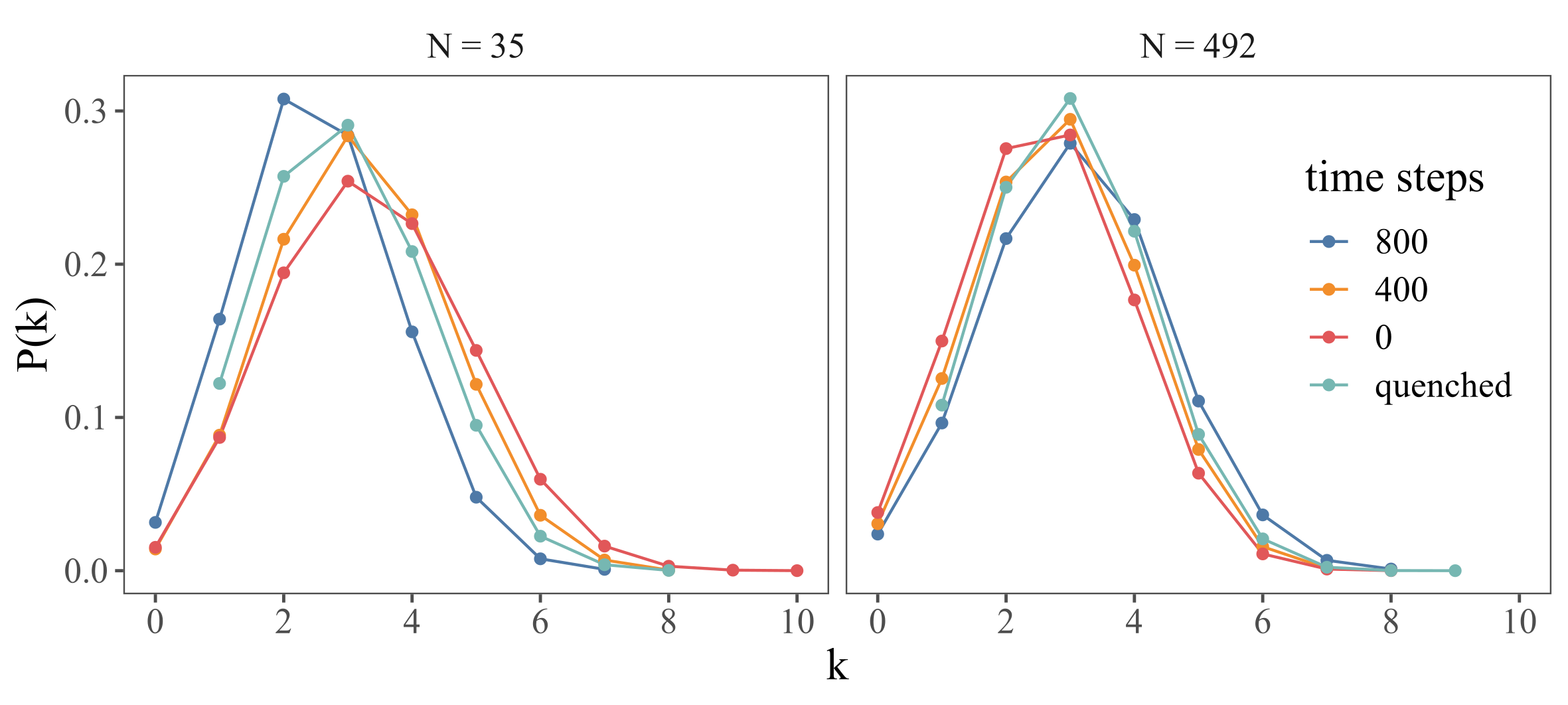}
    \includegraphics[width=1\columnwidth]{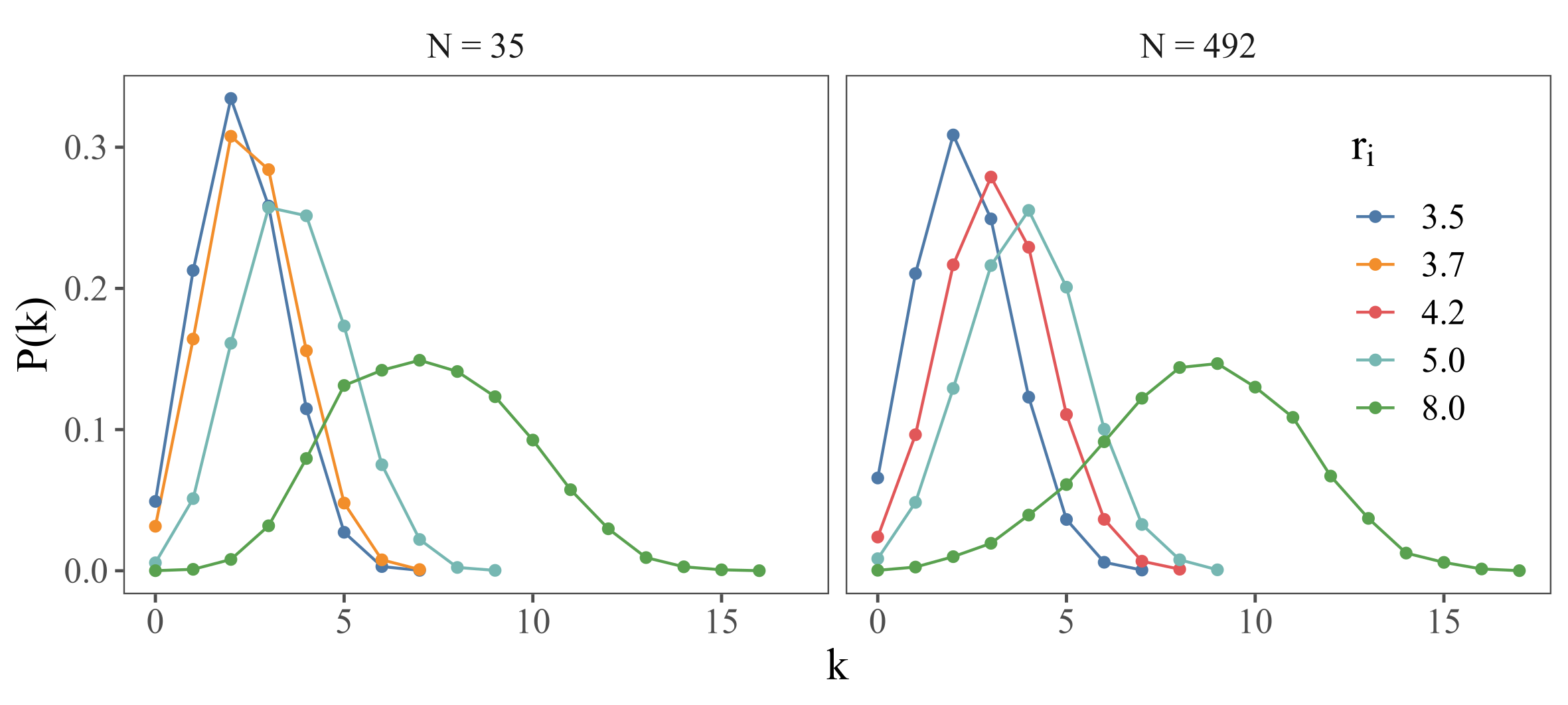}
    \fi
     \caption{Top: Degree distribution $P(k)$ at the percolation threshold $r_{\tt int}^*$. 
     Data obtained from Kilombo simulations integrating over different time-steps, 
     $\Delta t= 0,400,800$ loop iterations and for $1000$ quenched bot configurations. 
     Bottom: Degree distributions obtained for different interaction radius $r_{\tt int}$ 
     and a fixed value of $\Delta t= 800$ loop iterations. 
     Left panels correspond to $N = 35$, and right panels to $N = 495$.
     }\label{fig:Degree_dist_Kilobots}
\end{figure}

In network theory, a node's degree $k$ represents its number of connections with other nodes, while the degree distribution $P(k)$ indicates the probability 
of a randomly chosen node having degree $k$~\cite{Newman10}. 
Both degree and degree distribution are crucial for understanding dynamic processes on networks, such as information spread in the Kilobots' infrared communication network.

Figure~\ref{fig:Degree_dist_Kilobots} illustrates the degree distribution observed 
for the communication network built-up in Kilombo simulations after integrating over various exploratory time-steps ($\Delta t$).
Two different system sizes with the same Kilobot number density are considered.
The bell-shaped curves roughly resemble a Poisson distribution 
($P(k) = e^{-\lambda} \frac{\lambda^k}{k!}$), where $\lambda$ represents the average 
degree $\langle k \rangle$.

Additionally, we compute the average degree ($\langle k \rangle$) of the 
communication network integrated over different $\Delta t$ values in Kilombo 
simulations and for quenched random configurations, and the same system sizes 
($N=35$ and $N=492$). 

%\break

\begin{figure}[b!]
    \centering
    \iffigures
    \includegraphics[width=1\columnwidth]{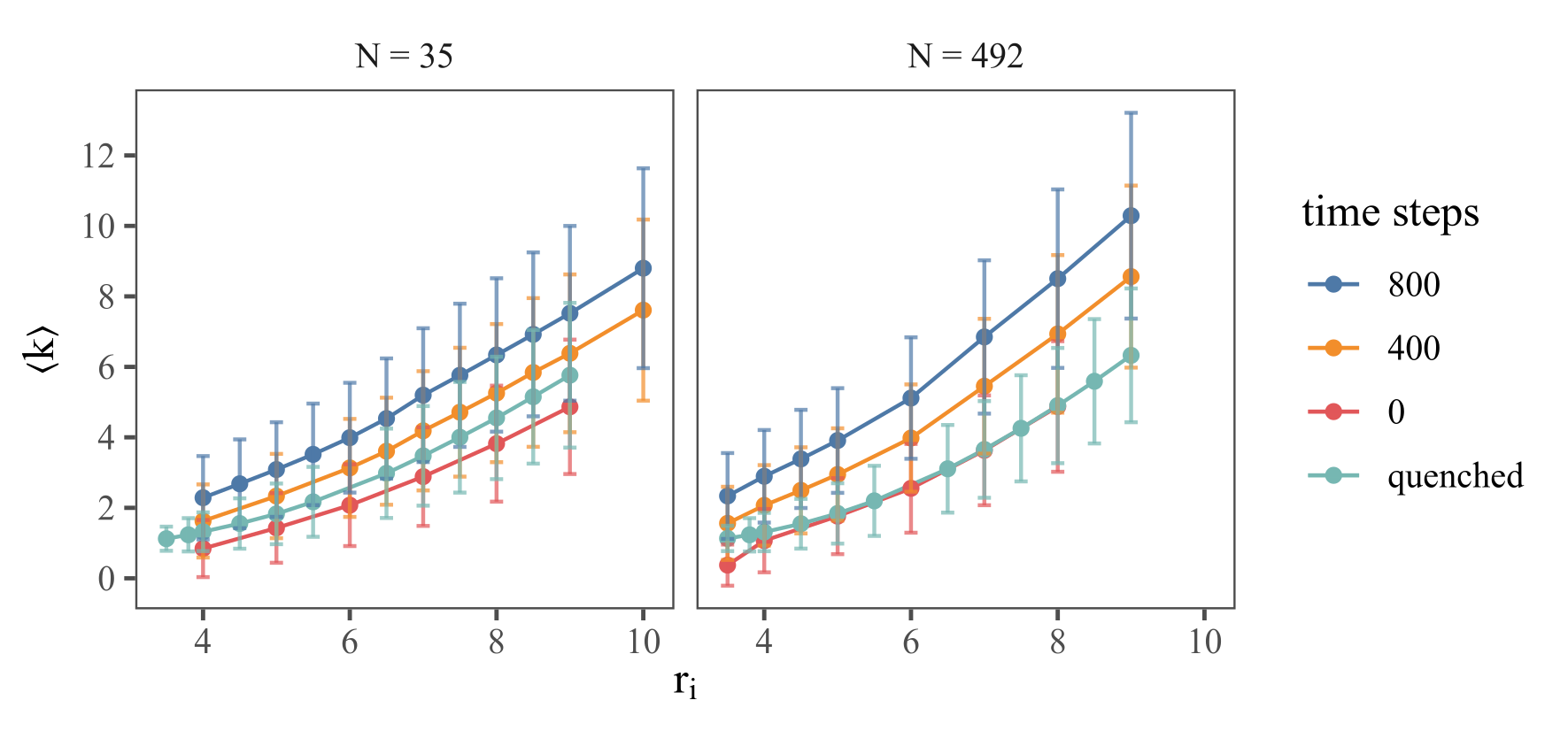}
    \fi
     \caption{Average degree $\langle k \rangle$ of the communication network as a function of $r_{\tt int}$. Results are obtained from Kilombo simulations integrating over different time steps $\Delta t= 0,400,800$ loop iterations, and for $1000$ quenched configurations. Left panel: $N = 35$. Right panel: $N = 495$.}  
    \label{fig:AvgDegree_Kilobots}
\end{figure}

Figure~\ref{fig:AvgDegree_Kilobots} shows $\langle k \rangle$ as a function of 
$r_{\tt int}$ for the same time windows.
Increasing $\Delta t$ and/or interaction radius $r_{\tt int}$ results in higher values 
of $\langle k \rangle$, which eventually overcome the threshold average degree 
$\langle k\rangle^*=1$ required for the presence of a giant component in this network 
according to the celebrated Molloy and Reed criterion~\cite{Molloy1995}.
These findings align with a network interpretation of the percolation transition.

For the smaller system ($N=35$), quenched Kilobot configurations exhibit a slightly 
larger average degree compared to single snapshots of Kilombo simulations ($\Delta t=0$),
potentially due to the accumulation of some bots at the arena wall. 
This effect diminishes  for the larger system size, yielding similar $\langle k \rangle$ 
values for both quenched and $\Delta t=0$ configurations. 

%\break

% \bibliographystyle{apsrev4-2}
% \bibliography{sample}

\begin{thebibliography}{78}%
\makeatletter
\providecommand \@ifxundefined [1]{%
 \@ifx{#1\undefined}
}%
\providecommand \@ifnum [1]{%
 \ifnum #1\expandafter \@firstoftwo
 \else \expandafter \@secondoftwo
 \fi
}%
\providecommand \@ifx [1]{%
 \ifx #1\expandafter \@firstoftwo
 \else \expandafter \@secondoftwo
 \fi
}%
\providecommand \natexlab [1]{#1}%
\providecommand \enquote  [1]{``#1''}%
\providecommand \bibnamefont  [1]{#1}%
\providecommand \bibfnamefont [1]{#1}%
\providecommand \citenamefont [1]{#1}%
\providecommand \href@noop [0]{\@secondoftwo}%
\providecommand \href [0]{\begingroup \@sanitize@url \@href}%
\providecommand \@href[1]{\@@startlink{#1}\@@href}%
\providecommand \@@href[1]{\endgroup#1\@@endlink}%
\providecommand \@sanitize@url [0]{\catcode `\\12\catcode `\$12\catcode
  `\&12\catcode `\#12\catcode `\^12\catcode `\_12\catcode `\%12\relax}%
\providecommand \@@startlink[1]{}%
\providecommand \@@endlink[0]{}%
\providecommand \url  [0]{\begingroup\@sanitize@url \@url }%
\providecommand \@url [1]{\endgroup\@href {#1}{\urlprefix }}%
\providecommand \urlprefix  [0]{URL }%
\providecommand \Eprint [0]{\href }%
\providecommand \doibase [0]{https://doi.org/}%
\providecommand \selectlanguage [0]{\@gobble}%
\providecommand \bibinfo  [0]{\@secondoftwo}%
\providecommand \bibfield  [0]{\@secondoftwo}%
\providecommand \translation [1]{[#1]}%
\providecommand \BibitemOpen [0]{}%
\providecommand \bibitemStop [0]{}%
\providecommand \bibitemNoStop [0]{.\EOS\space}%
\providecommand \EOS [0]{\spacefactor3000\relax}%
\providecommand \BibitemShut  [1]{\csname bibitem#1\endcsname}%
\let\auto@bib@innerbib\@empty
%</preamble>
\bibitem [{\citenamefont {Bose}\ \emph {et~al.}(2017)\citenamefont {Bose},
  \citenamefont {Reina},\ and\ \citenamefont {Marshall}}]{bose2017}%
  \BibitemOpen
  \bibfield  {author} {\bibinfo {author} {\bibfnamefont {T.}~\bibnamefont
  {Bose}}, \bibinfo {author} {\bibfnamefont {A.}~\bibnamefont {Reina}},\ and\
  \bibinfo {author} {\bibfnamefont {J.~A.}\ \bibnamefont {Marshall}},\
  }\bibfield  {title} {\bibinfo {title} {Collective decision-making},\ }\href
  {https://doi.org/https://doi.org/10.1016/j.cobeha.2017.03.004} {\bibfield
  {journal} {\bibinfo  {journal} {Current Opinion in Behavioral Sciences}\
  }\textbf {\bibinfo {volume} {16}},\ \bibinfo {pages} {30} (\bibinfo {year}
  {2017})}\BibitemShut {NoStop}%
\bibitem [{\citenamefont {Dyer}\ \emph {et~al.}(2009)\citenamefont {Dyer},
  \citenamefont {Johansson}, \citenamefont {Helbing}, \citenamefont {Couzin},\
  and\ \citenamefont {Krause}}]{dyer2009}%
  \BibitemOpen
  \bibfield  {author} {\bibinfo {author} {\bibfnamefont {J.~R.}\ \bibnamefont
  {Dyer}}, \bibinfo {author} {\bibfnamefont {A.}~\bibnamefont {Johansson}},
  \bibinfo {author} {\bibfnamefont {D.}~\bibnamefont {Helbing}}, \bibinfo
  {author} {\bibfnamefont {I.~D.}\ \bibnamefont {Couzin}},\ and\ \bibinfo
  {author} {\bibfnamefont {J.}~\bibnamefont {Krause}},\ }\bibfield  {title}
  {\bibinfo {title} {Leadership, consensus decision making and collective
  behaviour in humans},\ }\href {https://doi.org/10.1098/rstb.2008.0233}
  {\bibfield  {journal} {\bibinfo  {journal} {Philosophical Transactions of the
  Royal Society B: Biological Sciences}\ }\textbf {\bibinfo {volume} {364}},\
  \bibinfo {pages} {781} (\bibinfo {year} {2009})}\BibitemShut {NoStop}%
\bibitem [{\citenamefont {Sasaki}\ and\ \citenamefont
  {Pratt}(2018)}]{sasaki2018}%
  \BibitemOpen
  \bibfield  {author} {\bibinfo {author} {\bibfnamefont {T.}~\bibnamefont
  {Sasaki}}\ and\ \bibinfo {author} {\bibfnamefont {S.~C.}\ \bibnamefont
  {Pratt}},\ }\bibfield  {title} {\bibinfo {title} {The psychology of
  superorganisms: Collective decision making by insect societies},\ }\href
  {https://doi.org/10.1146/annurev-ento-020117-043249} {\bibfield  {journal}
  {\bibinfo  {journal} {Annual Review of Entomology}\ }\textbf {\bibinfo
  {volume} {63}},\ \bibinfo {pages} {259} (\bibinfo {year} {2018})}\BibitemShut
  {NoStop}%
\bibitem [{\citenamefont {von Frisch}(1954)}]{frisch1954}%
  \BibitemOpen
  \bibfield  {author} {\bibinfo {author} {\bibfnamefont {K.}~\bibnamefont {von
  Frisch}},\ }\href {https://doi.org/10.1007/978-3-7091-4697-2} {\emph
  {\bibinfo {title} {The dancing bees}}}\ (\bibinfo  {publisher} {Springer
  Vienna},\ \bibinfo {year} {1954})\BibitemShut {NoStop}%
\bibitem [{\citenamefont {Seeley}(2010)}]{honeybee_democracy}%
  \BibitemOpen
  \bibfield  {author} {\bibinfo {author} {\bibfnamefont {T.~D.}\ \bibnamefont
  {Seeley}},\ }\href {https://doi.org/https://doi.org/10.1515/9781400835959}
  {\emph {\bibinfo {title} {Honeybee Democracy}}}\ (\bibinfo  {publisher}
  {Princeton University Press},\ \bibinfo {year} {2010})\BibitemShut {NoStop}%
\bibitem [{\citenamefont {Couzin}\ \emph {et~al.}(2005)\citenamefont {Couzin},
  \citenamefont {Krause}, \citenamefont {Franks},\ and\ \citenamefont
  {Levin}}]{CouzinNature2005}%
  \BibitemOpen
  \bibfield  {author} {\bibinfo {author} {\bibfnamefont {I.~D.}\ \bibnamefont
  {Couzin}}, \bibinfo {author} {\bibfnamefont {J.}~\bibnamefont {Krause}},
  \bibinfo {author} {\bibfnamefont {N.~R.}\ \bibnamefont {Franks}},\ and\
  \bibinfo {author} {\bibfnamefont {S.~A.}\ \bibnamefont {Levin}},\ }\bibfield
  {title} {\bibinfo {title} {Effective leadership and decision-making in animal
  groups on the move},\ }\href {https://doi.org/10.1038/nature03236} {\bibfield
   {journal} {\bibinfo  {journal} {Nature}\ }\textbf {\bibinfo {volume}
  {433}},\ \bibinfo {pages} {513} (\bibinfo {year} {2005})}\BibitemShut
  {NoStop}%
\bibitem [{\citenamefont {Dong}\ \emph {et~al.}(2023)\citenamefont {Dong},
  \citenamefont {Lin}, \citenamefont {Nieh},\ and\ \citenamefont
  {Tan}}]{DongScience2023}%
  \BibitemOpen
  \bibfield  {author} {\bibinfo {author} {\bibfnamefont {S.}~\bibnamefont
  {Dong}}, \bibinfo {author} {\bibfnamefont {T.}~\bibnamefont {Lin}}, \bibinfo
  {author} {\bibfnamefont {J.~C.}\ \bibnamefont {Nieh}},\ and\ \bibinfo
  {author} {\bibfnamefont {K.}~\bibnamefont {Tan}},\ }\bibfield  {title}
  {\bibinfo {title} {Social signal learning of the waggle dance in honey
  bees},\ }\href {https://doi.org/10.1126/science.ade1702} {\bibfield
  {journal} {\bibinfo  {journal} {Science}\ }\textbf {\bibinfo {volume}
  {379}},\ \bibinfo {pages} {1015} (\bibinfo {year} {2023})}\BibitemShut
  {NoStop}%
\bibitem [{\citenamefont {Britton}\ \emph {et~al.}(2002)\citenamefont
  {Britton}, \citenamefont {Franks}, \citenamefont {Pratt},\ and\ \citenamefont
  {Seeley}}]{britton2002}%
  \BibitemOpen
  \bibfield  {author} {\bibinfo {author} {\bibfnamefont {N.~F.}\ \bibnamefont
  {Britton}}, \bibinfo {author} {\bibfnamefont {N.~R.}\ \bibnamefont {Franks}},
  \bibinfo {author} {\bibfnamefont {S.~C.}\ \bibnamefont {Pratt}},\ and\
  \bibinfo {author} {\bibfnamefont {T.~D.}\ \bibnamefont {Seeley}},\ }\bibfield
   {title} {\bibinfo {title} {Deciding on a new home: how do honeybees
  agree?},\ }\href {https://doi.org/10.1098/rspb.2002.2001} {\bibfield
  {journal} {\bibinfo  {journal} {Proceedings of the Royal Society of London.
  Series B: Biological Sciences}\ }\textbf {\bibinfo {volume} {269}},\ \bibinfo
  {pages} {1383} (\bibinfo {year} {2002})}\BibitemShut {NoStop}%
\bibitem [{\citenamefont {Passino}\ and\ \citenamefont
  {Seeley}(2006)}]{passino2006}%
  \BibitemOpen
  \bibfield  {author} {\bibinfo {author} {\bibfnamefont {K.~M.}\ \bibnamefont
  {Passino}}\ and\ \bibinfo {author} {\bibfnamefont {T.~D.}\ \bibnamefont
  {Seeley}},\ }\bibfield  {title} {\bibinfo {title} {Modeling and analysis of
  nest-site selection by honeybee swarms: the speed and accuracy trade-off},\
  }\href {https://doi.org/https://doi.org/10.1007/s00265-005-0067-y} {\bibfield
   {journal} {\bibinfo  {journal} {Behavioral Ecology and Sociobiology}\
  }\textbf {\bibinfo {volume} {59}},\ \bibinfo {pages} {427} (\bibinfo {year}
  {2006})}\BibitemShut {NoStop}%
\bibitem [{\citenamefont {List}\ \emph {et~al.}(2009)\citenamefont {List},
  \citenamefont {Elsholtz},\ and\ \citenamefont {Seeley}}]{list2009}%
  \BibitemOpen
  \bibfield  {author} {\bibinfo {author} {\bibfnamefont {C.}~\bibnamefont
  {List}}, \bibinfo {author} {\bibfnamefont {C.}~\bibnamefont {Elsholtz}},\
  and\ \bibinfo {author} {\bibfnamefont {T.~D.}\ \bibnamefont {Seeley}},\
  }\bibfield  {title} {\bibinfo {title} {Independence and interdependence in
  collective decision making: an agent-based model of nest-site choice by
  honeybee swarms},\ }\href {https://doi.org/10.1098/rstb.2008.0277} {\bibfield
   {journal} {\bibinfo  {journal} {Philosophical Transactions of the Royal
  Society B: Biological Sciences}\ }\textbf {\bibinfo {volume} {364}},\
  \bibinfo {pages} {755} (\bibinfo {year} {2009})}\BibitemShut {NoStop}%
\bibitem [{\citenamefont {Pais}\ \emph {et~al.}(2013)\citenamefont {Pais},
  \citenamefont {Hogan}, \citenamefont {Schlegel}, \citenamefont {Franks},
  \citenamefont {Leonard},\ and\ \citenamefont {Marshall}}]{pais2013}%
  \BibitemOpen
  \bibfield  {author} {\bibinfo {author} {\bibfnamefont {D.}~\bibnamefont
  {Pais}}, \bibinfo {author} {\bibfnamefont {P.~M.}\ \bibnamefont {Hogan}},
  \bibinfo {author} {\bibfnamefont {T.}~\bibnamefont {Schlegel}}, \bibinfo
  {author} {\bibfnamefont {N.~R.}\ \bibnamefont {Franks}}, \bibinfo {author}
  {\bibfnamefont {N.~E.}\ \bibnamefont {Leonard}},\ and\ \bibinfo {author}
  {\bibfnamefont {J.~A.~R.}\ \bibnamefont {Marshall}},\ }\bibfield  {title}
  {\bibinfo {title} {A mechanism for value-sensitive decision-making},\ }\href
  {https://doi.org/10.1371/journal.pone.0073216} {\bibfield  {journal}
  {\bibinfo  {journal} {PLOS ONE}\ }\textbf {\bibinfo {volume} {8}},\ \bibinfo
  {pages} {1} (\bibinfo {year} {2013})}\BibitemShut {NoStop}%
\bibitem [{\citenamefont {Reina}\ \emph {et~al.}(2017)\citenamefont {Reina},
  \citenamefont {Marshall}, \citenamefont {Trianni},\ and\ \citenamefont
  {Bose}}]{reina2017}%
  \BibitemOpen
  \bibfield  {author} {\bibinfo {author} {\bibfnamefont {A.}~\bibnamefont
  {Reina}}, \bibinfo {author} {\bibfnamefont {J.~A.~R.}\ \bibnamefont
  {Marshall}}, \bibinfo {author} {\bibfnamefont {V.}~\bibnamefont {Trianni}},\
  and\ \bibinfo {author} {\bibfnamefont {T.}~\bibnamefont {Bose}},\ }\bibfield
  {title} {\bibinfo {title} {Model of the best-of-$n$ nest-site selection
  process in honeybees},\ }\href {https://doi.org/10.1103/PhysRevE.95.052411}
  {\bibfield  {journal} {\bibinfo  {journal} {Phys. Rev. E}\ }\textbf {\bibinfo
  {volume} {95}},\ \bibinfo {pages} {052411} (\bibinfo {year}
  {2017})}\BibitemShut {NoStop}%
\bibitem [{\citenamefont {Gray}\ \emph {et~al.}(2018)\citenamefont {Gray},
  \citenamefont {Franci}, \citenamefont {Srivastava},\ and\ \citenamefont
  {Leonard}}]{gray_multiagent_2018}%
  \BibitemOpen
  \bibfield  {author} {\bibinfo {author} {\bibfnamefont {R.}~\bibnamefont
  {Gray}}, \bibinfo {author} {\bibfnamefont {A.}~\bibnamefont {Franci}},
  \bibinfo {author} {\bibfnamefont {V.}~\bibnamefont {Srivastava}},\ and\
  \bibinfo {author} {\bibfnamefont {N.~E.}\ \bibnamefont {Leonard}},\
  }\bibfield  {title} {\bibinfo {title} {Multiagent {Decision}-{Making}
  {Dynamics} {Inspired} by {Honeybees}},\ }\href
  {https://doi.org/10.1109/TCNS.2018.2796301} {\bibfield  {journal} {\bibinfo
  {journal} {IEEE Transactions on Control of Network Systems}\ }\textbf
  {\bibinfo {volume} {5}},\ \bibinfo {pages} {793} (\bibinfo {year} {2018})},\
  \bibinfo {note} {conference Name: IEEE Transactions on Control of Network
  Systems}\BibitemShut {NoStop}%
\bibitem [{\citenamefont {Cavagna}\ \emph {et~al.}(2010)\citenamefont
  {Cavagna}, \citenamefont {Cimarelli}, \citenamefont {Giardina}, \citenamefont
  {Parisi}, \citenamefont {Santagati}, \citenamefont {Stefanini},\ and\
  \citenamefont {Viale}}]{cavagna2010scale}%
  \BibitemOpen
  \bibfield  {author} {\bibinfo {author} {\bibfnamefont {A.}~\bibnamefont
  {Cavagna}}, \bibinfo {author} {\bibfnamefont {A.}~\bibnamefont {Cimarelli}},
  \bibinfo {author} {\bibfnamefont {I.}~\bibnamefont {Giardina}}, \bibinfo
  {author} {\bibfnamefont {G.}~\bibnamefont {Parisi}}, \bibinfo {author}
  {\bibfnamefont {R.}~\bibnamefont {Santagati}}, \bibinfo {author}
  {\bibfnamefont {F.}~\bibnamefont {Stefanini}},\ and\ \bibinfo {author}
  {\bibfnamefont {M.}~\bibnamefont {Viale}},\ }\bibfield  {title} {\bibinfo
  {title} {Scale-free correlations in starling flocks},\ }\href
  {https://doi.org/10.1073/pnas.1005766107} {\bibfield  {journal} {\bibinfo
  {journal} {Proceedings of the National Academy of Sciences}\ }\textbf
  {\bibinfo {volume} {107}},\ \bibinfo {pages} {11865} (\bibinfo {year}
  {2010})}\BibitemShut {NoStop}%
\bibitem [{\citenamefont {Rosenthal}\ \emph {et~al.}(2015)\citenamefont
  {Rosenthal}, \citenamefont {Twomey}, \citenamefont {Hartnett}, \citenamefont
  {Wu},\ and\ \citenamefont {Couzin}}]{Rosenthal2015}%
  \BibitemOpen
  \bibfield  {author} {\bibinfo {author} {\bibfnamefont {S.~B.}\ \bibnamefont
  {Rosenthal}}, \bibinfo {author} {\bibfnamefont {C.~R.}\ \bibnamefont
  {Twomey}}, \bibinfo {author} {\bibfnamefont {A.~T.}\ \bibnamefont
  {Hartnett}}, \bibinfo {author} {\bibfnamefont {H.~S.}\ \bibnamefont {Wu}},\
  and\ \bibinfo {author} {\bibfnamefont {I.~D.}\ \bibnamefont {Couzin}},\
  }\bibfield  {title} {\bibinfo {title} {Revealing the hidden networks of
  interaction in mobile animal groups allows prediction of complex behavioral
  contagion},\ }\href {https://doi.org/10.1073/pnas.1420068112} {\bibfield
  {journal} {\bibinfo  {journal} {Proceedings of the National Academy of
  Sciences}\ }\textbf {\bibinfo {volume} {112}},\ \bibinfo {pages} {4690}
  (\bibinfo {year} {2015})}\BibitemShut {NoStop}%
\bibitem [{\citenamefont {Chen}\ \emph {et~al.}(2016)\citenamefont {Chen},
  \citenamefont {Vicsek}, \citenamefont {Liu}, \citenamefont {Zhou},\ and\
  \citenamefont {Zhang}}]{Chen2016}%
  \BibitemOpen
  \bibfield  {author} {\bibinfo {author} {\bibfnamefont {D.}~\bibnamefont
  {Chen}}, \bibinfo {author} {\bibfnamefont {T.}~\bibnamefont {Vicsek}},
  \bibinfo {author} {\bibfnamefont {X.}~\bibnamefont {Liu}}, \bibinfo {author}
  {\bibfnamefont {T.}~\bibnamefont {Zhou}},\ and\ \bibinfo {author}
  {\bibfnamefont {H.-T.}\ \bibnamefont {Zhang}},\ }\bibfield  {title} {\bibinfo
  {title} {Switching hierarchical leadership mechanism in homing flight of
  pigeon flocks},\ }\href {https://doi.org/10.1209/0295-5075/114/60008}
  {\bibfield  {journal} {\bibinfo  {journal} {Europhysics Letters}\ }\textbf
  {\bibinfo {volume} {114}},\ \bibinfo {pages} {60008} (\bibinfo {year}
  {2016})}\BibitemShut {NoStop}%
\bibitem [{\citenamefont {Calovi}\ \emph {et~al.}(2018)\citenamefont {Calovi},
  \citenamefont {Litchinko}, \citenamefont {Lecheval}, \citenamefont {Lopez},
  \citenamefont {Pérez~Escudero}, \citenamefont {Chaté}, \citenamefont
  {Sire},\ and\ \citenamefont {Theraulaz}}]{Calovi2018}%
  \BibitemOpen
  \bibfield  {author} {\bibinfo {author} {\bibfnamefont {D.~S.}\ \bibnamefont
  {Calovi}}, \bibinfo {author} {\bibfnamefont {A.}~\bibnamefont {Litchinko}},
  \bibinfo {author} {\bibfnamefont {V.}~\bibnamefont {Lecheval}}, \bibinfo
  {author} {\bibfnamefont {U.}~\bibnamefont {Lopez}}, \bibinfo {author}
  {\bibfnamefont {A.}~\bibnamefont {Pérez~Escudero}}, \bibinfo {author}
  {\bibfnamefont {H.}~\bibnamefont {Chaté}}, \bibinfo {author} {\bibfnamefont
  {C.}~\bibnamefont {Sire}},\ and\ \bibinfo {author} {\bibfnamefont
  {G.}~\bibnamefont {Theraulaz}},\ }\bibfield  {title} {\bibinfo {title}
  {Disentangling and modeling interactions in fish with burst-and-coast
  swimming reveal distinct alignment and attraction behaviors},\ }\href
  {https://doi.org/10.1371/journal.pcbi.1005933} {\bibfield  {journal}
  {\bibinfo  {journal} {PLOS Computational Biology}\ }\textbf {\bibinfo
  {volume} {14}},\ \bibinfo {pages} {1} (\bibinfo {year} {2018})}\BibitemShut
  {NoStop}%
\bibitem [{\citenamefont {M{\'u}gica}\ \emph {et~al.}(2022)\citenamefont
  {M{\'u}gica}, \citenamefont {Torrents}, \citenamefont {Crist{\'i}n},
  \citenamefont {Puy}, \citenamefont {Miguel},\ and\ \citenamefont
  {Pastor-Satorras}}]{mugica2022}%
  \BibitemOpen
  \bibfield  {author} {\bibinfo {author} {\bibfnamefont {J.}~\bibnamefont
  {M{\'u}gica}}, \bibinfo {author} {\bibfnamefont {J.}~\bibnamefont
  {Torrents}}, \bibinfo {author} {\bibfnamefont {J.}~\bibnamefont
  {Crist{\'i}n}}, \bibinfo {author} {\bibfnamefont {A.}~\bibnamefont {Puy}},
  \bibinfo {author} {\bibfnamefont {M.~C.}\ \bibnamefont {Miguel}},\ and\
  \bibinfo {author} {\bibfnamefont {R.}~\bibnamefont {Pastor-Satorras}},\
  }\bibfield  {title} {\bibinfo {title} {Scale-free behavioral cascades and
  effective leadership in schooling fish},\ }\href
  {https://doi.org/10.1038/s41598-022-14337-0} {\bibfield  {journal} {\bibinfo
  {journal} {Scientific Reports}\ }\textbf {\bibinfo {volume} {12}},\ \bibinfo
  {pages} {10783} (\bibinfo {year} {2022})}\BibitemShut {NoStop}%
\bibitem [{\citenamefont {Seeley}\ and\ \citenamefont
  {Buhrman}(2001)}]{seeley2001}%
  \BibitemOpen
  \bibfield  {author} {\bibinfo {author} {\bibfnamefont {T.~D.}\ \bibnamefont
  {Seeley}}\ and\ \bibinfo {author} {\bibfnamefont {S.~C.}\ \bibnamefont
  {Buhrman}},\ }\bibfield  {title} {\bibinfo {title} {Nest-site selection in
  honey bees: how well do swarms implement the "best-of-n" decision rule?},\
  }\href {https://doi.org/10.1007/s002650000299} {\bibfield  {journal}
  {\bibinfo  {journal} {Behavioral Ecology and Sociobiology}\ }\textbf
  {\bibinfo {volume} {49}},\ \bibinfo {pages} {416} (\bibinfo {year}
  {2001})}\BibitemShut {NoStop}%
\bibitem [{\citenamefont {Valentini}\ \emph {et~al.}(2017)\citenamefont
  {Valentini}, \citenamefont {Ferrante},\ and\ \citenamefont
  {Dorigo}}]{valentini2017}%
  \BibitemOpen
  \bibfield  {author} {\bibinfo {author} {\bibfnamefont {G.}~\bibnamefont
  {Valentini}}, \bibinfo {author} {\bibfnamefont {E.}~\bibnamefont
  {Ferrante}},\ and\ \bibinfo {author} {\bibfnamefont {M.}~\bibnamefont
  {Dorigo}},\ }\bibfield  {title} {\bibinfo {title} {The best-of-n problem in
  robot swarms: Formalization, state of the art, and novel perspectives},\
  }\href {https://doi.org/10.3389/frobt.2017.00009} {\bibfield  {journal}
  {\bibinfo  {journal} {Frontiers in Robotics and AI}\ }\textbf {\bibinfo
  {volume} {4}} (\bibinfo {year} {2017})}\BibitemShut {NoStop}%
\bibitem [{\citenamefont {Reina}\ \emph {et~al.}(2024)\citenamefont {Reina},
  \citenamefont {Njougouo}, \citenamefont {Tuci},\ and\ \citenamefont
  {Carletti}}]{reina_voter_2023}%
  \BibitemOpen
  \bibfield  {author} {\bibinfo {author} {\bibfnamefont {A.}~\bibnamefont
  {Reina}}, \bibinfo {author} {\bibfnamefont {T.}~\bibnamefont {Njougouo}},
  \bibinfo {author} {\bibfnamefont {E.}~\bibnamefont {Tuci}},\ and\ \bibinfo
  {author} {\bibfnamefont {T.}~\bibnamefont {Carletti}},\ }\bibfield  {title}
  {\bibinfo {title} {Speed-accuracy trade-offs in best-of-$n$ collective
  decision making through heterogeneous mean-field modeling},\ }\href
  {https://doi.org/10.1103/PhysRevE.109.054307} {\bibfield  {journal} {\bibinfo
   {journal} {Phys. Rev. E}\ }\textbf {\bibinfo {volume} {109}},\ \bibinfo
  {pages} {054307} (\bibinfo {year} {2024})}\BibitemShut {NoStop}%
\bibitem [{\citenamefont {Bizyaeva}\ \emph {et~al.}(2023)\citenamefont
  {Bizyaeva}, \citenamefont {Franci},\ and\ \citenamefont
  {Leonard}}]{bizyaeva_nonlinear_2023}%
  \BibitemOpen
  \bibfield  {author} {\bibinfo {author} {\bibfnamefont {A.}~\bibnamefont
  {Bizyaeva}}, \bibinfo {author} {\bibfnamefont {A.}~\bibnamefont {Franci}},\
  and\ \bibinfo {author} {\bibfnamefont {N.~E.}\ \bibnamefont {Leonard}},\
  }\bibfield  {title} {\bibinfo {title} {Nonlinear {Opinion} {Dynamics} {With}
  {Tunable} {Sensitivity}},\ }\href {https://doi.org/10.1109/TAC.2022.3159527}
  {\bibfield  {journal} {\bibinfo  {journal} {IEEE Transactions on Automatic
  Control}\ }\textbf {\bibinfo {volume} {68}},\ \bibinfo {pages} {1415}
  (\bibinfo {year} {2023})},\ \bibinfo {note} {conference Name: IEEE
  Transactions on Automatic Control}\BibitemShut {NoStop}%
\bibitem [{\citenamefont {Garnier}\ \emph {et~al.}(2009)\citenamefont
  {Garnier}, \citenamefont {Gautrais}, \citenamefont {Asadpour}, \citenamefont
  {Jost},\ and\ \citenamefont {Theraulaz}}]{garnier2009}%
  \BibitemOpen
  \bibfield  {author} {\bibinfo {author} {\bibfnamefont {S.}~\bibnamefont
  {Garnier}}, \bibinfo {author} {\bibfnamefont {J.}~\bibnamefont {Gautrais}},
  \bibinfo {author} {\bibfnamefont {M.}~\bibnamefont {Asadpour}}, \bibinfo
  {author} {\bibfnamefont {C.}~\bibnamefont {Jost}},\ and\ \bibinfo {author}
  {\bibfnamefont {G.}~\bibnamefont {Theraulaz}},\ }\bibfield  {title} {\bibinfo
  {title} {Self-organized aggregation triggers collective decision making in a
  group of cockroach-like robots},\ }\href
  {https://doi.org/10.1177/1059712309103430} {\bibfield  {journal} {\bibinfo
  {journal} {Adaptive Behavior}\ }\textbf {\bibinfo {volume} {17}},\ \bibinfo
  {pages} {109} (\bibinfo {year} {2009})}\BibitemShut {NoStop}%
\bibitem [{\citenamefont {Hamann}\ \emph {et~al.}(2012)\citenamefont {Hamann},
  \citenamefont {Schmickl}, \citenamefont {Wörn},\ and\ \citenamefont
  {Crailsheim}}]{hamann_analysis_2012}%
  \BibitemOpen
  \bibfield  {author} {\bibinfo {author} {\bibfnamefont {H.}~\bibnamefont
  {Hamann}}, \bibinfo {author} {\bibfnamefont {T.}~\bibnamefont {Schmickl}},
  \bibinfo {author} {\bibfnamefont {H.}~\bibnamefont {Wörn}},\ and\ \bibinfo
  {author} {\bibfnamefont {K.}~\bibnamefont {Crailsheim}},\ }\bibfield  {title}
  {\bibinfo {title} {Analysis of emergent symmetry breaking in collective
  decision making},\ }\href {https://doi.org/10.1007/s00521-010-0368-6}
  {\bibfield  {journal} {\bibinfo  {journal} {Neural Computing and
  Applications}\ }\textbf {\bibinfo {volume} {21}},\ \bibinfo {pages} {207}
  (\bibinfo {year} {2012})}\BibitemShut {NoStop}%
\bibitem [{\citenamefont {Schmickl}\ \emph {et~al.}(2007)\citenamefont
  {Schmickl}, \citenamefont {M{\"o}slinger},\ and\ \citenamefont
  {Crailsheim}}]{schmickl2007}%
  \BibitemOpen
  \bibfield  {author} {\bibinfo {author} {\bibfnamefont {T.}~\bibnamefont
  {Schmickl}}, \bibinfo {author} {\bibfnamefont {C.}~\bibnamefont
  {M{\"o}slinger}},\ and\ \bibinfo {author} {\bibfnamefont {K.}~\bibnamefont
  {Crailsheim}},\ }\bibfield  {title} {\bibinfo {title} {Collective perception
  in a robot swarm},\ }in\ \href
  {https://doi.org/https://doi.org/10.1007/978-3-540-71541-2_10} {\emph
  {\bibinfo {booktitle} {Swarm Robotics}}},\ \bibinfo {editor} {edited by\
  \bibinfo {editor} {\bibfnamefont {E.}~\bibnamefont {{\c{S}}ahin}}, \bibinfo
  {editor} {\bibfnamefont {W.~M.}\ \bibnamefont {Spears}},\ and\ \bibinfo
  {editor} {\bibfnamefont {A.~F.~T.}\ \bibnamefont {Winfield}}}\ (\bibinfo
  {publisher} {Springer Berlin Heidelberg},\ \bibinfo {address} {Berlin,
  Heidelberg},\ \bibinfo {year} {2007})\ pp.\ \bibinfo {pages}
  {144--157}\BibitemShut {NoStop}%
\bibitem [{\citenamefont {Valentini}\ \emph {et~al.}(2014)\citenamefont
  {Valentini}, \citenamefont {Hamann},\ and\ \citenamefont
  {Dorigo}}]{valentini2014}%
  \BibitemOpen
  \bibfield  {author} {\bibinfo {author} {\bibfnamefont {G.}~\bibnamefont
  {Valentini}}, \bibinfo {author} {\bibfnamefont {H.}~\bibnamefont {Hamann}},\
  and\ \bibinfo {author} {\bibfnamefont {M.}~\bibnamefont {Dorigo}},\
  }\bibfield  {title} {\bibinfo {title} {Self-organized collective decision
  making: The weighted voter model},\ }in\ \href
  {https://doi.org/https://dl.acm.org/doi/10.5555/2615731.2615742} {\emph
  {\bibinfo {booktitle} {Proceedings of the 2014 International Conference on
  Autonomous Agents and Multi-Agent Systems}}},\ \bibinfo {series and number}
  {AAMAS '14}\ (\bibinfo  {publisher} {International Foundation for Autonomous
  Agents and Multiagent Systems},\ \bibinfo {address} {Richland, SC},\ \bibinfo
  {year} {2014})\ p.\ \bibinfo {pages} {45–52}\BibitemShut {NoStop}%
\bibitem [{\citenamefont {Zakir}\ \emph {et~al.}(2022)\citenamefont {Zakir},
  \citenamefont {Dorigo},\ and\ \citenamefont {Reina}}]{zakir_robot_2022}%
  \BibitemOpen
  \bibfield  {author} {\bibinfo {author} {\bibfnamefont {R.}~\bibnamefont
  {Zakir}}, \bibinfo {author} {\bibfnamefont {M.}~\bibnamefont {Dorigo}},\ and\
  \bibinfo {author} {\bibfnamefont {A.}~\bibnamefont {Reina}},\ }\bibfield
  {title} {\bibinfo {title} {Robot {Swarms} {Break} {Decision} {Deadlocks}
  in {Collective} {Perception} {Through} {Cross}-{Inhibition}},\ }in\ \href
  {https://doi.org/10.1007/978-3-031-20176-9_17} {\emph {\bibinfo {booktitle}
  {Swarm {Intelligence}: 13th {International} {Conference}, {ANTS} 2022,
  {Málaga}, {Spain}, {November} 2–4, 2022, {Proceedings}}}}\ (\bibinfo
  {publisher} {Springer-Verlag},\ \bibinfo {address} {Berlin, Heidelberg},\
  \bibinfo {year} {2022})\ pp.\ \bibinfo {pages} {209--221}\BibitemShut
  {NoStop}%
\bibitem [{\citenamefont {Holley}\ and\ \citenamefont
  {Liggett}(1975)}]{holley1975}%
  \BibitemOpen
  \bibfield  {author} {\bibinfo {author} {\bibfnamefont {R.~A.}\ \bibnamefont
  {Holley}}\ and\ \bibinfo {author} {\bibfnamefont {T.~M.}\ \bibnamefont
  {Liggett}},\ }\bibfield  {title} {\bibinfo {title} {{Ergodic Theorems for
  Weakly Interacting Infinite Systems and the Voter Model}},\ }\href
  {https://doi.org/10.1214/aop/1176996306} {\bibfield  {journal} {\bibinfo
  {journal} {The Annals of Probability}\ }\textbf {\bibinfo {volume} {3}},\
  \bibinfo {pages} {643 } (\bibinfo {year} {1975})}\BibitemShut {NoStop}%
\bibitem [{\citenamefont {Castellano}\ \emph {et~al.}(2009)\citenamefont
  {Castellano}, \citenamefont {Fortunato},\ and\ \citenamefont
  {Loreto}}]{castellano2009}%
  \BibitemOpen
  \bibfield  {author} {\bibinfo {author} {\bibfnamefont {C.}~\bibnamefont
  {Castellano}}, \bibinfo {author} {\bibfnamefont {S.}~\bibnamefont
  {Fortunato}},\ and\ \bibinfo {author} {\bibfnamefont {V.}~\bibnamefont
  {Loreto}},\ }\bibfield  {title} {\bibinfo {title} {Statistical physics of
  social dynamics},\ }\href {https://doi.org/10.1103/RevModPhys.81.591}
  {\bibfield  {journal} {\bibinfo  {journal} {Rev. Mod. Phys.}\ }\textbf
  {\bibinfo {volume} {81}},\ \bibinfo {pages} {591} (\bibinfo {year}
  {2009})}\BibitemShut {NoStop}%
\bibitem [{\citenamefont {{Galam, S.}}(2002)}]{galam2002}%
  \BibitemOpen
  \bibfield  {author} {\bibinfo {author} {\bibnamefont {{Galam, S.}}},\
  }\bibfield  {title} {\bibinfo {title} {Minority opinion spreading in random
  geometry},\ }\href {https://doi.org/10.1140/epjb/e20020045} {\bibfield
  {journal} {\bibinfo  {journal} {Eur. Phys. J. B}\ }\textbf {\bibinfo {volume}
  {25}},\ \bibinfo {pages} {403} (\bibinfo {year} {2002})}\BibitemShut
  {NoStop}%
\bibitem [{\citenamefont {Galam}(2008)}]{galam2008}%
  \BibitemOpen
  \bibfield  {author} {\bibinfo {author} {\bibfnamefont {S.}~\bibnamefont
  {Galam}},\ }\bibfield  {title} {\bibinfo {title} {Sociophysics: A review of
  galam models},\ }\href {https://doi.org/10.1142/S0129183108012297} {\bibfield
   {journal} {\bibinfo  {journal} {International Journal of Modern Physics C}\
  }\textbf {\bibinfo {volume} {19}},\ \bibinfo {pages} {409} (\bibinfo {year}
  {2008})}\BibitemShut {NoStop}%
\bibitem [{\citenamefont {Deffuant}\ \emph {et~al.}(2000)\citenamefont
  {Deffuant}, \citenamefont {Neau}, \citenamefont {Amblard},\ and\
  \citenamefont {Weisbuch}}]{deffuant2000}%
  \BibitemOpen
  \bibfield  {author} {\bibinfo {author} {\bibfnamefont {G.}~\bibnamefont
  {Deffuant}}, \bibinfo {author} {\bibfnamefont {D.}~\bibnamefont {Neau}},
  \bibinfo {author} {\bibfnamefont {F.}~\bibnamefont {Amblard}},\ and\ \bibinfo
  {author} {\bibfnamefont {G.}~\bibnamefont {Weisbuch}},\ }\bibfield  {title}
  {\bibinfo {title} {Mixing beliefs among interacting agents},\ }\href
  {https://doi.org/10.1142/S0219525900000078} {\bibfield  {journal} {\bibinfo
  {journal} {Advances in Complex Systems}\ }\textbf {\bibinfo {volume} {03}},\
  \bibinfo {pages} {87} (\bibinfo {year} {2000})}\BibitemShut {NoStop}%
\bibitem [{\citenamefont {Redner}(2019)}]{redner_reality-inspired_2019}%
  \BibitemOpen
  \bibfield  {author} {\bibinfo {author} {\bibfnamefont {S.}~\bibnamefont
  {Redner}},\ }\bibfield  {title} {\bibinfo {title} {Reality-inspired voter
  models: {A} mini-review},\ }\href
  {https://doi.org/10.1016/j.crhy.2019.05.004} {\bibfield  {journal} {\bibinfo
  {journal} {Comptes Rendus Physique}\ }\textbf {\bibinfo {volume} {20}},\
  \bibinfo {pages} {275} (\bibinfo {year} {2019})}\BibitemShut {NoStop}%
\bibitem [{\citenamefont {De~Marzo}\ \emph {et~al.}(2020)\citenamefont
  {De~Marzo}, \citenamefont {Zaccaria},\ and\ \citenamefont
  {Castellano}}]{de_marzo_emergence_2020}%
  \BibitemOpen
  \bibfield  {author} {\bibinfo {author} {\bibfnamefont {G.}~\bibnamefont
  {De~Marzo}}, \bibinfo {author} {\bibfnamefont {A.}~\bibnamefont {Zaccaria}},\
  and\ \bibinfo {author} {\bibfnamefont {C.}~\bibnamefont {Castellano}},\
  }\bibfield  {title} {\bibinfo {title} {Emergence of polarization in a voter
  model with personalized information},\ }\href
  {https://doi.org/10.1103/PhysRevResearch.2.043117} {\bibfield  {journal}
  {\bibinfo  {journal} {Physical Review Research}\ }\textbf {\bibinfo {volume}
  {2}},\ \bibinfo {pages} {043117} (\bibinfo {year} {2020})},\ \bibinfo {note}
  {publisher: American Physical Society}\BibitemShut {NoStop}%
\bibitem [{\citenamefont {Franci}\ \emph {et~al.}(2021)\citenamefont {Franci},
  \citenamefont {Bizyaeva}, \citenamefont {Park},\ and\ \citenamefont
  {Leonard}}]{franci_analysis_2021}%
  \BibitemOpen
  \bibfield  {author} {\bibinfo {author} {\bibfnamefont {A.}~\bibnamefont
  {Franci}}, \bibinfo {author} {\bibfnamefont {A.}~\bibnamefont {Bizyaeva}},
  \bibinfo {author} {\bibfnamefont {S.}~\bibnamefont {Park}},\ and\ \bibinfo
  {author} {\bibfnamefont {N.~E.}\ \bibnamefont {Leonard}},\ }\bibfield
  {title} {\bibinfo {title} {Analysis and control of agreement and disagreement
  opinion cascades},\ }\href {https://doi.org/10.1007/s11721-021-00190-w}
  {\bibfield  {journal} {\bibinfo  {journal} {Swarm Intelligence}\ }\textbf
  {\bibinfo {volume} {15}},\ \bibinfo {pages} {47} (\bibinfo {year}
  {2021})}\BibitemShut {NoStop}%
\bibitem [{\citenamefont {Leonard}\ \emph {et~al.}(2024)\citenamefont
  {Leonard}, \citenamefont {Bizyaeva},\ and\ \citenamefont
  {Franci}}]{Leonard2024}%
  \BibitemOpen
  \bibfield  {author} {\bibinfo {author} {\bibfnamefont {N.~E.}\ \bibnamefont
  {Leonard}}, \bibinfo {author} {\bibfnamefont {A.}~\bibnamefont {Bizyaeva}},\
  and\ \bibinfo {author} {\bibfnamefont {A.}~\bibnamefont {Franci}},\
  }\bibfield  {title} {\bibinfo {title} {Fast and flexible multiagent
  decision-making},\ }\href
  {https://doi.org/https://doi.org/10.1146/annurev-control-090523-100059}
  {\bibfield  {journal} {\bibinfo  {journal} {Annual Review of Control,
  Robotics, and Autonomous Systems}\ }\textbf {\bibinfo {volume} {7}},\
  \bibinfo {pages} {19} (\bibinfo {year} {2024})}\BibitemShut {NoStop}%
\bibitem [{\citenamefont {Reina}\ \emph {et~al.}(2015)\citenamefont {Reina},
  \citenamefont {Valentini}, \citenamefont {Fernandez-Oto}, \citenamefont
  {Dorigo},\ and\ \citenamefont {Trianni}}]{reina_desing_pattern}%
  \BibitemOpen
  \bibfield  {author} {\bibinfo {author} {\bibfnamefont {A.}~\bibnamefont
  {Reina}}, \bibinfo {author} {\bibfnamefont {G.}~\bibnamefont {Valentini}},
  \bibinfo {author} {\bibfnamefont {C.}~\bibnamefont {Fernandez-Oto}}, \bibinfo
  {author} {\bibfnamefont {M.}~\bibnamefont {Dorigo}},\ and\ \bibinfo {author}
  {\bibfnamefont {V.}~\bibnamefont {Trianni}},\ }\bibfield  {title} {\bibinfo
  {title} {A design pattern for decentralised decision making},\ }\href
  {https://doi.org/10.1371/journal.pone.0140950} {\bibfield  {journal}
  {\bibinfo  {journal} {PloS one}\ }\textbf {\bibinfo {volume} {10}},\ \bibinfo
  {pages} {e0140950} (\bibinfo {year} {2015})}\BibitemShut {NoStop}%
\bibitem [{\citenamefont {Passino}\ \emph {et~al.}(2008)\citenamefont
  {Passino}, \citenamefont {Seeley},\ and\ \citenamefont
  {Visscher}}]{passino_swarm_2008}%
  \BibitemOpen
  \bibfield  {author} {\bibinfo {author} {\bibfnamefont {K.}~\bibnamefont
  {Passino}}, \bibinfo {author} {\bibfnamefont {T.}~\bibnamefont {Seeley}},\
  and\ \bibinfo {author} {\bibfnamefont {P.}~\bibnamefont {Visscher}},\
  }\bibfield  {title} {\bibinfo {title} {Swarm cognition in honey bees},\
  }\href {https://doi.org/10.1007/s00265-007-0468-1} {\bibfield  {journal}
  {\bibinfo  {journal} {Behavioral Ecology and Sociobiology}\ }\textbf
  {\bibinfo {volume} {62}},\ \bibinfo {pages} {401} (\bibinfo {year}
  {2008})}\BibitemShut {NoStop}%
\bibitem [{\citenamefont {Trianni}\ \emph {et~al.}(2016)\citenamefont
  {Trianni}, \citenamefont {De~Simone}, \citenamefont {Reina},\ and\
  \citenamefont {Baronchelli}}]{trianni2016_naming_game}%
  \BibitemOpen
  \bibfield  {author} {\bibinfo {author} {\bibfnamefont {V.}~\bibnamefont
  {Trianni}}, \bibinfo {author} {\bibfnamefont {D.}~\bibnamefont {De~Simone}},
  \bibinfo {author} {\bibfnamefont {A.}~\bibnamefont {Reina}},\ and\ \bibinfo
  {author} {\bibfnamefont {A.}~\bibnamefont {Baronchelli}},\ }\bibfield
  {title} {\bibinfo {title} {Emergence of consensus in a multi-robot network:
  From abstract models to empirical validation},\ }\href
  {https://doi.org/10.1109/LRA.2016.2519537} {\bibfield  {journal} {\bibinfo
  {journal} {IEEE Robotics and Automation Letters}\ }\textbf {\bibinfo {volume}
  {1}},\ \bibinfo {pages} {1} (\bibinfo {year} {2016})}\BibitemShut {NoStop}%
\bibitem [{\citenamefont {Valentini}\ \emph {et~al.}(2016)\citenamefont
  {Valentini}, \citenamefont {Ferrante}, \citenamefont {Hamann},\ and\
  \citenamefont {Dorigo}}]{valentini2016}%
  \BibitemOpen
  \bibfield  {author} {\bibinfo {author} {\bibfnamefont {G.}~\bibnamefont
  {Valentini}}, \bibinfo {author} {\bibfnamefont {E.}~\bibnamefont {Ferrante}},
  \bibinfo {author} {\bibfnamefont {H.}~\bibnamefont {Hamann}},\ and\ \bibinfo
  {author} {\bibfnamefont {M.}~\bibnamefont {Dorigo}},\ }\bibfield  {title}
  {\bibinfo {title} {Collective decision with 100 kilobots: speed versus
  accuracy in binary discrimination problems},\ }\href
  {https://doi.org/10.1007/s10458-015-9323-3} {\bibfield  {journal} {\bibinfo
  {journal} {Autonomous Agents and Multi-Agent Systems}\ }\textbf {\bibinfo
  {volume} {30}},\ \bibinfo {pages} {553} (\bibinfo {year} {2016})}\BibitemShut
  {NoStop}%
\bibitem [{\citenamefont {Reina}\ \emph
  {et~al.}(2018{\natexlab{a}})\citenamefont {Reina}, \citenamefont {Bose},
  \citenamefont {Trianni},\ and\ \citenamefont
  {Marshall}}]{Reina2018_spatiality}%
  \BibitemOpen
  \bibfield  {author} {\bibinfo {author} {\bibfnamefont {A.}~\bibnamefont
  {Reina}}, \bibinfo {author} {\bibfnamefont {T.}~\bibnamefont {Bose}},
  \bibinfo {author} {\bibfnamefont {V.}~\bibnamefont {Trianni}},\ and\ \bibinfo
  {author} {\bibfnamefont {J.~A.~R.}\ \bibnamefont {Marshall}},\ }\bibinfo
  {title} {Effects of spatiality on value-sensitive decisions made by robot
  swarms},\ in\ \href {https://doi.org/10.1007/978-3-319-73008-0_32} {\emph
  {\bibinfo {booktitle} {Distributed Autonomous Robotic Systems: The 13th
  International Symposium}}},\ \bibinfo {editor} {edited by\ \bibinfo {editor}
  {\bibfnamefont {R.}~\bibnamefont {Gro{\ss}}}, \bibinfo {editor}
  {\bibfnamefont {A.}~\bibnamefont {Kolling}}, \bibinfo {editor} {\bibfnamefont
  {S.}~\bibnamefont {Berman}}, \bibinfo {editor} {\bibfnamefont
  {E.}~\bibnamefont {Frazzoli}}, \bibinfo {editor} {\bibfnamefont
  {A.}~\bibnamefont {Martinoli}}, \bibinfo {editor} {\bibfnamefont
  {F.}~\bibnamefont {Matsuno}},\ and\ \bibinfo {editor} {\bibfnamefont
  {M.}~\bibnamefont {Gauci}}}\ (\bibinfo  {publisher} {Springer International
  Publishing},\ \bibinfo {address} {Cham},\ \bibinfo {year} {2018})\ pp.\
  \bibinfo {pages} {461--473}\BibitemShut {NoStop}%
\bibitem [{\citenamefont {Khaluf}\ \emph {et~al.}(2018)\citenamefont {Khaluf},
  \citenamefont {Rausch},\ and\ \citenamefont
  {Simoens}}]{khaluf_interaction_models_2018}%
  \BibitemOpen
  \bibfield  {author} {\bibinfo {author} {\bibfnamefont {Y.}~\bibnamefont
  {Khaluf}}, \bibinfo {author} {\bibfnamefont {I.}~\bibnamefont {Rausch}},\
  and\ \bibinfo {author} {\bibfnamefont {P.}~\bibnamefont {Simoens}},\
  }\bibfield  {title} {\bibinfo {title} {The impact of interaction models on
  the coherence of collective decision-making: A case study with simulated
  locusts},\ }in\ \href {https://doi.org/10.1007/978-3-030-00533-7_20} {\emph
  {\bibinfo {booktitle} {Swarm Intelligence}}},\ \bibinfo {editor} {edited by\
  \bibinfo {editor} {\bibfnamefont {M.}~\bibnamefont {Dorigo}}, \bibinfo
  {editor} {\bibfnamefont {M.}~\bibnamefont {Birattari}}, \bibinfo {editor}
  {\bibfnamefont {C.}~\bibnamefont {Blum}}, \bibinfo {editor} {\bibfnamefont
  {A.~L.}\ \bibnamefont {Christensen}}, \bibinfo {editor} {\bibfnamefont
  {A.}~\bibnamefont {Reina}},\ and\ \bibinfo {editor} {\bibfnamefont
  {V.}~\bibnamefont {Trianni}}}\ (\bibinfo  {publisher} {Springer International
  Publishing},\ \bibinfo {address} {Cham},\ \bibinfo {year} {2018})\ pp.\
  \bibinfo {pages} {252--263}\BibitemShut {NoStop}%
\bibitem [{\citenamefont {Dimidov}\ \emph {et~al.}(2016)\citenamefont
  {Dimidov}, \citenamefont {Oriolo},\ and\ \citenamefont
  {Trianni}}]{dimidov2016}%
  \BibitemOpen
  \bibfield  {author} {\bibinfo {author} {\bibfnamefont {C.}~\bibnamefont
  {Dimidov}}, \bibinfo {author} {\bibfnamefont {G.}~\bibnamefont {Oriolo}},\
  and\ \bibinfo {author} {\bibfnamefont {V.}~\bibnamefont {Trianni}},\
  }\bibfield  {title} {\bibinfo {title} {Random walks in swarm robotics: An
  experiment with kilobots},\ }in\ \href
  {https://doi.org/https://doi.org/10.1007/978-3-319-44427-7_16} {\emph
  {\bibinfo {booktitle} {Swarm Intelligence}}},\ \bibinfo {editor} {edited by\
  \bibinfo {editor} {\bibfnamefont {M.}~\bibnamefont {Dorigo}}, \bibinfo
  {editor} {\bibfnamefont {M.}~\bibnamefont {Birattari}}, \bibinfo {editor}
  {\bibfnamefont {X.}~\bibnamefont {Li}}, \bibinfo {editor} {\bibfnamefont
  {M.}~\bibnamefont {L{\'o}pez-Ib{\'a}{\~{n}}ez}}, \bibinfo {editor}
  {\bibfnamefont {K.}~\bibnamefont {Ohkura}}, \bibinfo {editor} {\bibfnamefont
  {C.}~\bibnamefont {Pinciroli}},\ and\ \bibinfo {editor} {\bibfnamefont
  {T.}~\bibnamefont {St{\"u}tzle}}}\ (\bibinfo  {publisher} {Springer
  International Publishing},\ \bibinfo {address} {Cham},\ \bibinfo {year}
  {2016})\ pp.\ \bibinfo {pages} {185--196}\BibitemShut {NoStop}%
\bibitem [{\citenamefont {Johnstone}(1997)}]{johnstone1997}%
  \BibitemOpen
  \bibfield  {author} {\bibinfo {author} {\bibfnamefont {R.~A.}\ \bibnamefont
  {Johnstone}},\ }\bibfield  {title} {\bibinfo {title} {The evolution of animal
  signals},\ }in\ \href {https://doi.org/10.5962/p.355474} {\emph {\bibinfo
  {booktitle} {Behavioural Ecology: An Evolutionary Approach}}},\ \bibinfo
  {editor} {edited by\ \bibinfo {editor} {\bibfnamefont {J.~R.}\ \bibnamefont
  {Krebs}}\ and\ \bibinfo {editor} {\bibfnamefont {N.~B.}\ \bibnamefont
  {Davies}}}\ (\bibinfo  {publisher} {Blackwell Publishing},\ \bibinfo
  {address} {Malden, USA},\ \bibinfo {year} {1997})\ \bibinfo {edition} {4th}\
  ed.,\ Chap.~\bibinfo {chapter} {7}, pp.\ \bibinfo {pages}
  {155--178}\BibitemShut {NoStop}%
\bibitem [{Far(2020)}]{Fard2020}%
  \BibitemOpen
  \href {https://doi.org/10.1162/isal_a_00262} {\emph {\bibinfo {title}
  {{Data-Driven Modeling of Resource Distribution in Honeybee Swarms}}}},\
  \bibinfo {series} {ALIFE 2023: Ghost in the Machine: Proceedings of the 2023
  Artificial Life Conference}, Vol.\ \bibinfo {volume} {ALIFE 2020: The 2020
  Conference on Artificial Life}\ (\bibinfo {year} {2020})\BibitemShut
  {NoStop}%
\bibitem [{\citenamefont {Seeley}\ \emph {et~al.}(2012)\citenamefont {Seeley},
  \citenamefont {Visscher}, \citenamefont {Schlegel}, \citenamefont {Hogan},
  \citenamefont {Franks},\ and\ \citenamefont {Marshall}}]{seeley2012}%
  \BibitemOpen
  \bibfield  {author} {\bibinfo {author} {\bibfnamefont {T.~D.}\ \bibnamefont
  {Seeley}}, \bibinfo {author} {\bibfnamefont {P.~K.}\ \bibnamefont
  {Visscher}}, \bibinfo {author} {\bibfnamefont {T.}~\bibnamefont {Schlegel}},
  \bibinfo {author} {\bibfnamefont {P.~M.}\ \bibnamefont {Hogan}}, \bibinfo
  {author} {\bibfnamefont {N.~R.}\ \bibnamefont {Franks}},\ and\ \bibinfo
  {author} {\bibfnamefont {J.~A.~R.}\ \bibnamefont {Marshall}},\ }\bibfield
  {title} {\bibinfo {title} {Stop signals provide cross inhibition in
  collective decision-making by honeybee swarms},\ }\href
  {https://doi.org/10.1126/science.1210361} {\bibfield  {journal} {\bibinfo
  {journal} {Science}\ }\textbf {\bibinfo {volume} {335}},\ \bibinfo {pages}
  {108} (\bibinfo {year} {2012})}\BibitemShut {NoStop}%
\bibitem [{\citenamefont {Beekman}\ and\ \citenamefont
  {Oldroyd}(2018)}]{beekman2018}%
  \BibitemOpen
  \bibfield  {author} {\bibinfo {author} {\bibfnamefont {M.}~\bibnamefont
  {Beekman}}\ and\ \bibinfo {author} {\bibfnamefont {B.~P.}\ \bibnamefont
  {Oldroyd}},\ }\bibfield  {title} {\bibinfo {title} {Different bees, different
  needs: how nest-site requirements have shaped the decision-making processes
  in homeless honeybees (apis spp.)},\ }\href
  {https://doi.org/10.1098/rstb.2017.0010} {\bibfield  {journal} {\bibinfo
  {journal} {Philosophical Transactions of the Royal Society B: Biological
  Sciences}\ }\textbf {\bibinfo {volume} {373}},\ \bibinfo {pages} {20170010}
  (\bibinfo {year} {2018})}\BibitemShut {NoStop}%
\bibitem [{\citenamefont {Dyer}(2002)}]{dyer2002}%
  \BibitemOpen
  \bibfield  {author} {\bibinfo {author} {\bibfnamefont {F.~C.}\ \bibnamefont
  {Dyer}},\ }\bibfield  {title} {\bibinfo {title} {The biology of the dance
  language},\ }\href {https://doi.org/10.1146/annurev.ento.47.091201.145306}
  {\bibfield  {journal} {\bibinfo  {journal} {Annual Review of Entomology}\
  }\textbf {\bibinfo {volume} {47}},\ \bibinfo {pages} {917} (\bibinfo {year}
  {2002})}\BibitemShut {NoStop}%
\bibitem [{\citenamefont {Seeley}(1997)}]{seeley1997}%
  \BibitemOpen
  \bibfield  {author} {\bibinfo {author} {\bibfnamefont {T.~D.}\ \bibnamefont
  {Seeley}},\ }\bibfield  {title} {\bibinfo {title} {Honey bee colonies are
  group‐level adaptive units},\ }\href {https://doi.org/10.1086/286048}
  {\bibfield  {journal} {\bibinfo  {journal} {The American Naturalist}\
  }\textbf {\bibinfo {volume} {150}},\ \bibinfo {pages} {S22} (\bibinfo {year}
  {1997})}\BibitemShut {NoStop}%
\bibitem [{\citenamefont {Myerscough}(2003)}]{myerscough2003}%
  \BibitemOpen
  \bibfield  {author} {\bibinfo {author} {\bibfnamefont {M.~R.}\ \bibnamefont
  {Myerscough}},\ }\bibfield  {title} {\bibinfo {title} {Dancing for a
  decision: a matrix model for nest–site choice by honeybees},\ }\href
  {https://doi.org/10.1098/rspb.2002.2293} {\bibfield  {journal} {\bibinfo
  {journal} {Proceedings of the Royal Society of London. Series B: Biological
  Sciences}\ }\textbf {\bibinfo {volume} {270}},\ \bibinfo {pages} {577}
  (\bibinfo {year} {2003})}\BibitemShut {NoStop}%
\bibitem [{\citenamefont {Perdriau}\ and\ \citenamefont
  {Myerscough}(2007)}]{perdriau2007}%
  \BibitemOpen
  \bibfield  {author} {\bibinfo {author} {\bibfnamefont {B.~S.}\ \bibnamefont
  {Perdriau}}\ and\ \bibinfo {author} {\bibfnamefont {M.~R.}\ \bibnamefont
  {Myerscough}},\ }\bibfield  {title} {\bibinfo {title} {Making good choices
  with variable information: a stochastic model for nest-site selection by
  honeybees},\ }\href {https://doi.org/10.1098/rsbl.2006.0599} {\bibfield
  {journal} {\bibinfo  {journal} {Biology Letters}\ }\textbf {\bibinfo {volume}
  {3}},\ \bibinfo {pages} {140} (\bibinfo {year} {2007})}\BibitemShut {NoStop}%
\bibitem [{\citenamefont {Galla}(2010)}]{galla2010}%
  \BibitemOpen
  \bibfield  {author} {\bibinfo {author} {\bibfnamefont {T.}~\bibnamefont
  {Galla}},\ }\bibfield  {title} {\bibinfo {title} {Independence and
  interdependence in the nest-site choice by honeybee swarms: Agent-based
  models, analytical approaches and pattern formation},\ }\href
  {https://doi.org/https://doi.org/10.1016/j.jtbi.2009.09.007} {\bibfield
  {journal} {\bibinfo  {journal} {Journal of Theoretical Biology}\ }\textbf
  {\bibinfo {volume} {262}},\ \bibinfo {pages} {186} (\bibinfo {year}
  {2010})}\BibitemShut {NoStop}%
\bibitem [{\citenamefont {Rubenstein}\ \emph {et~al.}(2012)\citenamefont
  {Rubenstein}, \citenamefont {Ahler},\ and\ \citenamefont
  {Nagpal}}]{rubenstein2012}%
  \BibitemOpen
  \bibfield  {author} {\bibinfo {author} {\bibfnamefont {M.}~\bibnamefont
  {Rubenstein}}, \bibinfo {author} {\bibfnamefont {C.}~\bibnamefont {Ahler}},\
  and\ \bibinfo {author} {\bibfnamefont {R.}~\bibnamefont {Nagpal}},\
  }\bibfield  {title} {\bibinfo {title} {Kilobot: A low cost scalable robot
  system for collective behaviors},\ }in\ \href
  {https://doi.org/10.1109/ICRA.2012.6224638} {\emph {\bibinfo {booktitle}
  {2012 IEEE International Conference on Robotics and Automation}}}\ (\bibinfo
  {year} {2012})\ pp.\ \bibinfo {pages} {3293--3298}\BibitemShut {NoStop}%
\bibitem [{\citenamefont {Talamali}\ \emph {et~al.}(2021)\citenamefont
  {Talamali}, \citenamefont {Saha}, \citenamefont {Marshall},\ and\
  \citenamefont {Reina}}]{talamali2021_less_more}%
  \BibitemOpen
  \bibfield  {author} {\bibinfo {author} {\bibfnamefont {M.~S.}\ \bibnamefont
  {Talamali}}, \bibinfo {author} {\bibfnamefont {A.}~\bibnamefont {Saha}},
  \bibinfo {author} {\bibfnamefont {J.~A.~R.}\ \bibnamefont {Marshall}},\ and\
  \bibinfo {author} {\bibfnamefont {A.}~\bibnamefont {Reina}},\ }\bibfield
  {title} {\bibinfo {title} {When less is more: {Robot} swarms adapt better to
  changes with constrained communication},\ }\href
  {https://doi.org/10.1126/scirobotics.abf1416} {\bibfield  {journal} {\bibinfo
   {journal} {Science Robotics}\ }\textbf {\bibinfo {volume} {6}},\ \bibinfo
  {pages} {eabf1416} (\bibinfo {year} {2021})},\ \bibinfo {note} {publisher:
  American Association for the Advancement of Science}\BibitemShut {NoStop}%
\bibitem [{\citenamefont {Raoufi}\ \emph {et~al.}(2023)\citenamefont {Raoufi},
  \citenamefont {Romanczuk},\ and\ \citenamefont {Hamann}}]{raoufi2023}%
  \BibitemOpen
  \bibfield  {author} {\bibinfo {author} {\bibfnamefont {M.}~\bibnamefont
  {Raoufi}}, \bibinfo {author} {\bibfnamefont {P.}~\bibnamefont {Romanczuk}},\
  and\ \bibinfo {author} {\bibfnamefont {H.}~\bibnamefont {Hamann}},\
  }\bibfield  {title} {\bibinfo {title} {Estimation of continuous environments
  by robot swarms: Correlated networks and decision-making},\ }in\ \href
  {https://doi.org/10.1109/ICRA48891.2023.10161354} {\emph {\bibinfo
  {booktitle} {2023 IEEE International Conference on Robotics and Automation
  (ICRA)}}}\ (\bibinfo {year} {2023})\ pp.\ \bibinfo {pages}
  {5486--5492}\BibitemShut {NoStop}%
\bibitem [{\citenamefont {Gauci}\ \emph {et~al.}(2018)\citenamefont {Gauci},
  \citenamefont {Nagpal},\ and\ \citenamefont {Rubenstein}}]{gauci2018}%
  \BibitemOpen
  \bibfield  {author} {\bibinfo {author} {\bibfnamefont {M.}~\bibnamefont
  {Gauci}}, \bibinfo {author} {\bibfnamefont {R.}~\bibnamefont {Nagpal}},\ and\
  \bibinfo {author} {\bibfnamefont {M.}~\bibnamefont {Rubenstein}},\ }\bibinfo
  {title} {Programmable self-disassembly for shape formation in large-scale
  robot collectives},\ in\ \href {https://doi.org/10.1007/978-3-319-73008-0_40}
  {\emph {\bibinfo {booktitle} {Distributed Autonomous Robotic Systems: The
  13th International Symposium}}},\ \bibinfo {editor} {edited by\ \bibinfo
  {editor} {\bibfnamefont {R.}~\bibnamefont {Gro{\ss}}}, \bibinfo {editor}
  {\bibfnamefont {A.}~\bibnamefont {Kolling}}, \bibinfo {editor} {\bibfnamefont
  {S.}~\bibnamefont {Berman}}, \bibinfo {editor} {\bibfnamefont
  {E.}~\bibnamefont {Frazzoli}}, \bibinfo {editor} {\bibfnamefont
  {A.}~\bibnamefont {Martinoli}}, \bibinfo {editor} {\bibfnamefont
  {F.}~\bibnamefont {Matsuno}},\ and\ \bibinfo {editor} {\bibfnamefont
  {M.}~\bibnamefont {Gauci}}}\ (\bibinfo  {publisher} {Springer International
  Publishing},\ \bibinfo {address} {Cham},\ \bibinfo {year} {2018})\ pp.\
  \bibinfo {pages} {573--586}\BibitemShut {NoStop}%
\bibitem [{\citenamefont {Slavkov}\ \emph {et~al.}(2018)\citenamefont
  {Slavkov}, \citenamefont {Carrillo-Zapata}, \citenamefont {Carranza},
  \citenamefont {Diego}, \citenamefont {Jansson}, \citenamefont {Kaandorp},
  \citenamefont {Hauert},\ and\ \citenamefont {Sharpe}}]{slavkov2018}%
  \BibitemOpen
  \bibfield  {author} {\bibinfo {author} {\bibfnamefont {I.}~\bibnamefont
  {Slavkov}}, \bibinfo {author} {\bibfnamefont {D.}~\bibnamefont
  {Carrillo-Zapata}}, \bibinfo {author} {\bibfnamefont {N.}~\bibnamefont
  {Carranza}}, \bibinfo {author} {\bibfnamefont {X.}~\bibnamefont {Diego}},
  \bibinfo {author} {\bibfnamefont {F.}~\bibnamefont {Jansson}}, \bibinfo
  {author} {\bibfnamefont {J.}~\bibnamefont {Kaandorp}}, \bibinfo {author}
  {\bibfnamefont {S.}~\bibnamefont {Hauert}},\ and\ \bibinfo {author}
  {\bibfnamefont {J.}~\bibnamefont {Sharpe}},\ }\bibfield  {title} {\bibinfo
  {title} {Morphogenesis in robot swarms},\ }\href
  {https://doi.org/10.1126/scirobotics.aau9178} {\bibfield  {journal} {\bibinfo
   {journal} {Science Robotics}\ }\textbf {\bibinfo {volume} {3}},\ \bibinfo
  {pages} {eaau9178} (\bibinfo {year} {2018})}\BibitemShut {NoStop}%
\bibitem [{\citenamefont {Talamali}\ \emph {et~al.}(2020)\citenamefont
  {Talamali}, \citenamefont {Bose}, \citenamefont {Haire}, \citenamefont {Xu},
  \citenamefont {Marshall},\ and\ \citenamefont
  {Reina}}]{talamali2020_foraging}%
  \BibitemOpen
  \bibfield  {author} {\bibinfo {author} {\bibfnamefont {M.}~\bibnamefont
  {Talamali}}, \bibinfo {author} {\bibfnamefont {T.}~\bibnamefont {Bose}},
  \bibinfo {author} {\bibfnamefont {M.}~\bibnamefont {Haire}}, \bibinfo
  {author} {\bibfnamefont {X.}~\bibnamefont {Xu}}, \bibinfo {author}
  {\bibfnamefont {J.}~\bibnamefont {Marshall}},\ and\ \bibinfo {author}
  {\bibfnamefont {A.}~\bibnamefont {Reina}},\ }\bibfield  {title} {\bibinfo
  {title} {Sophisticated collective foraging with minimalist agents: a swarm
  robotics test},\ }\href {https://doi.org/10.1007/s11721-019-00176-9}
  {\bibfield  {journal} {\bibinfo  {journal} {Swarm Intelligence}\ }\textbf
  {\bibinfo {volume} {14}} (\bibinfo {year} {2020})}\BibitemShut {NoStop}%
\bibitem [{\citenamefont {Rubenstein}\ \emph {et~al.}(2013)\citenamefont
  {Rubenstein}, \citenamefont {Cabrera}, \citenamefont {Werfel}, \citenamefont
  {Habibi}, \citenamefont {McLurkin},\ and\ \citenamefont
  {Nagpal}}]{rubenstein2013}%
  \BibitemOpen
  \bibfield  {author} {\bibinfo {author} {\bibfnamefont {M.}~\bibnamefont
  {Rubenstein}}, \bibinfo {author} {\bibfnamefont {A.}~\bibnamefont {Cabrera}},
  \bibinfo {author} {\bibfnamefont {J.}~\bibnamefont {Werfel}}, \bibinfo
  {author} {\bibfnamefont {G.}~\bibnamefont {Habibi}}, \bibinfo {author}
  {\bibfnamefont {J.}~\bibnamefont {McLurkin}},\ and\ \bibinfo {author}
  {\bibfnamefont {R.}~\bibnamefont {Nagpal}},\ }\bibfield  {title} {\bibinfo
  {title} {Collective transport of complex objects by simple robots: Theory and
  experiments},\ }in\ \href
  {https://doi.org/https://dl.acm.org/doi/10.5555/2484920.2484932} {\emph
  {\bibinfo {booktitle} {Proceedings of the 2013 International Conference on
  Autonomous Agents and Multi-Agent Systems}}},\ \bibinfo {series and number}
  {AAMAS '13}\ (\bibinfo  {publisher} {International Foundation for Autonomous
  Agents and Multiagent Systems},\ \bibinfo {address} {Richland, SC},\ \bibinfo
  {year} {2013})\ p.\ \bibinfo {pages} {47–54}\BibitemShut {NoStop}%
\bibitem [{\citenamefont {Zion}\ \emph {et~al.}(2023)\citenamefont {Zion},
  \citenamefont {Fersula}, \citenamefont {Bredeche},\ and\ \citenamefont
  {Dauchot}}]{BenZion2023}%
  \BibitemOpen
  \bibfield  {author} {\bibinfo {author} {\bibfnamefont {M.~Y.~B.}\
  \bibnamefont {Zion}}, \bibinfo {author} {\bibfnamefont {J.}~\bibnamefont
  {Fersula}}, \bibinfo {author} {\bibfnamefont {N.}~\bibnamefont {Bredeche}},\
  and\ \bibinfo {author} {\bibfnamefont {O.}~\bibnamefont {Dauchot}},\
  }\bibfield  {title} {\bibinfo {title} {Morphological computation and
  decentralized learning in a swarm of sterically interacting robots},\ }\href
  {https://doi.org/10.1126/scirobotics.abo6140} {\bibfield  {journal} {\bibinfo
   {journal} {Science Robotics}\ }\textbf {\bibinfo {volume} {8}},\ \bibinfo
  {pages} {eabo6140} (\bibinfo {year} {2023})}\BibitemShut {NoStop}%
\bibitem [{\citenamefont {Jansson}\ \emph {et~al.}(2016)\citenamefont
  {Jansson}, \citenamefont {Hartley}, \citenamefont {Hinsch}, \citenamefont
  {Slavkov}, \citenamefont {Carranza}, \citenamefont {Olsson}, \citenamefont
  {Dries}, \citenamefont {Grönqvist}, \citenamefont {Marée}, \citenamefont
  {Sharpe}, \citenamefont {Kaandorp},\ and\ \citenamefont
  {Grieneisen}}]{jansson2015}%
  \BibitemOpen
  \bibfield  {author} {\bibinfo {author} {\bibfnamefont {F.}~\bibnamefont
  {Jansson}}, \bibinfo {author} {\bibfnamefont {M.}~\bibnamefont {Hartley}},
  \bibinfo {author} {\bibfnamefont {M.}~\bibnamefont {Hinsch}}, \bibinfo
  {author} {\bibfnamefont {I.}~\bibnamefont {Slavkov}}, \bibinfo {author}
  {\bibfnamefont {N.}~\bibnamefont {Carranza}}, \bibinfo {author}
  {\bibfnamefont {T.~S.~G.}\ \bibnamefont {Olsson}}, \bibinfo {author}
  {\bibfnamefont {R.~M.}\ \bibnamefont {Dries}}, \bibinfo {author}
  {\bibfnamefont {J.~H.}\ \bibnamefont {Grönqvist}}, \bibinfo {author}
  {\bibfnamefont {A.~F.~M.}\ \bibnamefont {Marée}}, \bibinfo {author}
  {\bibfnamefont {J.}~\bibnamefont {Sharpe}}, \bibinfo {author} {\bibfnamefont
  {J.~A.}\ \bibnamefont {Kaandorp}},\ and\ \bibinfo {author} {\bibfnamefont
  {V.~A.}\ \bibnamefont {Grieneisen}},\ }\href
  {https://doi.org/https://doi.org/10.48550/arXiv.1511.04285} {\bibinfo {title}
  {Kilombo: a kilobot simulator to enable effective research in swarm
  robotics}} (\bibinfo {year} {2016}),\ \Eprint
  {https://arxiv.org/abs/1511.04285} {arXiv:1511.04285 [cs.RO]} \BibitemShut
  {NoStop}%
\bibitem [{\citenamefont {Sumpter}(2006)}]{Sumpter2006}%
  \BibitemOpen
  \bibfield  {author} {\bibinfo {author} {\bibfnamefont {D.}~\bibnamefont
  {Sumpter}},\ }\bibfield  {title} {\bibinfo {title} {The principles of
  collective animal behaviour},\ }\href
  {https://doi.org/10.1098/rstb.2005.1733} {\bibfield  {journal} {\bibinfo
  {journal} {Philosophical Transactions of the Royal Society B: Biological
  Sciences}\ }\textbf {\bibinfo {volume} {361}},\ \bibinfo {pages} {5}
  (\bibinfo {year} {2006})}\BibitemShut {NoStop}%
\bibitem [{\citenamefont {Judd}(1994)}]{judd1994}%
  \BibitemOpen
  \bibfield  {author} {\bibinfo {author} {\bibfnamefont {T.~M.}\ \bibnamefont
  {Judd}},\ }\bibfield  {title} {\bibinfo {title} {The waggle dance of the
  honey bee: Which bees following a dancer successfully acquire the
  information?},\ }\href {https://doi.org/10.1007/BF01989363} {\bibfield
  {journal} {\bibinfo  {journal} {Journal of Insect Behavior}\ }\textbf
  {\bibinfo {volume} {8}},\ \bibinfo {pages} {343} (\bibinfo {year}
  {1994})}\BibitemShut {NoStop}%
\bibitem [{\citenamefont {March-Pons}\ \emph {et~al.}(2024)\citenamefont
  {March-Pons}, \citenamefont {Ferrero},\ and\ \citenamefont
  {Miguel}}]{marchpons2024consensus}%
  \BibitemOpen
  \bibfield  {author} {\bibinfo {author} {\bibfnamefont {D.}~\bibnamefont
  {March-Pons}}, \bibinfo {author} {\bibfnamefont {E.~E.}\ \bibnamefont
  {Ferrero}},\ and\ \bibinfo {author} {\bibfnamefont {M.~C.}\ \bibnamefont
  {Miguel}},\ }\href@noop {} {\bibinfo {title} {Consensus formation in
  quality-sensitive interdependent agent systems}} (\bibinfo {year} {2024}),\
  \Eprint {https://arxiv.org/abs/2403.14856} {arXiv:2403.14856
  [cond-mat.dis-nn]} \BibitemShut {NoStop}%
\bibitem [{\citenamefont {Reina}\ \emph
  {et~al.}(2018{\natexlab{b}})\citenamefont {Reina}, \citenamefont {Bose},
  \citenamefont {Trianni},\ and\ \citenamefont
  {Marshall}}]{reina_psychophysical_2018}%
  \BibitemOpen
  \bibfield  {author} {\bibinfo {author} {\bibfnamefont {A.}~\bibnamefont
  {Reina}}, \bibinfo {author} {\bibfnamefont {T.}~\bibnamefont {Bose}},
  \bibinfo {author} {\bibfnamefont {V.}~\bibnamefont {Trianni}},\ and\ \bibinfo
  {author} {\bibfnamefont {J.~A.~R.}\ \bibnamefont {Marshall}},\ }\bibfield
  {title} {\bibinfo {title} {Psychophysical {Laws} and the {Superorganism}},\
  }\href {https://doi.org/10.1038/s41598-018-22616-y} {\bibfield  {journal}
  {\bibinfo  {journal} {Scientific Reports}\ }\textbf {\bibinfo {volume} {8}},\
  \bibinfo {pages} {4387} (\bibinfo {year} {2018}{\natexlab{b}})},\ \bibinfo
  {note} {number: 1 Publisher: Nature Publishing Group}\BibitemShut {NoStop}%
\bibitem [{\citenamefont {Reina}\ \emph {et~al.}(2023)\citenamefont {Reina},
  \citenamefont {Zakir}, \citenamefont {De~Masi},\ and\ \citenamefont
  {Ferrante}}]{reina_cross_inhibition_2023}%
  \BibitemOpen
  \bibfield  {author} {\bibinfo {author} {\bibfnamefont {A.}~\bibnamefont
  {Reina}}, \bibinfo {author} {\bibfnamefont {R.}~\bibnamefont {Zakir}},
  \bibinfo {author} {\bibfnamefont {G.}~\bibnamefont {De~Masi}},\ and\ \bibinfo
  {author} {\bibfnamefont {E.}~\bibnamefont {Ferrante}},\ }\bibfield  {title}
  {\bibinfo {title} {Cross-inhibition leads to group consensus despite the
  presence of strongly opinionated minorities and asocial behaviour},\ }\href
  {https://doi.org/10.1038/s42005-023-01345-3} {\bibfield  {journal} {\bibinfo
  {journal} {Communications Physics}\ }\textbf {\bibinfo {volume} {6}},\
  \bibinfo {pages} {1} (\bibinfo {year} {2023})},\ \bibinfo {note} {publisher:
  Nature Publishing Group}\BibitemShut {NoStop}%
\bibitem [{\citenamefont {Peruani}\ and\ \citenamefont
  {Sibona}(2008)}]{peruani_dynamics_2008}%
  \BibitemOpen
  \bibfield  {author} {\bibinfo {author} {\bibfnamefont {F.}~\bibnamefont
  {Peruani}}\ and\ \bibinfo {author} {\bibfnamefont {G.~J.}\ \bibnamefont
  {Sibona}},\ }\bibfield  {title} {\bibinfo {title} {Dynamics and {Steady}
  {States} in {Excitable} {Mobile} {Agent} {Systems}},\ }\href
  {https://doi.org/10.1103/PhysRevLett.100.168103} {\bibfield  {journal}
  {\bibinfo  {journal} {Physical Review Letters}\ }\textbf {\bibinfo {volume}
  {100}},\ \bibinfo {pages} {168103} (\bibinfo {year} {2008})},\ \bibinfo
  {note} {publisher: American Physical Society}\BibitemShut {NoStop}%
\bibitem [{\citenamefont {Levis}\ \emph {et~al.}(2017)\citenamefont {Levis},
  \citenamefont {Pagonabarraga},\ and\ \citenamefont
  {Díaz-Guilera}}]{levis_synchronization_2017}%
  \BibitemOpen
  \bibfield  {author} {\bibinfo {author} {\bibfnamefont {D.}~\bibnamefont
  {Levis}}, \bibinfo {author} {\bibfnamefont {I.}~\bibnamefont
  {Pagonabarraga}},\ and\ \bibinfo {author} {\bibfnamefont {A.}~\bibnamefont
  {Díaz-Guilera}},\ }\bibfield  {title} {\bibinfo {title} {Synchronization in
  {Dynamical} {Networks} of {Locally} {Coupled} {Self}-{Propelled}
  {Oscillators}},\ }\href {https://doi.org/10.1103/PhysRevX.7.011028}
  {\bibfield  {journal} {\bibinfo  {journal} {Physical Review X}\ }\textbf
  {\bibinfo {volume} {7}},\ \bibinfo {pages} {011028} (\bibinfo {year}
  {2017})},\ \bibinfo {note} {publisher: American Physical Society}\BibitemShut
  {NoStop}%
\bibitem [{\citenamefont {Starnini}\ \emph {et~al.}(2016)\citenamefont
  {Starnini}, \citenamefont {Frasca},\ and\ \citenamefont
  {Baronchelli}}]{starnini_emergence_2016}%
  \BibitemOpen
  \bibfield  {author} {\bibinfo {author} {\bibfnamefont {M.}~\bibnamefont
  {Starnini}}, \bibinfo {author} {\bibfnamefont {M.}~\bibnamefont {Frasca}},\
  and\ \bibinfo {author} {\bibfnamefont {A.}~\bibnamefont {Baronchelli}},\
  }\bibfield  {title} {\bibinfo {title} {Emergence of metapopulations and echo
  chambers in mobile agents},\ }\href {https://doi.org/10.1038/srep31834}
  {\bibfield  {journal} {\bibinfo  {journal} {Scientific Reports}\ }\textbf
  {\bibinfo {volume} {6}},\ \bibinfo {pages} {31834} (\bibinfo {year}
  {2016})},\ \bibinfo {note} {publisher: Nature Publishing Group}\BibitemShut
  {NoStop}%
\bibitem [{\citenamefont {Paoluzzi}\ \emph {et~al.}(2020)\citenamefont
  {Paoluzzi}, \citenamefont {Leoni},\ and\ \citenamefont
  {Marchetti}}]{paoluzzi_information_2020}%
  \BibitemOpen
  \bibfield  {author} {\bibinfo {author} {\bibfnamefont {M.}~\bibnamefont
  {Paoluzzi}}, \bibinfo {author} {\bibfnamefont {M.}~\bibnamefont {Leoni}},\
  and\ \bibinfo {author} {\bibfnamefont {M.~C.}\ \bibnamefont {Marchetti}},\
  }\bibfield  {title} {\bibinfo {title} {Information and motility exchange in
  collectives of active particles},\ }\href
  {https://doi.org/10.1039/D0SM00204F} {\bibfield  {journal} {\bibinfo
  {journal} {Soft Matter}\ }\textbf {\bibinfo {volume} {16}},\ \bibinfo {pages}
  {6317} (\bibinfo {year} {2020})},\ \bibinfo {note} {publisher: The Royal
  Society of Chemistry}\BibitemShut {NoStop}%
\bibitem [{\citenamefont {Newman}(2010)}]{Newman10}%
  \BibitemOpen
  \bibfield  {author} {\bibinfo {author} {\bibfnamefont {M.}~\bibnamefont
  {Newman}},\ }\href
  {https://doi.org/10.1093/acprof:oso/9780199206650.001.0001} {\emph {\bibinfo
  {title} {Networks: An Introduction}}}\ (\bibinfo  {publisher} {Oxford
  University Press, Inc.},\ \bibinfo {address} {USA},\ \bibinfo {year}
  {2010})\BibitemShut {NoStop}%
\bibitem [{\citenamefont {Barthélemy}(2011)}]{BARTHELEMY20111}%
  \BibitemOpen
  \bibfield  {author} {\bibinfo {author} {\bibfnamefont {M.}~\bibnamefont
  {Barthélemy}},\ }\bibfield  {title} {\bibinfo {title} {Spatial networks},\
  }\href {https://doi.org/https://doi.org/10.1016/j.physrep.2010.11.002}
  {\bibfield  {journal} {\bibinfo  {journal} {Physics Reports}\ }\textbf
  {\bibinfo {volume} {499}},\ \bibinfo {pages} {1} (\bibinfo {year}
  {2011})}\BibitemShut {NoStop}%
\bibitem [{\citenamefont {Stauffer}\ and\ \citenamefont
  {Aharony}(2018)}]{stauffer2018}%
  \BibitemOpen
  \bibfield  {author} {\bibinfo {author} {\bibfnamefont {D.}~\bibnamefont
  {Stauffer}}\ and\ \bibinfo {author} {\bibfnamefont {A.}~\bibnamefont
  {Aharony}},\ }\href {https://doi.org/https://doi.org/10.1201/9781315274386}
  {\emph {\bibinfo {title} {Introduction to percolation theory}}}\ (\bibinfo
  {publisher} {Taylor \& Francis},\ \bibinfo {year} {2018})\BibitemShut
  {NoStop}%
\bibitem [{\citenamefont {Aust}\ \emph {et~al.}(2022)\citenamefont {Aust},
  \citenamefont {Talamali}, \citenamefont {Dorigo}, \citenamefont {Hamann},\
  and\ \citenamefont {Reina}}]{aust_hidden_2022}%
  \BibitemOpen
  \bibfield  {author} {\bibinfo {author} {\bibfnamefont {T.}~\bibnamefont
  {Aust}}, \bibinfo {author} {\bibfnamefont {M.~S.}\ \bibnamefont {Talamali}},
  \bibinfo {author} {\bibfnamefont {M.}~\bibnamefont {Dorigo}}, \bibinfo
  {author} {\bibfnamefont {H.}~\bibnamefont {Hamann}},\ and\ \bibinfo {author}
  {\bibfnamefont {A.}~\bibnamefont {Reina}},\ }\bibfield  {title} {\bibinfo
  {title} {The {Hidden} {Benefits} of {Limited} {Communication} and {Slow}
  {Sensing} in {Collective} {Monitoring} of {Dynamic} {Environments}},\ }in\
  \href {https://doi.org/10.1007/978-3-031-20176-9_19} {\emph {\bibinfo
  {booktitle} {Swarm {Intelligence}}}},\ \bibinfo {series and number} {Lecture
  {Notes} in {Computer} {Science}},\ \bibinfo {editor} {edited by\ \bibinfo
  {editor} {\bibfnamefont {M.}~\bibnamefont {Dorigo}}, \bibinfo {editor}
  {\bibfnamefont {H.}~\bibnamefont {Hamann}}, \bibinfo {editor} {\bibfnamefont
  {M.}~\bibnamefont {López-Ibáñez}}, \bibinfo {editor} {\bibfnamefont
  {J.}~\bibnamefont {García-Nieto}}, \bibinfo {editor} {\bibfnamefont
  {A.}~\bibnamefont {Engelbrecht}}, \bibinfo {editor} {\bibfnamefont
  {C.}~\bibnamefont {Pinciroli}}, \bibinfo {editor} {\bibfnamefont
  {V.}~\bibnamefont {Strobel}},\ and\ \bibinfo {editor} {\bibfnamefont
  {C.}~\bibnamefont {Camacho-Villalón}}}\ (\bibinfo  {publisher} {Springer
  International Publishing},\ \bibinfo {address} {Cham},\ \bibinfo {year}
  {2022})\ pp.\ \bibinfo {pages} {234--247}\BibitemShut {NoStop}%
\bibitem [{sup()}]{suppmat}%
  \BibitemOpen
  \href@noop {} {}\bibinfo {howpublished}
  {\url{URL_will_be_inserted_by_publisher}},\ \bibinfo {note} {supplementary
  material for this work which contains videos of our experiments}\BibitemShut
  {NoStop}%
\bibitem [{kil()}]{kilocounter}%
  \BibitemOpen
  \href@noop {} {\bibinfo {title} {Kilocounter}},\ \bibinfo {howpublished}
  {\url{https://github.com/ivan-paz/kiloColors/blob/main/RGBKiloCounter/}},\
  \bibinfo {note} {blob detection and color counting software.}\BibitemShut
  {Stop}%
\bibitem [{\citenamefont {Mertens}\ and\ \citenamefont
  {Moore}(2012)}]{mertens2012}%
  \BibitemOpen
  \bibfield  {author} {\bibinfo {author} {\bibfnamefont {S.}~\bibnamefont
  {Mertens}}\ and\ \bibinfo {author} {\bibfnamefont {C.}~\bibnamefont
  {Moore}},\ }\bibfield  {title} {\bibinfo {title} {Continuum percolation
  thresholds in two dimensions},\ }\href
  {https://doi.org/10.1103/PhysRevE.86.061109} {\bibfield  {journal} {\bibinfo
  {journal} {Phys. Rev. E}\ }\textbf {\bibinfo {volume} {86}},\ \bibinfo
  {pages} {061109} (\bibinfo {year} {2012})}\BibitemShut {NoStop}%
\bibitem [{\citenamefont {Molloy}\ and\ \citenamefont
  {Reed}(1995)}]{Molloy1995}%
  \BibitemOpen
  \bibfield  {author} {\bibinfo {author} {\bibfnamefont {M.}~\bibnamefont
  {Molloy}}\ and\ \bibinfo {author} {\bibfnamefont {B.}~\bibnamefont {Reed}},\
  }\bibfield  {title} {\bibinfo {title} {A critical point for random graphs
  with a given degree sequence},\ }\href
  {https://doi.org/https://doi.org/10.1002/rsa.3240060204} {\bibfield
  {journal} {\bibinfo  {journal} {Random Structures \& Algorithms}\ }\textbf
  {\bibinfo {volume} {6}},\ \bibinfo {pages} {161} (\bibinfo {year}
  {1995})}\BibitemShut {NoStop}%
\end{thebibliography}


\end{document}